\documentclass[preprint]{aastex6}
\newcommand{\degree}{\ensuremath{\,^\circ}}

\newcommand{\gyr}{\ensuremath{\,{\rm Gyr}}}
\newcommand{\khz}{\ensuremath{\,{\rm kHz}}}
\newcommand{\mhz}{\ensuremath{\,{\rm MHz}}}
\newcommand{\ghz}{\ensuremath{\,{\rm GHz}}}

\newcommand{\K}{\ensuremath{\,{\rm K}}}
\newcommand{\mk}{\ensuremath{\,{\rm mK}}}
\newcommand{\m}{\ensuremath{\,{\rm m}}}

\newcommand{\percc}{\ensuremath{\,{\rm cm^{-3}}}}

\newcommand{\kpc}{\ensuremath{\,{\rm kpc}}}

\newcommand{\kms}{\ensuremath{\,{\rm km\, s^{-1}}}}
\newcommand{\msun}{\ensuremath{\,M_\odot}}

\newcommand{\hr}{\,hr}

\newcommand{\jy}{\,Jy}
\newcommand{\jyb}{\ensuremath{\rm \,Jy\,beam^{-1}}}

\newcommand{\mjyb}{\ensuremath{\rm \,mJy\,beam^{-1}}}

\newcommand{\te}{\ensuremath{T_{\rm e}}}

\newcommand{\Ne}{\ensuremath{n_{\rm e}}}

\newcommand{\hrrl}[2]{H{#1#2}\ensuremath{\alpha}}
\newcommand{\herrl}[2]{He{#1#2}\ensuremath{\alpha}}
\newcommand{\expo}[1]{\ensuremath{10^{#1}}}
\newcommand{\nexpo}[2]{\ensuremath{#1 \times 10^{#2}}}

\newcommand{\hii}{H\,{\sc ii}}

\newcommand{\deut}{\ensuremath{^{2}{\rm H}}}
\newcommand{\he}[1]{\ensuremath{^{#1}{\rm He}}}
\newcommand{\hep}[1]{\ensuremath{^{#1}{\rm He}^{+}}}

\newcommand{\li}[1]{\ensuremath{{}^#1{\rm Li}}}

\newcommand{\her}[1]{\ensuremath{{}^#1{\rm He}/{\rm H}}}
\newcommand{\hepr}[1]{\ensuremath{{}^#1{\rm He}^{+}/{\rm H^+}}}
\newcommand{\heppr}[1]{\ensuremath{{}^#1{\rm He}^{++}/{\rm H^+}}}
\newcommand{\cratio}{\ensuremath{{}^{12}{\rm C}/^{13}{\rm C}}}
\newcommand{\threec}[1]{3C\thinspace #1}
\newcommand{\ic}[1]{IC\thinspace #1}
\newcommand{\ngc}[1]{NGC\thinspace #1}

\newcommand{\lsim}{\ensuremath{\lesssim}}
\newcommand\urltilda{\kern -.15em\lower .7ex\hbox{\~{}}\kern .04em}

\begin{document}

\title{Green Bank Telescope Observations of \hep3: \hii\ Regions}

\author{Dana S. Balser\altaffilmark{1} \& T. M. Bania\altaffilmark{2}}

\altaffiltext{1}{National Radio Astronomy Observatory, 520 Edgemont
  Rd., Charlottesville, VA 22903, USA.}

\altaffiltext{2}{Institute for Astrophysical Research, Astronomy
  Department, Boston University, 725 Commonwealth Ave., Boston, MA
  02215, USA.}

\begin{abstract}

  During the era of primordial nucleosynthesis the light elements
  \deut, \he3, \he4, and \li7\ were produced in significant amounts
  and these abundances have since been modified primarily by stars.
  Observations of \hep3\ in \hii\ regions located throughout the Milky
  Way disk reveal very little variation in the \her3\ abundance
  ratio---the ``\he3\ Plateau''---indicating that the net effect of
  \he3\ production in stars is negligible.  This is in contrast to
  much higher \her3\ abundance ratios found in some planetary
  nebulae. This discrepancy is known as the ``\he3\ Problem''.
  Stellar evolution models that include thermohaline mixing can
  resolve the \he3\ Problem by drastically reducing the net \he3\
  production in most stars.  These models predict a small negative
  \her3\ abundance gradient across the Galactic disk.  Here we use the
  Green Bank Telescope to observe \hep3\ in five \hii\ regions with
  high accuracy to confirm the predictions of stellar and Galactic
  chemical evolution models that include thermohaline mixing.  We
  detect \hep3\ in all the sources and derive the \hepr3\ abundance
  ratio using model \hii\ regions and the numerical radiative transfer
  code NEBULA.  The over 35 radio recombination lines (RRLs) that are
  simultaneously observed, together with the \hep3\ transition provide
  stringent constraints for these models. We apply an ionization
  correction using observations of \he4\ RRLs.  We determine a \her3\
  abundance gradient as a function of Galactocentric radius of $-0.116
  \pm\ 0.022\, \times \,$\expo{-5}$\,$kpc$^{-1}$, consistent with
  stellar evolution models including thermohaline mixing that predict
  a small net contribution of \he3\ from solar mass stars.

\end{abstract}


\section{The \he3\ Problem}\label{sec:intro}

Standard stellar evolution models\footnote{Here we define ``standard''
  stellar evolution models as those that only consider convection as a
  physical process to mix material inside of stars.} predict the
production of significant amounts of \he3\ in low-mass stars ($M <
3$\msun), with peak abundances of $^{3}{\rm He/H} \sim \nexpo{\rm
  few}{-3}$ by number \citep{iben67a, iben67b, rood72}.  As the star
ascends the red giant branch, the convective zone subsumes the
enriched material which is expected to be expelled into the
interstellar medium (ISM) via stellar winds and planetary nebulae
\citep{rood76, vassiliadis93, dearborn96, weiss96, forestini97}.
Using yields from these standard stellar models, \citet{rood76}
predicted present day abundances of $^{3}{\rm He/H} \sim
\nexpo{4}{-5}$ which they interpreted as stemming from more of a
stellar than primordial origin.

Early measurements of \hep3\ in Galactic \hii\ regions found large
source-to-source variations, $^{3}{\rm He/H} = \nexpo{1-15}{-5}$, that
were difficult to reconcile with Galactic chemical evolution
\citep{rood79, rood84, bania87, balser94}.  Deriving abundance ratios
is a two-step process: (1) accurately measuring the spectral lines of
interest; and (2) calculating the abundance ratio which may require a
model of the source.  \citet{balser99a} showed that the most accurate
\her3\ abundances could be determined from nebulae that are
morphologically simple; that is, \hii\ regions with a homogeneous
density.  Using only these ``simple'' sources, \citet{bania02} found
that the \her3\ abundance was relatively constant across the Galactic
disk revealing a ``\he3\ Plateau''.  They suggested that the net
production/destruction of \he3\ by stars is close to zero and that the
\he3\ Plateau level corresponded to the primordial abundance produced
during Big Bang Nucleosynthesis (BBN).  This was later confirmed by
combining results from the {\it Wilkinson Microwave Anisotropy Probe}
(WMAP) with BBN models resulting in a primordial abundance of
$(^{3}{\rm He/H})_{\rm p} = \nexpo{1.00 \pm\ 0.07}{-5}$
\citep{romano03, cyburt08}.

Galactic chemical evolution (GCE) models assuming standard stellar
yields predict significantly larger \her3\ abundance ratios over the
history of the Galaxy than are observed in \hii\ regions
\citep{galli95, galli97, olive95}.  The \her3\ abundance ratio should
increase with time and be higher in locations with more star
formation.  Most models therefore predict a negative \her3\ radial
abundance gradient within the Galactic disk since the star formation
rate is higher in the central regions of the Milky Way.  Thus the
inner Galaxy should have substantially more stellar processing than
the outer disk.  Detection of \hep3\ in a few planetary nebulae (PNe)
yielded abundances of $^{3}{\rm He/H} \sim \expo{-3}$, consistent with
standard stellar evolution theory \citep{rood92, balser97, balser06b}.
But the \he3\ Plateau revealed by \hii\ region observations is
inconsistent with this picture.  Moreover, in-situ measurements of the
Jovian atmosphere with the {\it Galileo} probe yielded $^{3}{\rm
  He}/^{4}{\rm He} = \nexpo{1.66 \pm 0.05}{-4}$ \citep{mahaffy98}.
This corresponds to a protosolar abundance of $^{3}{\rm He/H} =
\nexpo{1.5 \pm 0.2}{-5}$, indicating very little production of \he3\
over the past 4.5\gyr.  These discrepancies are called the ``The \he3\
Problem'' \citep[e.g.,][]{galli97}.

\citet{rood84} suggested that some extra-mixing process may reduce the
\he3\ abundance, and this might also explain the depletion of \li7\ in
main-sequence stars and the low \cratio\ abundance ratios in low-mass
red giant branch (RGB) stars \citep[also see][]{hogan95, charbonnel95,
  weiss96}.  \citet{sweigart79} proposed that meridional circulation
on the RGB could lead to reduced \cratio\ ratios in field stars.
\citet{boothroyd99} developed an ``ad hoc'' mixing mechanism in
low-mass stars to further process \he3.  GCE models that included this
extra-mixing in about 90\% of low-mass stars were shown to be
consistent with observations \citep{galli97, tosi98, palla00,
  chiappini02}.  \citet{zahn92} developed a more consistent theory
including the interaction between meridional circulation and
turbulence in rotating, non-magnetic stars.  \citet{charbonnel95} used
this rotation-induced mixing and showed that this could explain the
\cratio, \li7, and \he3\ anomalies in RGB stars. \citet{charbonnel98a}
argued that this rotationally induced extra mixing occurs in low-mass
stars above the luminosity function bump produced when the hydrogen
burning shell crosses the chemical discontinuity left by the
retreating convective envelope early in the RGB phase.  Furthermore,
about 96\% of low-mass RGB field or cluster stars have anomalously low
\cratio\ ratios \citep{charbonnel98b}.  Unfortunately, more realistic
stellar evolution models that treat the transport of angular momentum
by meridional circulation and shear turbulence self consistently do
not produce enough mixing around the luminosity bump to account for
the observed surface abundance variations \citep{palacios06}.

\citet{eggleton06} described another type of mixing that was important
by modeling a red giant in three-dimensions, whereby the \he3(\he3,
2p)\he4\ reaction creates a molecular weight inversion.  They
interpreted this mixing as a Rayleigh-Taylor instability just above
the hydrogen-burning shell.  This convective instability occurs when
heavier material lies above lighter material. In contrast,
\citet{charbonnel07a} interpreted this mixing as a thermohaline
instability, a double-diffusive instability, that occurs in oceans and
is also called thermohaline convection \citep{stern60}.  As the
molecular weight gradient increases, the temperature has a stabilizing
effect since the time scale for thermal diffusion is shorter than the
time it takes for the material to mix.  This analysis developed by
\citet{charbonnel07a} explains the \cratio\ abundance anomalies on the
RGB and the \he3\ abundance ratios observed in \hii\ regions
\citep[but also see][]{ denissenkov10, denissenkov11, henkel17}.
Currently, the best stellar evolutionary models that include both the
thermohaline instability and rotation-induced mixing were developed
for low and intermediate-mass stars \citep{charbonnel10, lagarde11}.
\citet{lagarde12} used these yields together with GCE models to
predict a modest enrichment of \he3\ with time and \her3\ abundance
ratios about a factor of two higher in the central regions of the
Milky Way relative to the outer regions.

\section{GBT Observations and Data Reduction}\label{sec:obs}

\subsection{H\,\textsl{\textsc{ii}} Region Sample}\label{sec:sample}

Our goal is to derive accurate \her3\ abundance ratios for a sub-set
of our \hii\ region sources to confirm the slight radial \he3\
gradient predicted by \citet{lagarde12}.  \hii\ region models have
shown that morphologically simple sources yield the most accurate
\her3\ abundance ratio determinations \citep{balser99a, bania07}.
They are nebulae that are well approximated by a uniform density
sphere.  We therefore selected five \hii\ regions from the sample in
\citet{bania02}, a sample that was chosen to be morphologically
simple, have relatively bright \hep3\ lines, and are located over a
range of Galactocentric radii.  Table~\ref{tab:prop} summarizes this
sample of \hii\ regions and lists the source name, equatorial
coordinates, local standard of rest (LSR) \footnote{Here we use the
  kinematic LSR defined by a solar motion of 20.0\kms\ toward
  ($\alpha$, $\delta$) = ($18^{\rm h}$, $+30$\degree) [1900.0]
  \citep{gordon76}.}  velocity, $V_{\rm LSR}$, Heliocentric distance,
$D_{\rm sun}$, and Galactocentric radius, $R_{\rm gal}$.

\begin{deluxetable}{lrrrrr}
\tablecaption{Galactic \hii\ Region Properties \label{tab:prop}}
\tablehead{
\colhead{} & \colhead{R.A. (J2000)} & \colhead{Decl. (J2000)} &
\colhead{$V_{\rm LSR}$} & \colhead{$D_{\rm sun}$} & \colhead{$R_{\rm gal}$} \\
\colhead{Source} & \colhead{(hh:mm:ss.ss)} & \colhead{(dd:mm:ss)} &
\colhead{(\kms)} & \colhead{(kpc)} & \colhead{(kpc)}
}
\startdata
S206       & 04:03:15.87 &   $+$51:18:54 & $-$25.4 & 3.3 & 11.5 \\
S209       & 04:11:06.74 &   $+$51:09:44 & $-$49.3 & 8.2 & 16.2 \\
M16        & 18:18:52.65 &   $-$13:50:05 & $+$26.3 & 2.0 &  6.6 \\  
G29.9      & 18:46:09.28 &   $-$02:41:47 & $+$96.7 & 5.8 &  4.4 \\ 
\ngc{7538} & 23:13:32.05 &   $+$61:30:12 & $-$59.9 & 2.8 &  9.9 \\
\enddata
%\tablecomments{} 
%\tablenotetext{a}{Continuum flux density at a frequency of 8.7\ghz.}
\end{deluxetable}

\subsection{Data Acquisition}\label{sec:acq}

We made observations of the \hep3\ spectral transition with the Green
Bank Telescope (GBT) at X-band (8-10\ghz) between 2012 March 02 and
2012 August 10 (GBT/12A-114).  The GBT half-power beam-width (HPBW) is
87\arcsec\ at the \hep3\ spectral transition frequency of 8665.65\mhz.
We employ total power position switching by observing an Off position
for 6 minutes and then the target (On) position for 6 minutes, for a
total time of 12 minutes.  The Off position is offset 6 minutes in
R.A. relative to the On position so that the telescope tracks the same
sky path.  The GBT auto-correlation spectrometer (ACS) is configured
with 8 spectral windows (SPWs) at two orthogonal, circular
polarizations for a total of 16 SPWs.  Each SPW had a bandwidth of
50\mhz\ and a spectral resolution of 12.2\khz, or a velocity
resolution of 0.42\kms\ at 8665.65\mhz.  Thus each total power On/Off
pair consists of 16 independent spectra.  We placed the \hep3\
transition in 8 SPWs (4 tunings at 2 polarizations) with center
frequencies: 8665.3, 8662.3, 8659.3, and 8656.3\mhz.  The \hep3\ line
was thus shifted by 3\mhz\ in each tuning.  The goal is to reduce spectral
baseline structure by averaging the four spectra since the detailed
baseline structure is a function of the center frequency (see
\S{\ref{sec:baseline}} for details).  We also tuned to various RRLs by
centering the remaining SPWs to 8586.56, 8440.0, 8918.0, and
8474.0\mhz.  This tuning strategy includes a series of high-order RRLs
(e.g., Hn$\alpha$, Hn$\beta$, Hn$\gamma$, etc.) that can be used to
monitor system performance and to constrain \hii\ region models.  It
also includes adjacent RRLs (e.g., H114$\beta$ and H115$\beta$) for
redundancy.

Project GBT/12A-114, consisting of 93 observing sessions, was designed
to be a filler project where scheduling blocks of a few hours could be
efficiently used at 9\ghz\ to accumulate the integration time needed
to detect the weak \hep3\ transition with a good signal-to-noise ratio
(SNR).  The observations were performed by the GBT operators and
inspected by the authors within days of the observations.  The
pointing and focus were updated every 2\hr\ by observing a calibrator
located within 15\degree\ of the target position.  Noise was injected
into the signal path with an intensity of 5-10\% of the total system
temperature ($T_{\rm sys}$) to calibrate the intensity scale.  For the
GBT X-band system the noise diodes are $T_{\rm cal} \sim 2$\K.  We
observed the flux density calibrator 3C286 using the digital continuum
receiver (DCR) to check the flux density calibration.  We use the
\citet{peng00} flux densities for 3C286 and assume a telescope gain of
$2\,{\rm K}\,{\rm Jy}^{-1}$ \citep{ghigo01}.  We deem the intensity
scale to be accurate to within 5-10\%.

\subsection{Data Reduction and Analysis}\label{sec:reduce}

The data were reduced and analyzed using the single-dish software
package TMBIDL \citep{bania16}.\footnote{V7.0, see
  https://github.com/tvwenger/tmbidl.}  Each spectrum was visually
inspected and discarded if significant spectral baseline structure or
radio frequency interference (RFI) was present.  Narrow band RFI that
did not contaminate the spectral lines was excised and included in the
average.  Spectra were averaged in a hierarchical way to assess any
problems with the spectral baselines.  For example, we performed the
following tests: (1) inspected the average spectrum for each observing
epoch to search for any anomalies; (2) divided the entire data set
into several groups to make sure the noise was integrating down as
expected; (3) compared the two orthogonal polarizations which should
be similar; and (4) compared the different \hep3\ SPWs.

Each SPW was divided into sub-bands with bandwidths between 10-25\mhz\
to better fit the spectral baselines and to directly compare with
\hii\ region models.  Table~\ref{tab:bands} lists the properties of
the 8 distinct sub-bands.  The main spectral transition is given
together with the center frequency, bandwidth, and all the other
transitions within the sub-band.  The spectral baseline was modeled by
fitting a polynomial, typically of order 3-5, to subtract any sky
continuum emission or baseline structure from the spectrum.  Each
spectrum was smoothed to a velocity resolution of 3\kms.  Spectral
line profiles were fit by a Gaussian function using a
Levenberg-Markwardt \citep{markwardt09} least squares method to derive
the peak intensity, the full-width at half-maximum (FWHM) line width,
and the LSR velocity.

\begin{deluxetable}{lccl}
\tablecaption{Spectra Line Sub-bands \label{tab:bands}}
\tablehead{
\colhead{Main} & \colhead{Rest Freq.} & \colhead{Bandwidth} & \colhead{} \\
\colhead{Transition} & \colhead{(MHz)} & \colhead{(MHz)} & \colhead{Other Transitions}
}
\startdata
\hep3\         & 8665.65 & 15.0 & H171$\eta$, H213$\xi$, H222$\pi$   \\
H91$\alpha$    & 8584.82 & 20.0 & H154$\epsilon$, H179$\theta$, H198$\lambda$, H227$\rho$, H231$\sigma$, H249$\psi$ \\
H114$\beta$    & 8649.10 & 15.0 & H203$\mu$, H238$\nu$, H245$\chi$ \\
H115$\beta$    & 8427.32 & 25.0 & H155$\epsilon$, H187$\iota$, H210$\nu$, H215$\xi$ \\
H130$\gamma$   & 8678.12 & 10.0 & H208$\nu$, H234$\tau$ \\
H131$\gamma$   & 8483.08 & 25.0 & H164$\zeta$, H193$\kappa$, H199$\lambda$, H228$\rho$, H232$\sigma$, H236$\tau$ \\
H144$\delta$   & 8455.38 & 10.0 & H180$\theta$, H247$\chi$ \\
H152$\epsilon$ & 8920.33 & 15.0 & H206$\nu$, H224$\rho$, H228$\sigma$, H239$\phi$ \\
\enddata
%\tablecomments{} 
%\tablenotetext{a}{}
\end{deluxetable}

\subsection{Spectral Baseline Structure}\label{sec:baseline}

The limiting factor in the spectral sensitivity of most single-dish
telescopes is instrumental baseline structure caused primarily by
reflections from the super structure (e.g, secondary focus, feed legs,
etc.) that produce standing waves within the spectrum.  The properties
of these standing waves are a function of the total system noise and
frequency and thus they are difficult to model.  \citet{balser94}
showed that since the phase of the standing waves depends on the sky
frequency, averaging observations of the target over different
observing {\it seasons} will reduce, but not eliminate, these baseline
effects.  This is because the Earth's orbital motion shifts the
observed sky frequency and, thus, the phase of the standing waves.

The GBT was specifically designed with a clear aperture to
significantly reduce reflections from the secondary structure and
therefore improve the image fidelity and spectral sensitivity.
Nevertheless, baseline structure still, unfortunately, exists and is
primarily located within the electronics
\citep{fisher03}.\footnote{See
  http://library.nrao.edu/public/memos/edir/EDIR\_312.pdf.}  The GBT
X-band receiver is a heterodyne receiver wherein radio waves from the
sky are mixed with a local oscillator (LO) to convert the signal to an
intermediate frequency (IF).  Unfortunately, the GBT IF system
contains analog signals over a long path length, $\sim 1\,$mile, and
across many electronic components that can produce reflections and,
therefore, standing waves.  The spectral baselines are significantly
better for the GBT than for traditionally designed, on-axis telescopes
but they are still the limiting factor in measuring accurate line
parameters for weak, broad spectral lines.

Using a strategy similar to \citet{balser94}, we simultaneously
observed the \hep3\ line in four SPWs, shifting the center sky
frequency relative to the center IF frequency by 3, 6, and 9\mhz.  The
goal was to reduce the baseline structure by averaging these four
SPWs.  To test the efficacy of this approach, we observed two
calibrators \threec{286} and \threec{84} with 9\ghz\ flux densities of
$\sim 5$\jy\ and 25\jy, respectively.  The telescope gain is $2\,{\rm
  K}\,{\rm Jy}^{-1}$, so these flux densities correspond to continuum
antenna temperatures of $T_{\rm c} = 10$\K\ and 50\K, respectively.
{\it N.B., we chose these bright calibrators to amplify any
  instrumental spectral artifacts and therefore the baseline structure
  is worse than in our target sources which have continuum antenna
  temperatures ${\it < 10\,}$K.}  The observing procedures and ACS
configuration were the same as the target observations.  The advantage
of using these extragalactic sources is that they do not have any
measurable spectral lines at these frequencies and have a relatively
flat spectrum across the 50\mhz\ bandwidth.

\begin{figure}
\includegraphics[angle=90,scale=0.35]{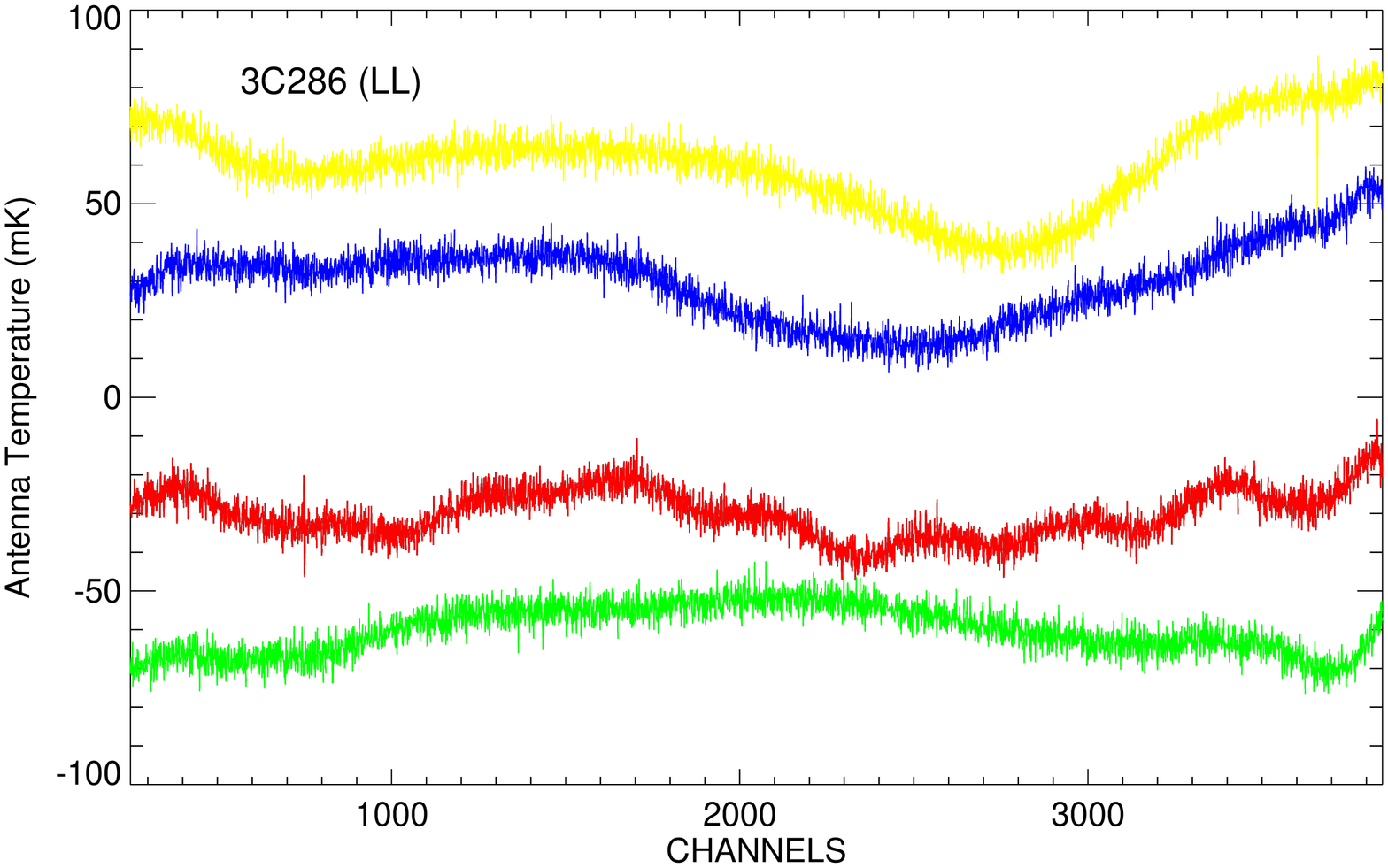} 
\includegraphics[angle=90,scale=0.35]{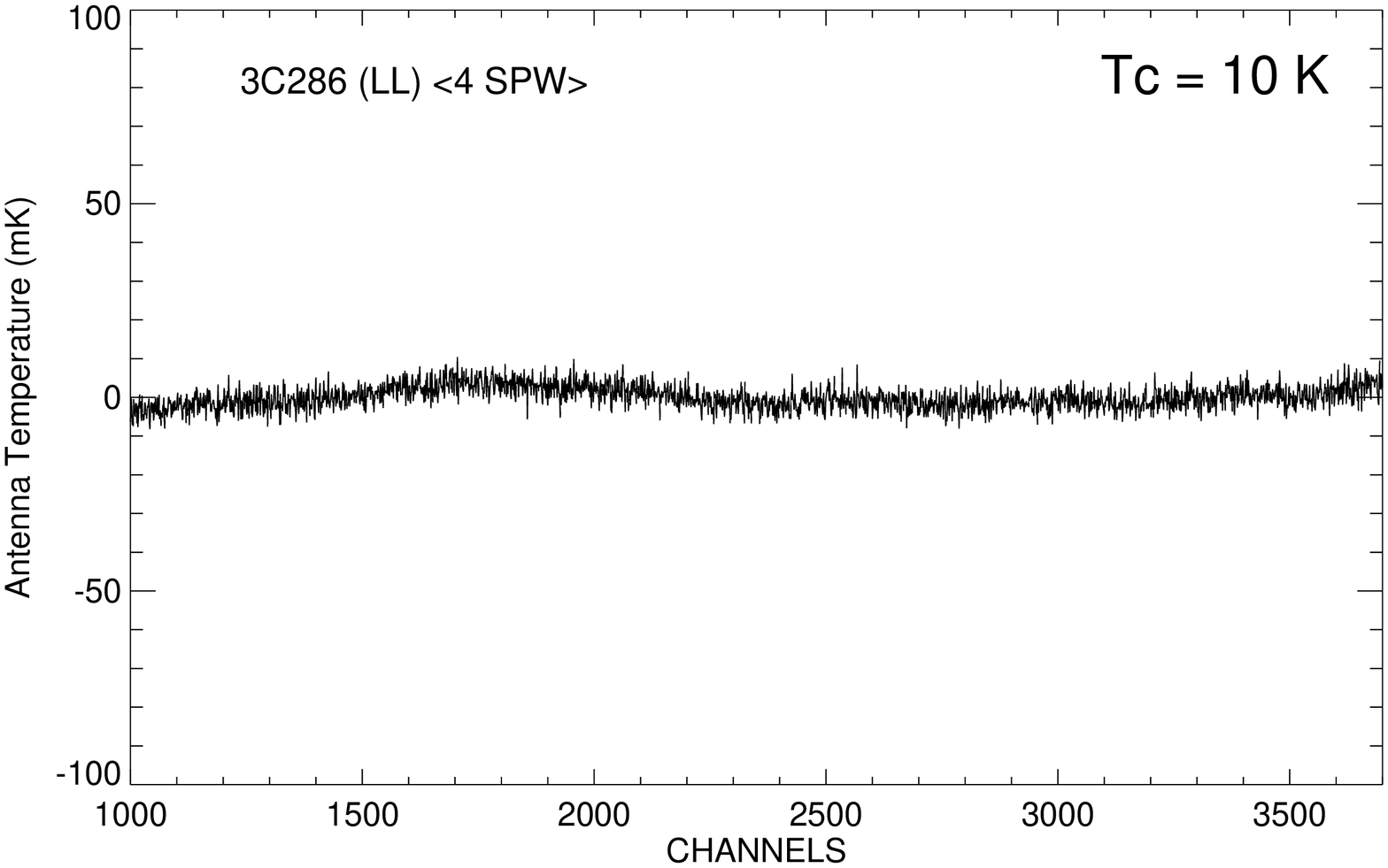} 
\includegraphics[angle=90,scale=0.35]{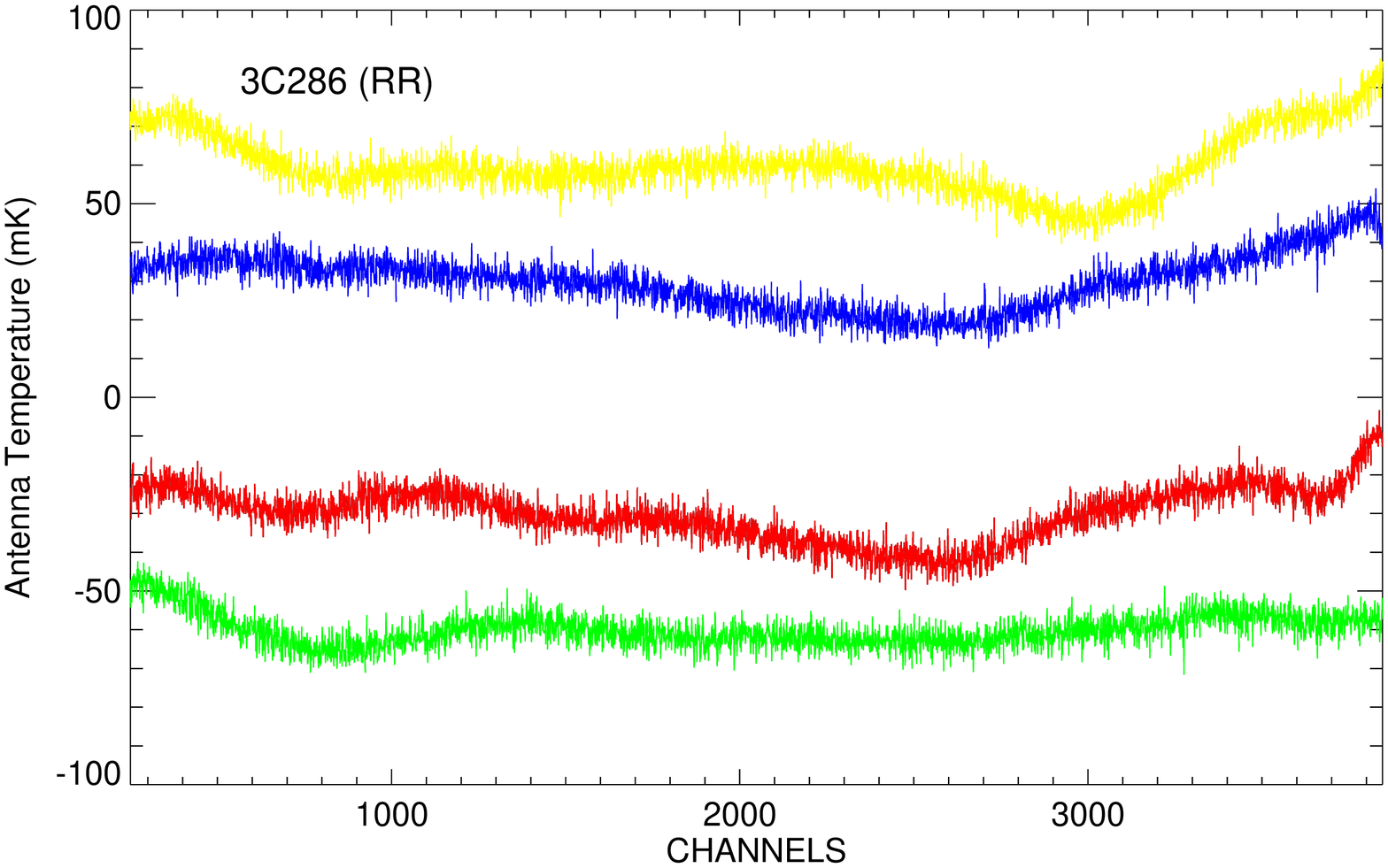} 
\includegraphics[angle=90,scale=0.35]{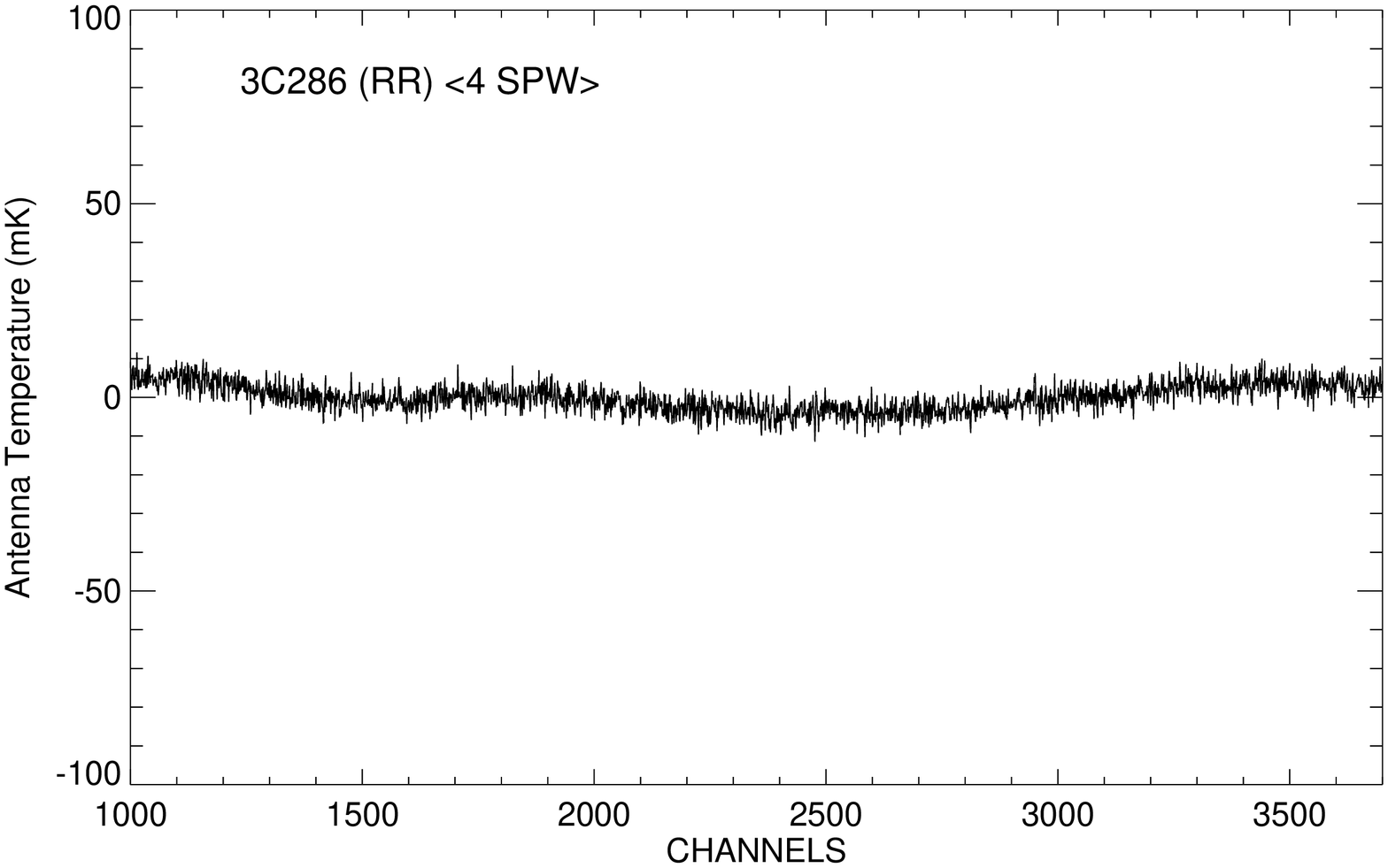} 
\caption{\threec{286} spectra where the antenna temperature is plotted
  as a function of channel number.  The spectra have been offset for
  clarity.  There are 4096 channels across a 50\mhz\ bandwidth. A
  linear baseline was removed from each spectrum.  {\it Left Panel:}
  All four \hep3\ SPW spectra are shown as different colors for LL
  (top) and RR (bottom) circular polarizations.  The sky frequency is
  shifted by 3, 6, and 9\mhz\ (or 245, 490, and 735 channels) for each
  consecutive spectrum.  {\it Right Panel:} The average \hep3\
  spectrum for LL (top) and RR (bottom) circular polarizations.  The
  spectra were shifted to align them in frequency before averaging and
  therefore fewer channels are shown.}
\label{fig:3c286}
\end{figure}

\begin{figure}
\includegraphics[angle=90,scale=0.35]{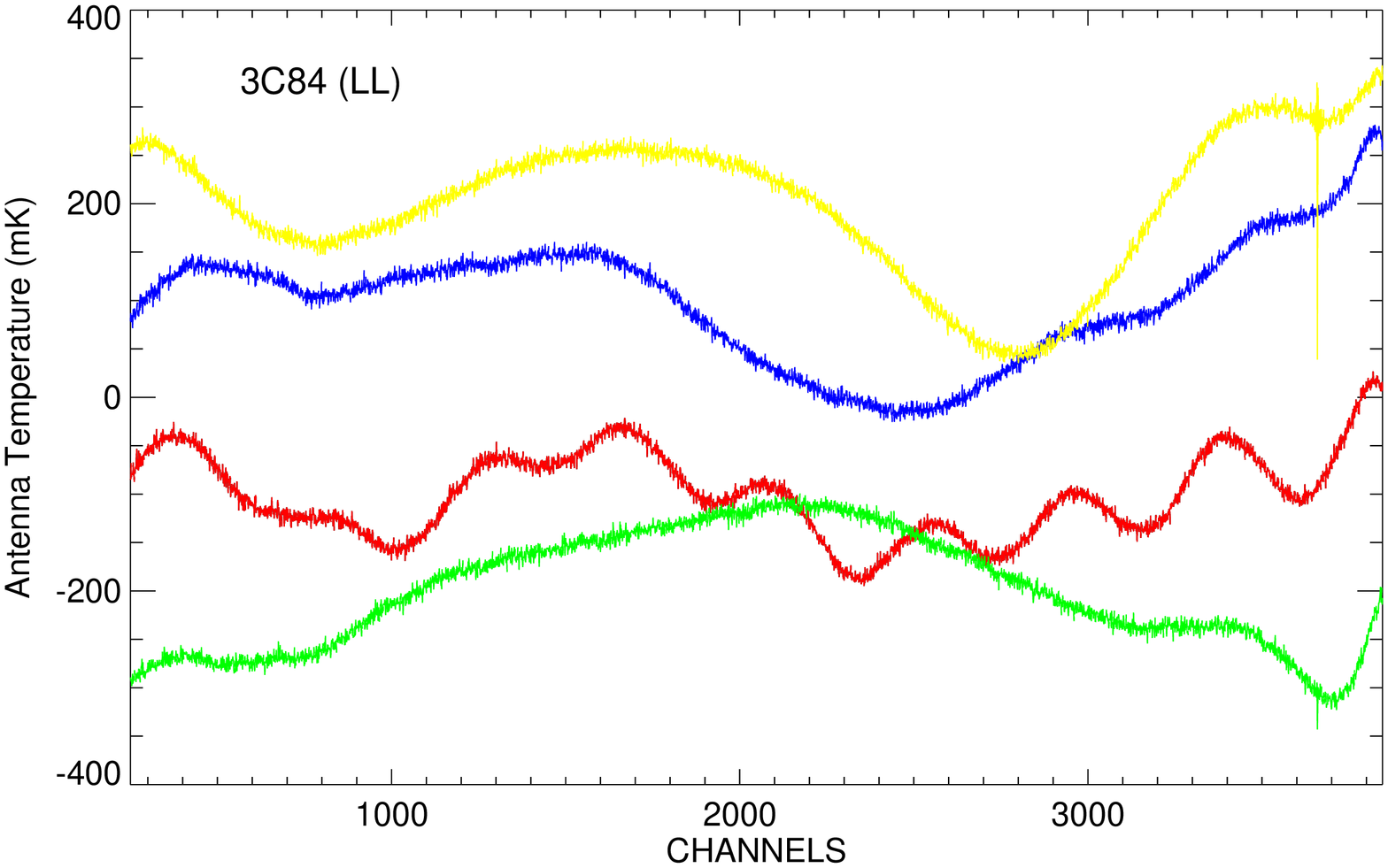} 
\includegraphics[angle=90,scale=0.35]{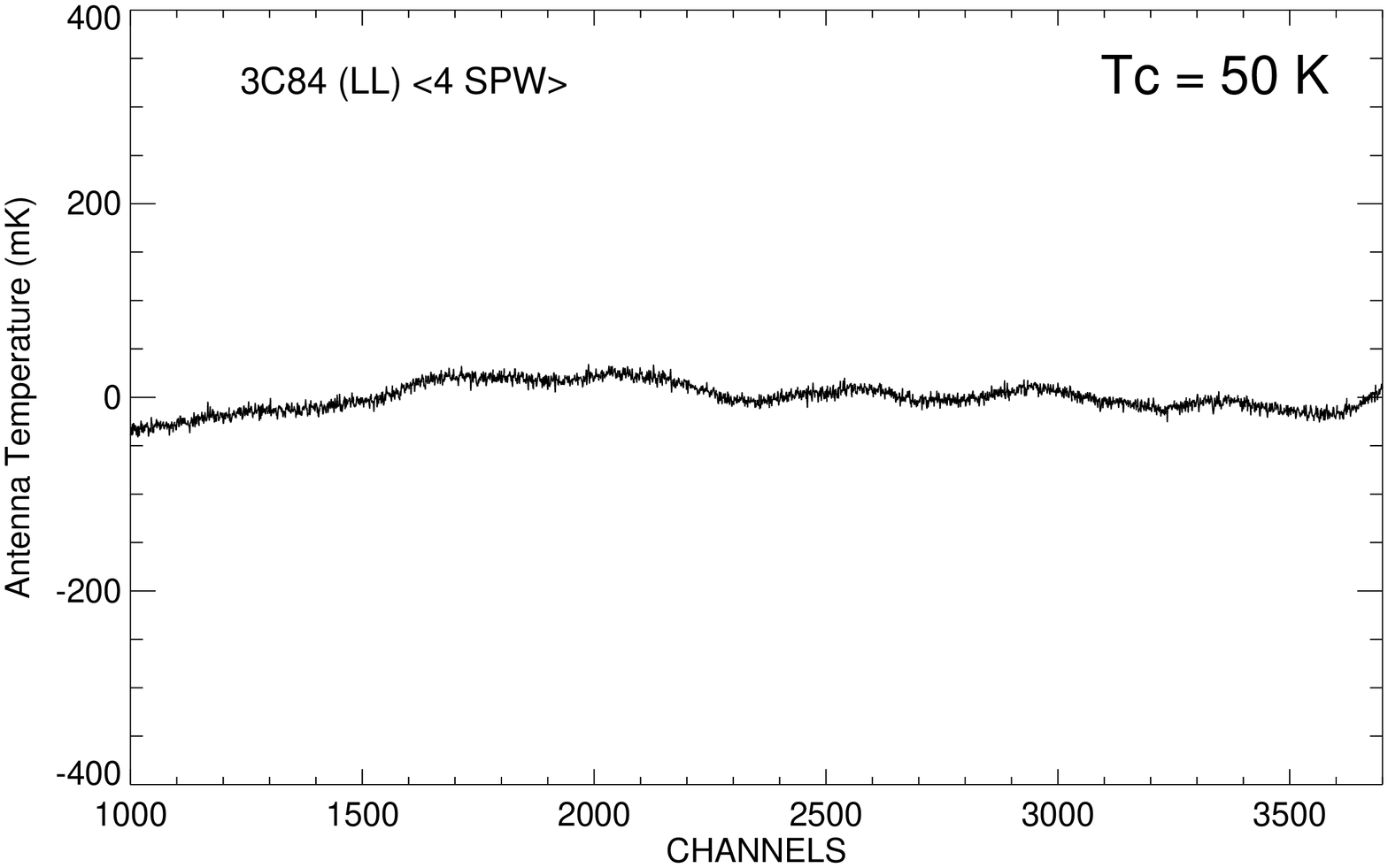} 
\includegraphics[angle=90,scale=0.35]{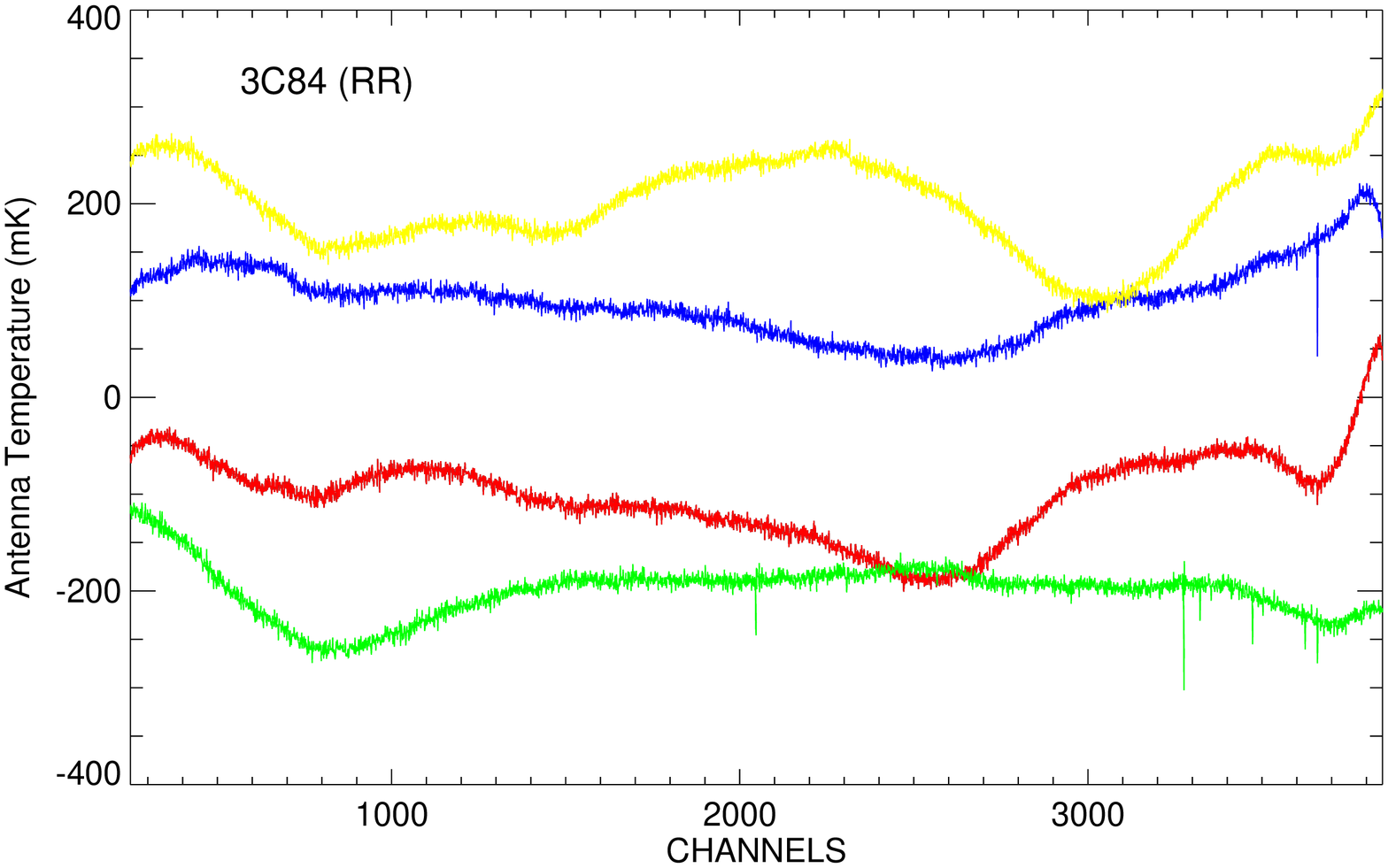} 
\includegraphics[angle=90,scale=0.35]{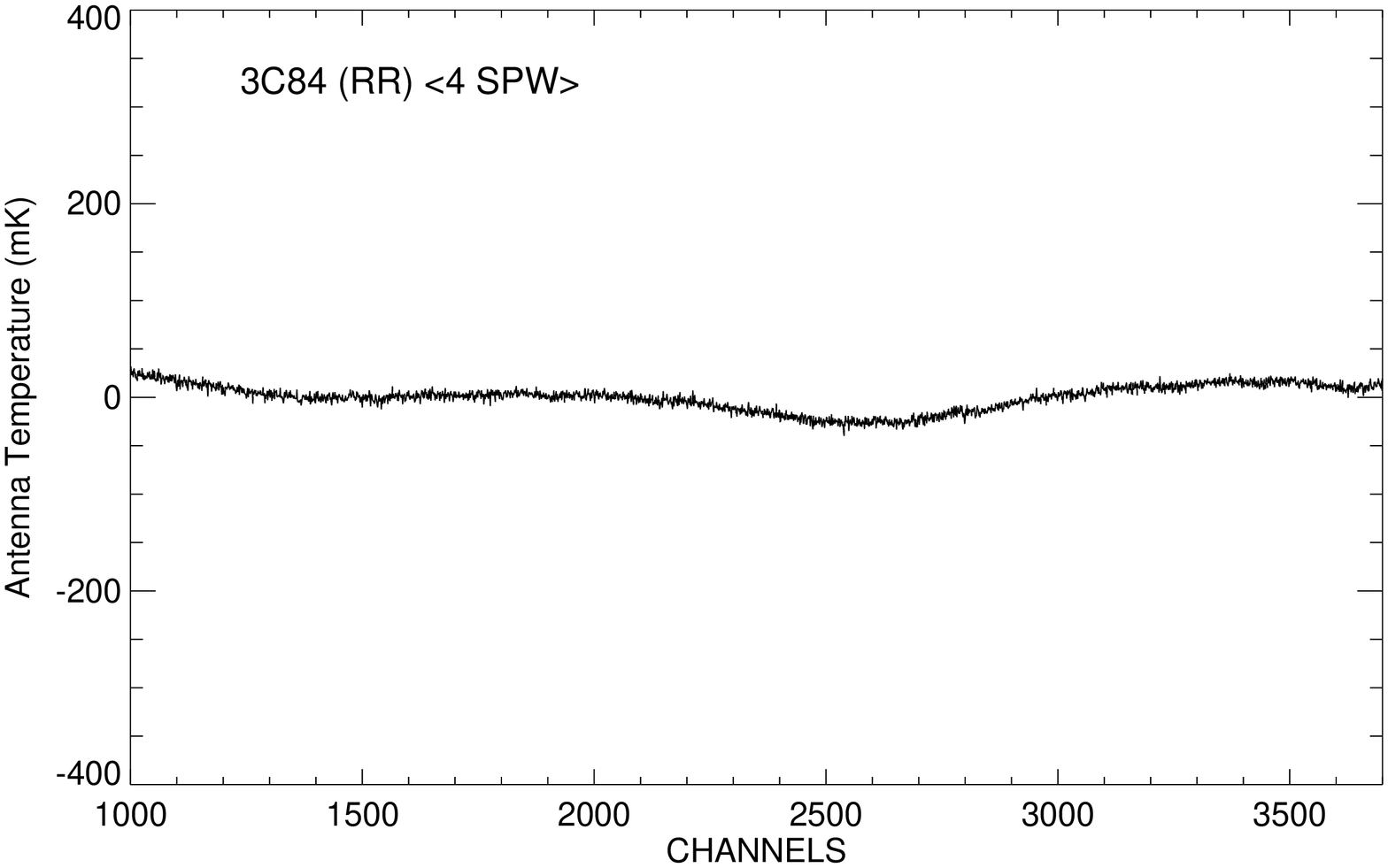} 
\caption{\threec{84} spectra where the antenna temperature is plotted
  as a function of channel number.  See Figure~\ref{fig:3c286} for
  details.  The spectral baseline frequency structure seen in the left
  panels scales roughly with source continuum intensity.  The
  \threec{84} continuum intensity is 5 times that of \threec{286} and
  the rms noise for \threec{84} is about 7 times larger than that of
  \threec{286}.  
}
\label{fig:3c84}
\end{figure}

The results are shown in Figure~\ref{fig:3c286} for \threec{286} and
Figure~\ref{fig:3c84} for \threec{84}.  Data were averaged over
several observing sessions for each of the \hep3\ SPWs separately
(left panels).  We fit a polynomial with order 1 to each spectrum to
remove any slope or intensity offset from the spectrum.  These four
spectra were shifted to align them in frequency and then averaged
(right panels).  Both circular polarizations are shown separately.
The baseline structure is clearly different for each \hep3\ SPW and
between the orthogonal polarizations.  The averaged spectrum is
clearly flatter.  Furthermore, the amplitude of the baseline features
roughly scales with continuum intensity (c.f., Figure~\ref{fig:3c286}
versus Figure~\ref{fig:3c84}).  The root-mean-square (rms) noise
across 3C84 spectra is about 7 times larger than for 3C286 spectra.
Averaging over the four \hep3\ SPWs does not reduce the random
(thermal) noise because the signals are correlated, but the
instrumental (systematic) noise is reduced.

\section{\he3\ Abundances}

For each source we generate spectra for all 8 sub-bands using the
procedures discussed in \S{\ref{sec:obs}}, and perform the analyses
described in \S{\ref{sec:reduce}} for the \hep3\ transition.
Table~\ref{tab:he3} summarizes the results.  Listed are the source
name, the Gaussian fit parameters and their associated 1-$\sigma$
errors, the rms noise in the line-free region, and the integration
time.  The Gaussian fits give the peak antenna temperature, the FWHM
line width, and the LSR center velocity.  The quantity of interest,
however, is the \her3\ abundance ratio by number.  That is, to compare
our results with theory we need to derive the abundance of \he3\
relative to H.  This requires a model since the \hep3\ hyperfine line
intensity depends on the \hep3\ column density, whereas both the
free-free continuum and H RRL intensities depend on the emission
measure or the integral of the density squared.  The free-free thermal
continuum intensity is used to derive the H abundance and so is a
critical step in the derivation of the \her3\ abundance ratio.  We
must also account for any neutral helium within the \hii\ region; that
is, an ionization correction is necessary to convert the \hepr3\ to
a \her3\ abundance ratio by number.

\begin{deluxetable}{lrrrrrrrr}
%\tabletypesize{\scriptsize}
\tablecaption{\hep3\ Spectral Line Parameters \label{tab:he3}}
\tablewidth{0pt}
\tablehead{
\colhead{} & \colhead{$T_{\rm L}$} & \colhead{$\sigma\,T_{\rm L}$} & 
\colhead{$\Delta{V}$} & \colhead{$\sigma\,\Delta{V}$} & 
\colhead{$V_{\rm LSR}$} & \colhead{$\sigma\,V_{\rm LSR}$} & 
\colhead{rms} &  \colhead{$t_{\rm intg}$} \\ 
\colhead{Source} & \colhead{(mK)} & \colhead{(mK)} & 
\colhead{(\kms)} & \colhead{(\kms)} & \colhead{(\kms)} & \colhead{(\kms)} & 
\colhead{(mK)} & \colhead{(hr)}
} 
\startdata 
S206        & 3.796 & 0.076 & 20.88 & 0.55 & $-25.37$ & 0.21 & 0.25 & 91.81 \\
S209        & 1.963 & 0.147 & 22.40 & 1.94 & $-50.83$ & 0.83 & 0.46 & 25.64 \\
M16         & 4.749 & 0.150 & 18.63 & 0.90 & $+27.21$ & 0.35 & 0.89 &  7.04 \\
G29.9       & 4.843 & 0.092 & 27.31 & 0.93 & $+98.64$ & 0.38 & 0.49 & 44.75 \\
\ngc{7538}  & 3.814 & 0.190 & 25.90 & 1.80 & $-61.83$ & 0.69 & 0.51 & 98.21 \\
\enddata 
\end{deluxetable}

Here we use the numerical program NEBULA \citep{balser18}\footnote{
  See http://ascl.net/1809.009.} to perform the radiative transfer of
the \hep3\ line, RRLs, and the free-free continuum emission through a
model nebula.  A detailed description of NEBULA is given in
\citet{balser95}.  Briefly, the model nebula is composed of only H and
He within a three-dimension Cartesian grid with arbitrary density,
temperature, and ionization structure.  Each numerical cell consists
of the following quantities: electron temperature, \te, electron
density, \Ne, \hepr4\ abundance ratio, \heppr4\ abundance ratio, and
the \hepr3\ abundance ratio.  Here we assume \heppr4\ = 0.0 for all
sources since the radiation field in Galactic \hii\ regions is not
hard enough to doubly ionize He.

The radiative transfer is performed from the back of the grid to the
front to produce the brightness distribution on the sky.  To simulate
an observation NEBULA calculates model spectra by convolving the
brightness distribution with a Gaussian beam by the GBT's HPBW at the
\hep3\ frequency.  The \hep3\ line is assumed to be in local
thermodynamic equilibrium (LTE), but non-LTE effects and pressure
broadening from electron impacts can be included for the RRLs.  All
spectra are broadened by thermal and microturbulent motions.  In
practice, each \hii\ region is modeled as a set of compact, uniform
spheres constrained by Very Large Array (VLA) continuum images
surrounded by a single halo component constrained by 140 Foot
telescope continuum data to account for the total flux density
\citep[e.g., see][]{balser95b, balser99a}.  Additional, small scale
structure is modeled by introducing a filling factor where gas is
moved into higher density, small-scale clumps with no gas between
clumps.

Table~\ref{tab:models} summarizes the adopted NEBULA models for each
source. Detailed information is given for each spherical, homogeneous
component that includes the J2000 position, the linear size or
diameter, $D$, the electron temperature, $T_{\rm e}$, the electron
density, $n_{\rm e}$, the \hepr4\ and \hepr3\ abundance ratios by number, and
the filling factor.  The last component listed for each source is the
halo component.

\begin{deluxetable}{lccrrcccc}
%\tabletypesize{\scriptsize}
\tablecaption{\hii\ Region Models \label{tab:models}}
\tablewidth{0pt}
\tablehead{
\colhead{} & \colhead{R.A.} & \colhead{decl.} & 
\colhead{$D$} & \colhead{$T_{\rm e}$} & \colhead{$n_{\rm e}$} & 
\colhead{} & \colhead{} & \colhead{Filling} \\
\colhead{} & \colhead{(J2000)} & \colhead{(J2000)} & 
\colhead{(pc)} &  \colhead{(\K)} & \colhead{(\percc)} & 
\colhead{\hepr4} & \colhead{\hepr3} & \colhead{Factor}
} 
\startdata 
\multicolumn{9}{c}{\underline{\bf S206}} \\
A    & 04:03:13.867 & $+$51:18:32.02 &  0.865 &  9000 & \nexpo{5.498}{2} & 0.085 & \nexpo{1.89}{-5} & 1.00 \\
B    & 04:03:13.753 & $+$51:18:59.40 &  1.604 &  9000 & \nexpo{3.129}{2} & 0.085 & \nexpo{1.89}{-5} & 1.00 \\
C    & 04:03:22.022 & $+$51:17:39.26 &  0.902 &  9000 & \nexpo{2.742}{2} & 0.085 & \nexpo{1.89}{-5} & 1.00 \\
D    & 04:03:19.237 & $+$51:18:03.70 &  1.097 &  9000 & \nexpo{2.575}{2} & 0.085 & \nexpo{1.89}{-5} & 1.00 \\
E    & 04:03:14.336 & $+$51:18:05.08 &  1.166 &  9000 & \nexpo{2.474}{2} & 0.085 & \nexpo{1.89}{-5} & 1.00 \\
Halo & 04:03:15.870 & $+$51:18:54.00 &  6.025 &  9000 & \nexpo{6.500}{1} & 0.085 & \nexpo{1.89}{-5} & 1.00 \\
\multicolumn{9}{c}{\underline{\bf S209}}\\
A    & 04:11:06.133 & $+$51:10:13.22 &  2.567 & 10500 & \nexpo{3.330}{2} & 0.070 & \nexpo{1.00}{-5} & 1.00 \\
B    & 04:11:05.656 & $+$51:09:31.09 &  3.395 & 10500 & \nexpo{2.065}{2} & 0.070 & \nexpo{1.00}{-5} & 1.00 \\
C    & 04:11:08.602 & $+$51:10:34.67 &  3.747 & 10500 & \nexpo{1.965}{2} & 0.070 & \nexpo{1.00}{-5} & 1.00 \\
D    & 04:11:08.931 & $+$51:08:57.27 &  3.063 & 10500 & \nexpo{1.668}{2} & 0.070 & \nexpo{1.00}{-5} & 1.00 \\
Halo & 04:11:06.740 & $+$51:09:44.00 & 15.084 & 10500 & \nexpo{1.200}{1} & 0.070 & \nexpo{1.00}{-5} & 1.00 \\
\multicolumn{9}{c}{\underline{\bf M16}}\\
A    & 18:18:50.252 & $-$13:48:50.39 &  0.308 &  6000 & \nexpo{5.839}{2} & 0.080 & \nexpo{1.95}{-5} & 1.00 \\
B    & 18:18:53.671 & $-$13:49:37.35 &  0.336 &  6000 & \nexpo{4.408}{2} & 0.080 & \nexpo{1.95}{-5} & 1.00 \\
C    & 18:18:53.666 & $-$13:49:57.76 &  0.392 &  6000 & \nexpo{3.615}{2} & 0.080 & \nexpo{1.95}{-5} & 1.00 \\
D    & 18:18:55.543 & $-$13:51:47.90 &  0.315 &  6000 & \nexpo{4.413}{2} & 0.080 & \nexpo{1.95}{-5} & 1.00 \\
Halo & 18:18:52.650 & $-$13:50:05.00 &  8.680 &  6000 & \nexpo{8.125}{1} & 0.080 & \nexpo{1.95}{-5} & 1.00 \\
\multicolumn{9}{c}{\underline{\bf G29.9}}\\
A    & 18:46:09.470 & $-$02.41.23.75 &  1.656 &  6500 & \nexpo{4.801}{2} & 0.070 & \nexpo{1.60}{-5} & 0.01 \\
B    & 18:46:10.880 & $-$02.41.58.91 &  1.469 &  6500 & \nexpo{4.409}{2} & 0.070 & \nexpo{1.60}{-5} & 0.01 \\
Halo & 18:46:09.280 & $-$02:41:47.00 & 10.070 &  6500 & \nexpo{1.100}{2} & 0.070 & \nexpo{1.60}{-5} & 1.00 \\
\multicolumn{9}{c}{\underline{\bf NGC$\thinspace$7538}}\\
A    & 23:13:30.694 & $+$61:30:03.31 &  1.025 &  8000 & \nexpo{9.580}{2} & 0.083 & \nexpo{1.86}{-5} & 0.15 \\
B    & 23:13:45.439 & $+$61:28:19.51 &  0.196 &  8000 & \nexpo{4.009}{3} & 0.083 & \nexpo{1.86}{-5} & 0.15 \\
C    & 23:13:30.663 & $+$61:29:30.14 &  1.144 &  8000 & \nexpo{6.419}{2} & 0.083 & \nexpo{1.86}{-5} & 0.15 \\
D    & 23:13:37.894 & $+$61:29:13.57 &  0.949 &  8000 & \nexpo{5.807}{2} & 0.083 & \nexpo{1.86}{-5} & 0.15 \\
Halo & 23:13:32.050 & $+$61:30:12.00 &  3.767 &  8000 & \nexpo{1.660}{2} & 0.083 & \nexpo{1.86}{-5} & 1.00 \\
\enddata 
%\tablenotetext{a}{}
\end{deluxetable}

These \hii\ region NEBULA models are constrained using the following
procedure:

\begin{enumerate}

\item {\bf Calculate ${\bf T}_{\bf \rm e}$.} Use the single-dish
  \hrrl91\ line-to-continuum ratio to calculate $T_{\rm e}$ assuming
  an optically thin nebula in LTE.  Any non-LTE effects, including
  pressure broadening, should be small for the \hrrl91\ RRL in \hii\
  regions \citep[see][]{shaver80a, shaver80b}.  Assume a constant
  electron temperature for all components.

\item {\bf Calculate $^{\bf 4}$He$^{\bf +}$/H$^{\bf +}$.}  Use the
  line areas of the \herrl91\ and \hrrl91\ RRLs to calculate the
  \hepr4\ abundance ratio \citep[e.g.,][]{bania07}.  Assume a constant
  \hepr4\ abundance ratio for all components.

\item {\bf Model density structure.}  Use high spatial resolution
  continuum images (e.g., VLA data) near the \hep3\ frequency to model
  the density structure assuming spherical, homogeneous spheres
  \citep[e.g.,][]{balser95}.  This provides values for $n_{\rm e}$ and
  $D$ given $T_{\rm e}$, \hepr4, and $D_{\rm sun}$.  Since \hii\
  regions are typically resolved, use single-dish data to model any
  missing flux by assuming the missing emission is produced by a
  spherical, homogeneous halo.

\item {\bf Constrain halo density.}  Use NEBULA to generate the RRL
  brightness distributions on the sky, convolved with the GBT HPBW.
  Assume LTE with no pressure broadening.  If necessary, adjust the
  halo electron density to match the \hrrl91\ line intensity. The
  physical parameters of the halo component were derived from the
  single-dish data and thus stem from the total flux density of the
  source.  Since we have added compact components contained with the
  single-dish telescope's beam, the halo density needs to be slightly
  reduced.

\item {\bf Determine $^{\bf 3}$He$^{\bf +}$/H$^{\bf +}$ assuming LTE.}
  Use NEBULA to calculate model \hep3\ and RRL spectra from the sky
  brightness distributions assuming LTE. Compare these spectra with
  the GBT data using the many H and \he4\ RRLs observed.  If the model
  RRL spectra match the data then adjust the \hepr3\ abundance ratio
  until the line intensity is consistent with the data.  Generating
  each model spectrum takes several hours of computing time so we
  compare the model and observed spectra by eye and do not try to
  minimize the rms over a large grid of models.  Assume that \hepr3\
  is constant within the \hii\ region.

\item {\bf Determine $^{\bf 3}$He$^{\bf +}$/H$^{\bf +}$ assuming
    non-LTE.}  If the NEBULA model in step (5) is inconsistent with
  the GBT data, then check the assumption of LTE by running the NEBULA
  model again assuming non-LTE.  Since pressure broadening is very
  sensitive to the local electron density, this can result in an
  overprediction of the RRL intensity for lines with higher principal
  quantum numbers (e.g., H142$\delta$ compared to \hrrl91).  If the
  NEBULA models are still not consistent with the data include a
  filling factor.  Experience has shown that density structure not
  detected with existing interferometer data mostly resides within the
  most compact components.  That is, high spatial resolution data will
  reveal multiple clumps with a given component.  Adjust the filling
  factor within the compact components to approximate this structure
  until the model matches the observations \citep[see][]{balser99a}.
  By this process, set the model \hepr3\ abundance ratio, assumed to
  be the same for all components, to match the data.

\end{enumerate}

This procedure yields a single value of \hepr3\ for each source.  To
derive \her3, however, requires an estimate of the amount of neutral
helium within the \hii\ region.  We characterize this by defining an
ionization correction, $\kappa_{\rm i}$:
\begin{equation}\label{eq:ki}
\kappa_{\rm i} = y_{4}/y_{4}^{+} = y_{3}/y_{3}^{+},
\end{equation}
where the $y$-factors are the atomic and ionic abundance ratios by
number and $y_{4} \equiv \,\,$\her4, $y_{4}^{+} \equiv \,\,$\hepr4,
$y_{3} \equiv \,\,$\her3, and $y_{3}^{+} \equiv \,\,$\hepr3.  Here we
assume the \he3\ and \he4\ ionization zones are identical.  The
ionization correction is difficult to measure since there are no
spectral lines at radio frequencies available to directly probe
neutral helium, and thus determine $y_{4}$.  Since most of the \he4\
is expected to be produced during BBN, with only a relatively small
contribution from stellar evolution, many studies just assume $y_{4} =
0.1$.  \citet{balser06a} determined a small helium enrichment from
stars of $\Delta{Y}/\Delta{Z} = 1.41 \pm\ 0.62$ in the Milky Way,
where Y and Z are the helium and metal abundance fractions by mass
\citep[also see][]{carigi08, lagarde12}.  So we expect a small $y_{4}$
radial abundance gradient.  The relative abundance of metals with
different ionization states provides constraints on the shape of the
ionizing radiation field and therefore insight into how much neutral
helium exists with the \hii\ region.  For example, \citet{deharveng00}
used the O$^{++}$/O abundance ratio to probe neutral helium in a
sample of Galactic \hii\ regions.  Using these methods only two
Galactic \hii\ regions, M17 and S206, have been found to contain no
neutral helium \citep{balser06a, carigi08}.  Since there is no
evidence for a finite \heppr4\ abundance ratio in M17 and S206, we
assume that $y_{4} = y_{4}^{+}$ for these sources.  Here we use the
values of $y_{4}$ derived for M17 and S206 to determine a linear
relationship between $y_{4}$ and the Galactocentric radius:
\begin{equation}\label{eq:y4}
y_{4} = \nexpo{-1.75}{-3}\,R_{\rm gal} + \nexpo{1.05}{-1}, 
\end{equation}
and assume that an \hii\ region's $R_{\rm gal}$ sets the $y_{4}$
abundance ratio.  Measuring the $y_{4}^{+}$ abundance ratio then
yields the ionization correction for the nebula via
Equation~\ref{eq:ki}.

\begin{figure}
\includegraphics[angle=0,scale=0.9]{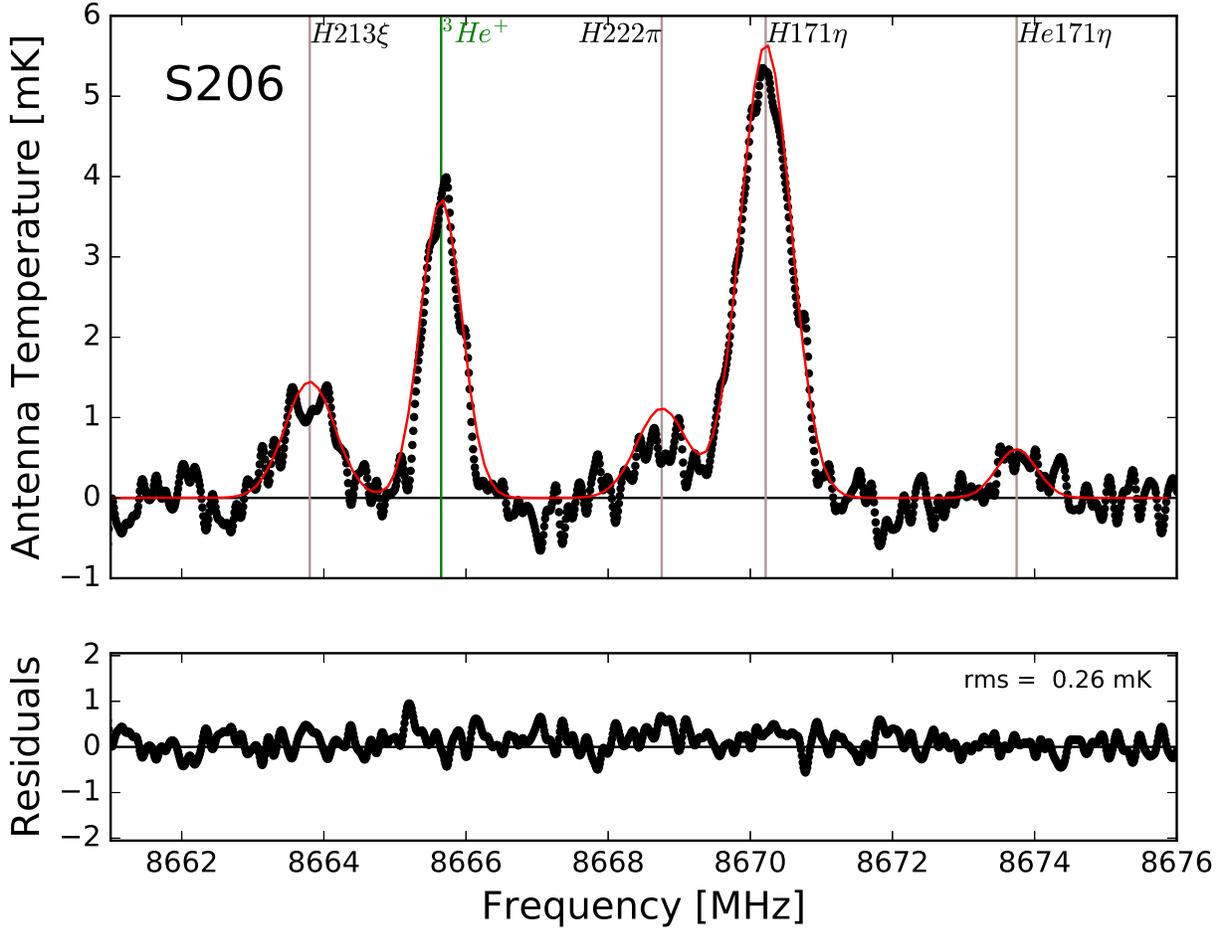} 
\caption{\hep3\ spectrum of S206.  {\it Top Panel:} The antenna
  temperature plotted as a function of rest frequency.  The black dots
  are the observed GBT data.  The solid red curve is the NEBULA model
  spectrum.  It is {\it not} a Gaussian fit to the GBT data.  The
  horizontal line is the zero level.  The vertical lines mark the
  location of the \hep3\ transition (green) and RRLs (gray).  {\it
    Bottom Panel:} The residuals, model $-$ data, from the top panel.
  The rms of these residuals calculated across the entire sub-band is
  shown.}
\label{fig:s206_he3}
\end{figure}

\begin{figure}
\includegraphics[angle=0,scale=0.45]{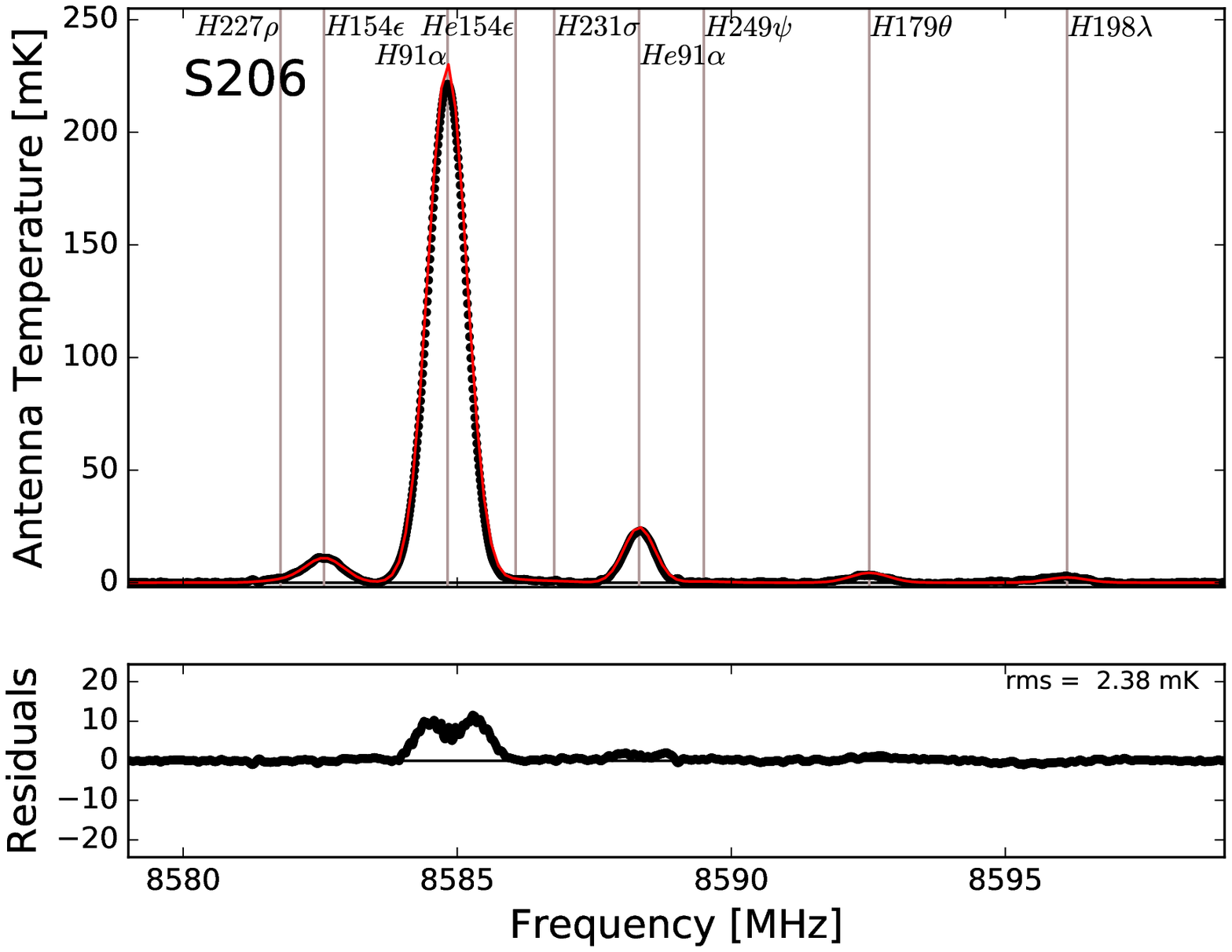} 
\includegraphics[angle=0,scale=0.45]{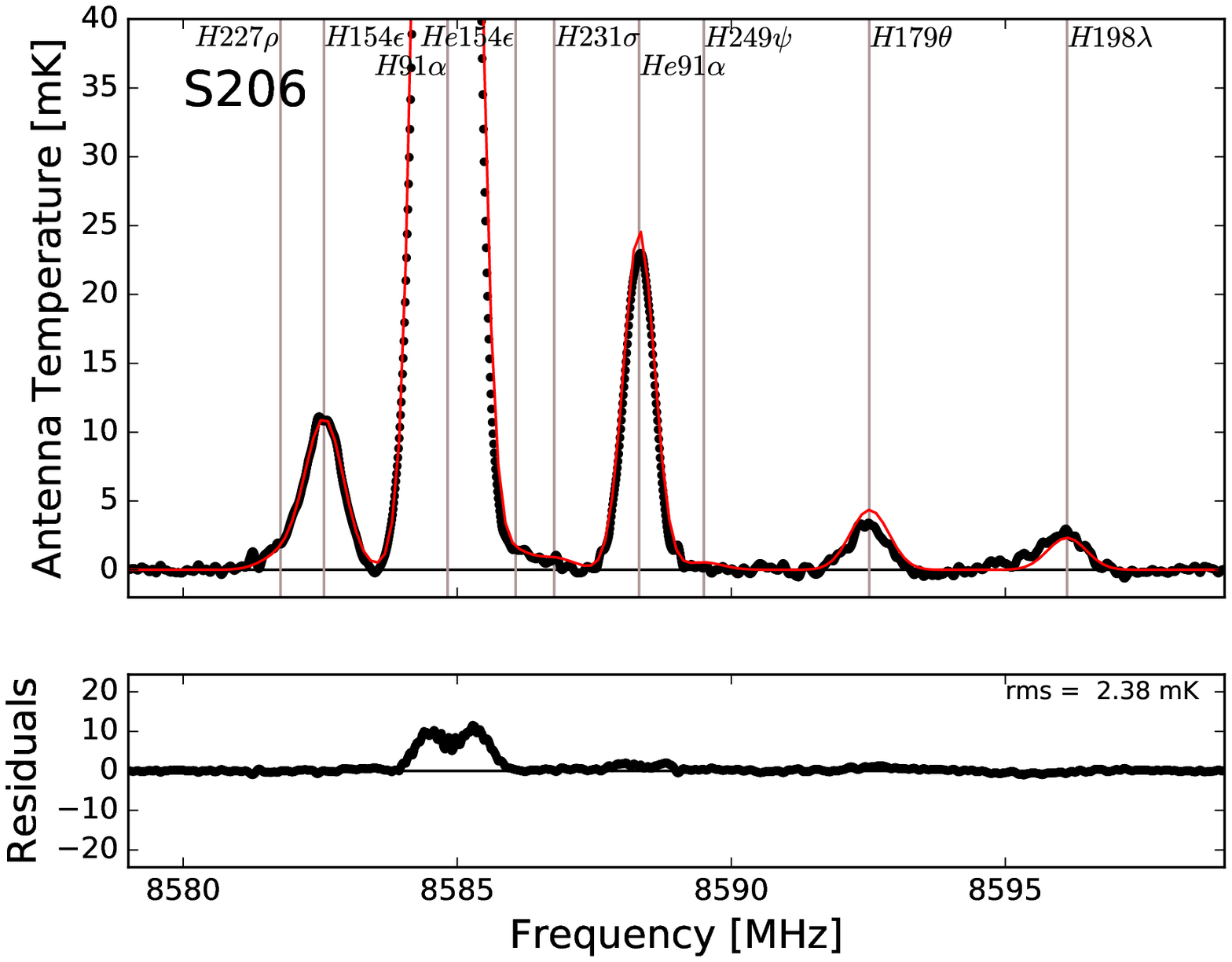} 
\includegraphics[angle=0,scale=0.45]{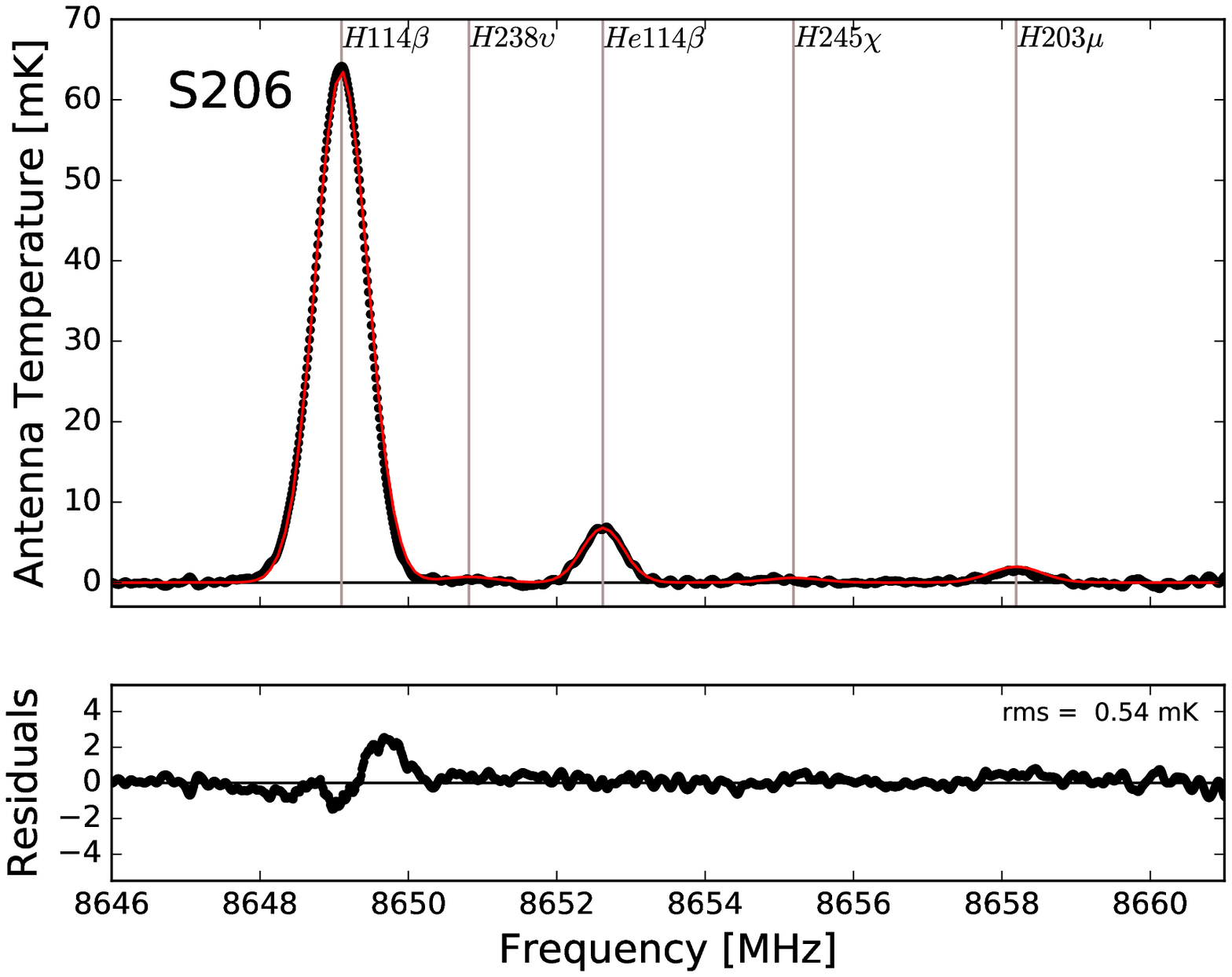} 
\includegraphics[angle=0,scale=0.45]{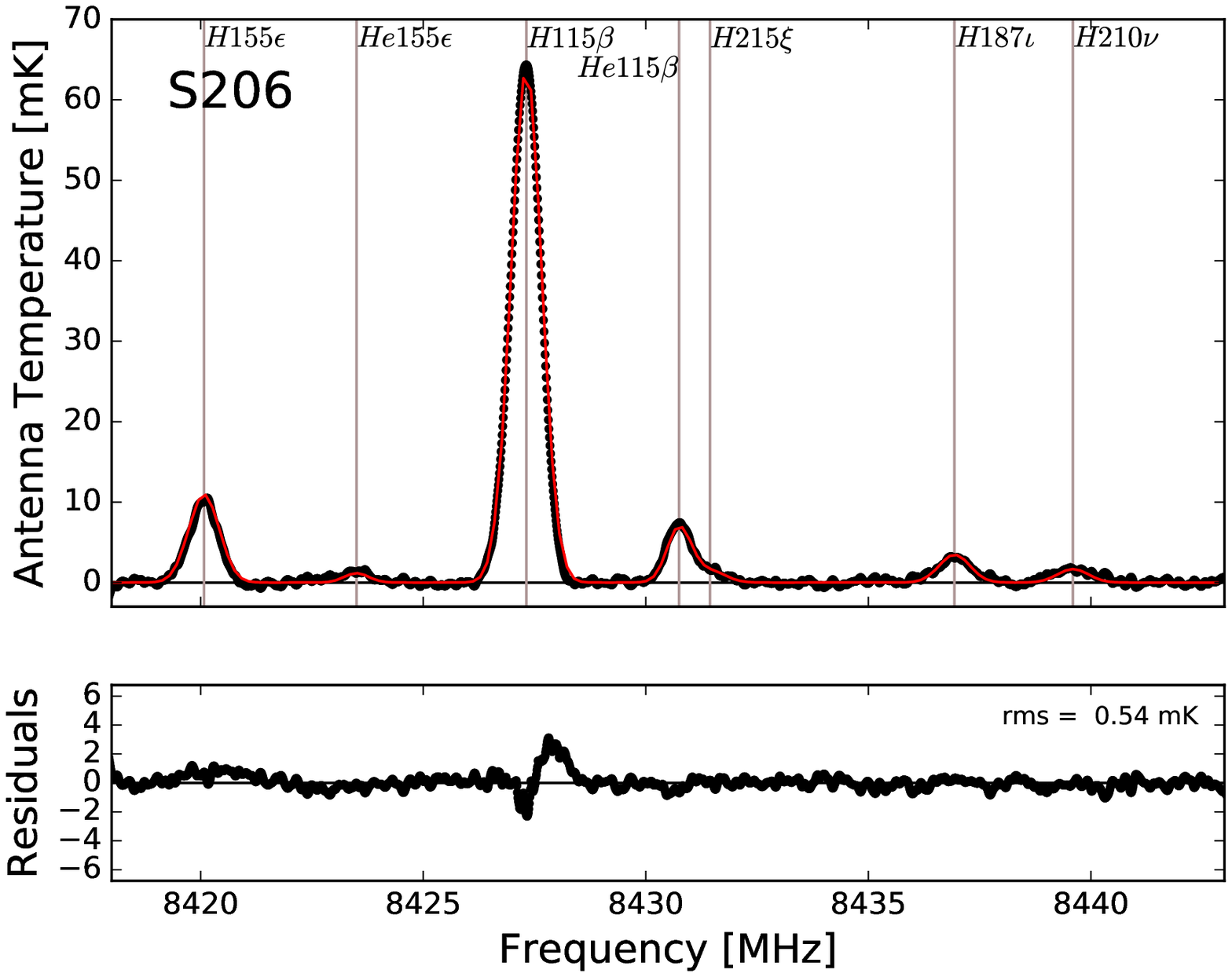} 
\caption{RRL spectra of S206 including the following sub-bands:
  H91$\alpha$ (top-left), expanded view of H91$\alpha$ (top-right),
  H114$\beta$ (bottom-left), and H115$\beta$ (bottom-right).  See
  Figure~\ref{fig:s206_he3} for details.}
\label{fig:s206_rrl1}
\end{figure}

\begin{figure}
\includegraphics[angle=0,scale=0.45]{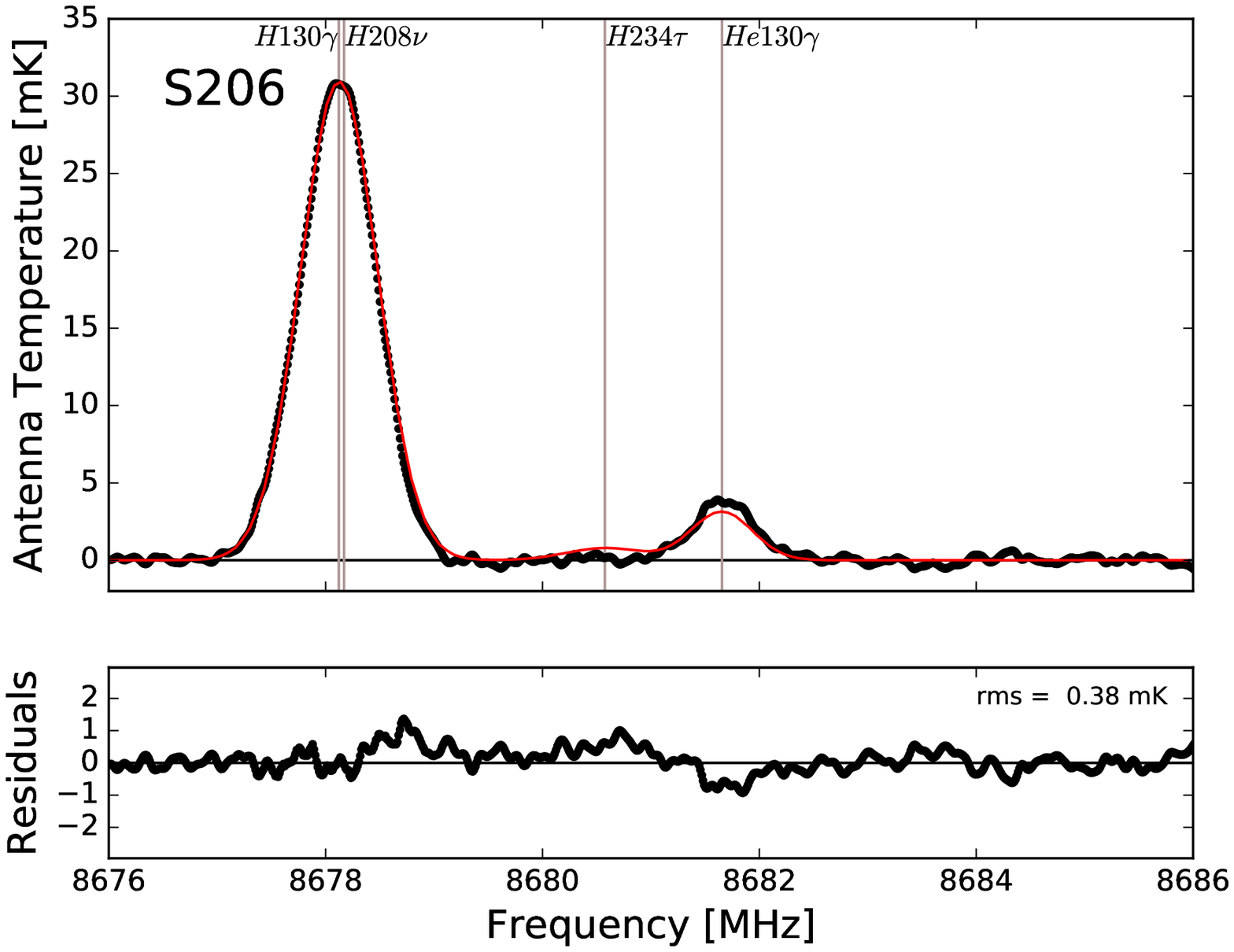} 
\includegraphics[angle=0,scale=0.45]{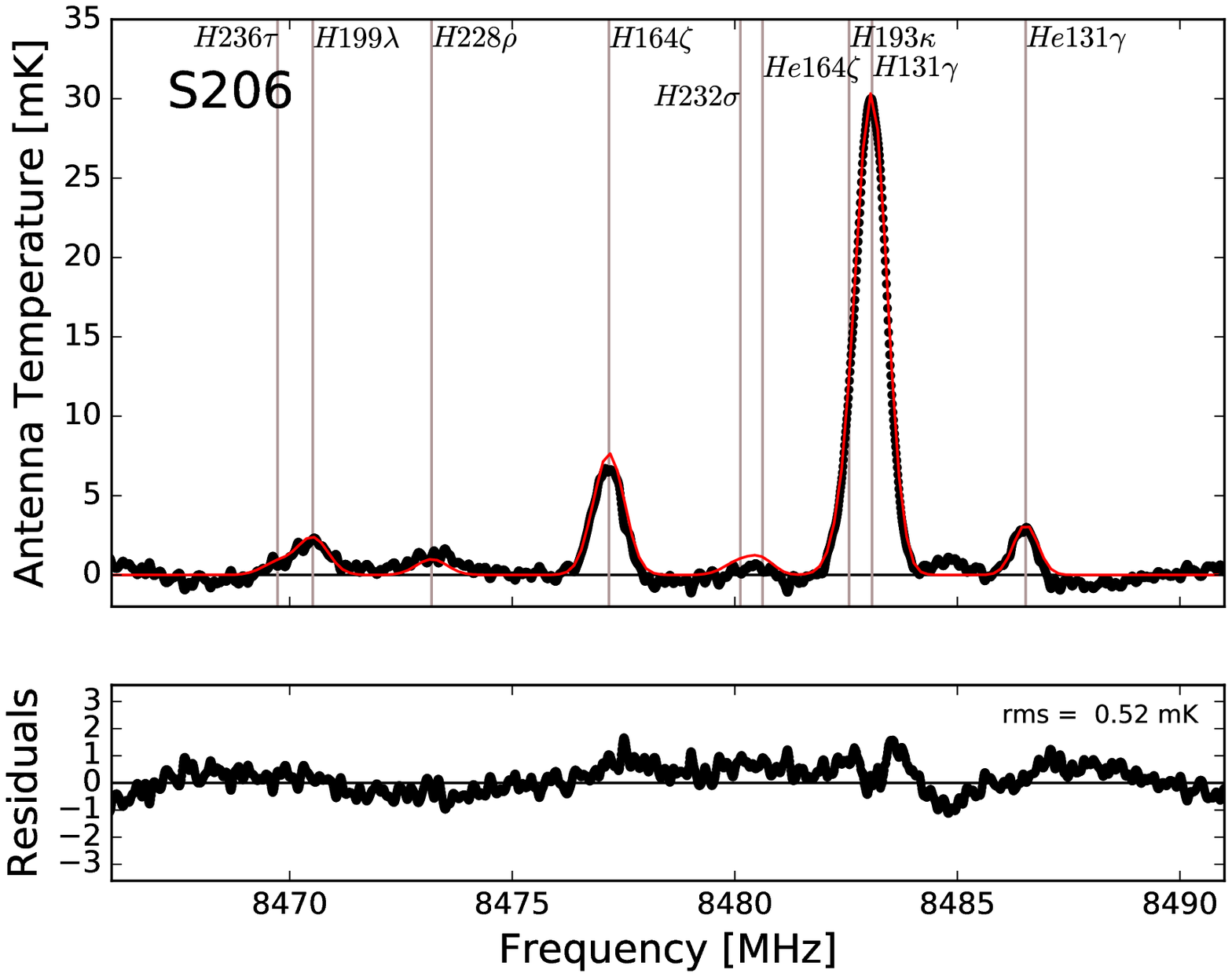} 
\includegraphics[angle=0,scale=0.45]{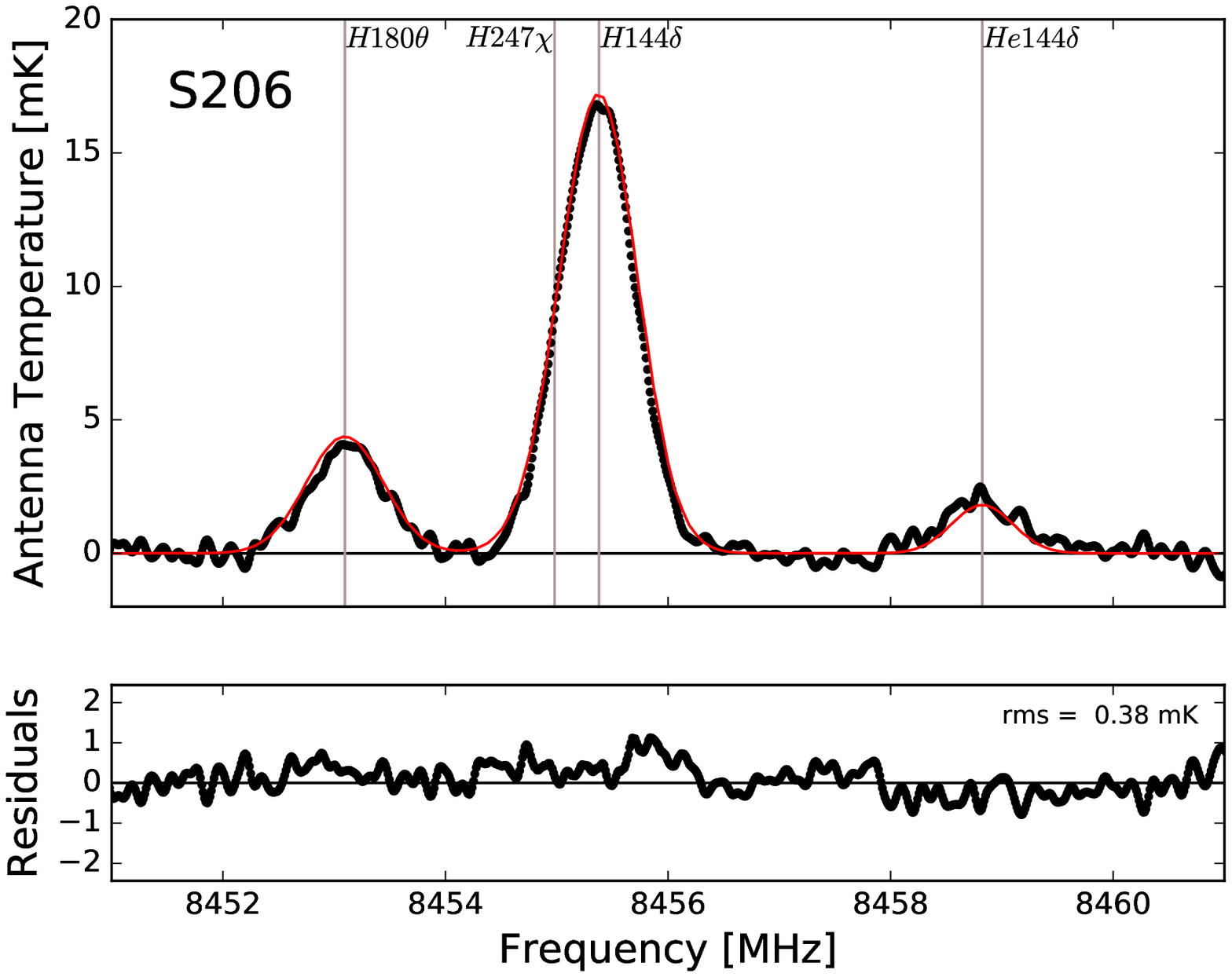} 
\includegraphics[angle=0,scale=0.45]{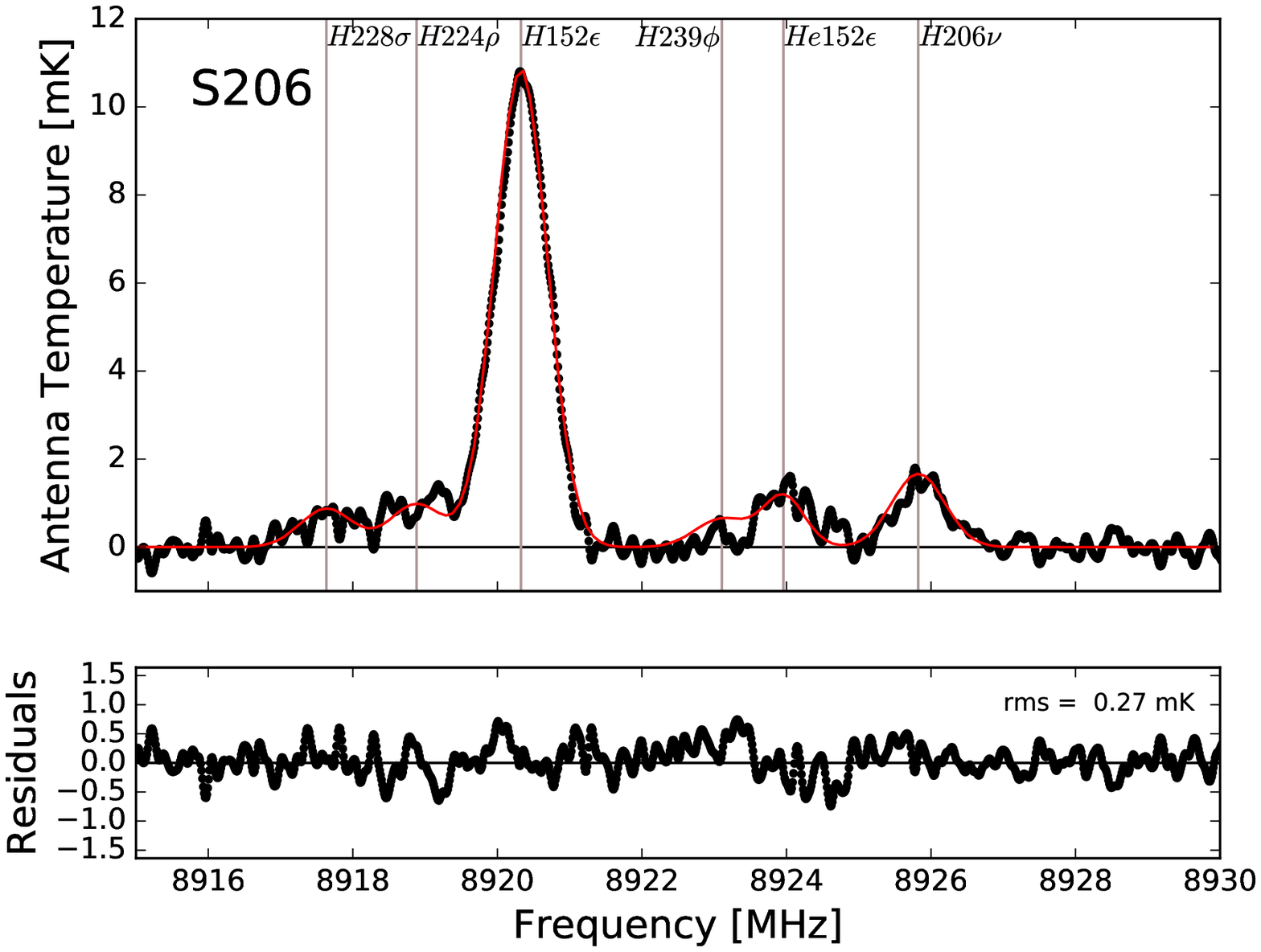} 
\caption{RRL spectra for S206 including the following sub-bands:
  H130$\gamma$ (top-left), H131$\gamma$ (top-right), H144$\delta$
  (bottom-left), and H152$\epsilon$ (bottom-right).  See
  Figure~\ref{fig:s206_he3} for details.}
\label{fig:s206_rrl2}
\end{figure}

The NEBULA modeling results are summarized in a series of plots that
compare the model spectra to the observed GBT spectra.  The \hep3\
spectrum for S206 is shown in Figure~\ref{fig:s206_he3}.  The top
panel plots the antenna temperature as a function of rest frequency.
The black dots form the GBT spectrum and the solid red line is the
NEBULA result.  {\bf The solid red line is NOT a numerical fit to the
  data but a synthetic spectrum resulting from the radiative transfer
  through a model nebula.} The vertical lines mark the location of the
\hep3\ transition (green) and various RRLs (gray).  The bottom panel
shows the residuals, (model $-$ data).  Figures~\ref{fig:s206_rrl1}
and \ref{fig:s206_rrl2} show the results for the other 7 sub-bands in
S206.

Previously, we compared the models and data by calculating the
difference between Gaussian fit line parameters for each transition
\citep[c.f.,][]{balser99a}.  The uncertainty in the adopted \her3\
abundance ratio came from the formal errors in the Gaussian fits to
the line and continuum data.  Here we take a different approach and
use the residuals over the entire spectrum to give uncertainty
estimates.  This has the advantage of directly including in the
derived uncertainty any deviation in the line shape from a pure
Gaussian and as well as any spectral baseline structure over the
entire spectrum.  Specifically, we set the 1-$\sigma$ error in the
\hepr3\ abundance ratio to be the value that produces a change in the
\hep3\ line intensity equal to the rms.  For the total uncertainty we
assume a 5\% error in the ionization correction that is added in
quadrature.

Overall, the model spectra fit the data remarkably well.  Inspection
of the residuals for each sub-band, however, reveals significant
deviations for the brightest RRLs.  For example, the residuals for the
\hrrl91\ transition in Figure~\ref{fig:s206_rrl1} have values as large
as 10\mk.  These larger residuals primarily result from the fact that
the various components do not share the same LSR velocity; that is,
there is velocity structure within the \hii\ region, that produces
multiple spectral components blended in velocity.  So the single-dish
sees a spectral line that cannot be fit by a single Gaussian.  This is
only revealed for the brightest spectral lines where the SNR is high.
Nevertheless, these effects are smaller than 5\% of the line area.
Results for the \hep3\ sub-band in the remaining sources are shown in
Figure~\ref{fig:four_he3}.  We give in the appendix comparisons
between NEBULA model and GBT observed spectra for all the RRL
transitions observed.  Below we discuss each source separately.

\begin{figure}
\includegraphics[angle=0,scale=0.45]{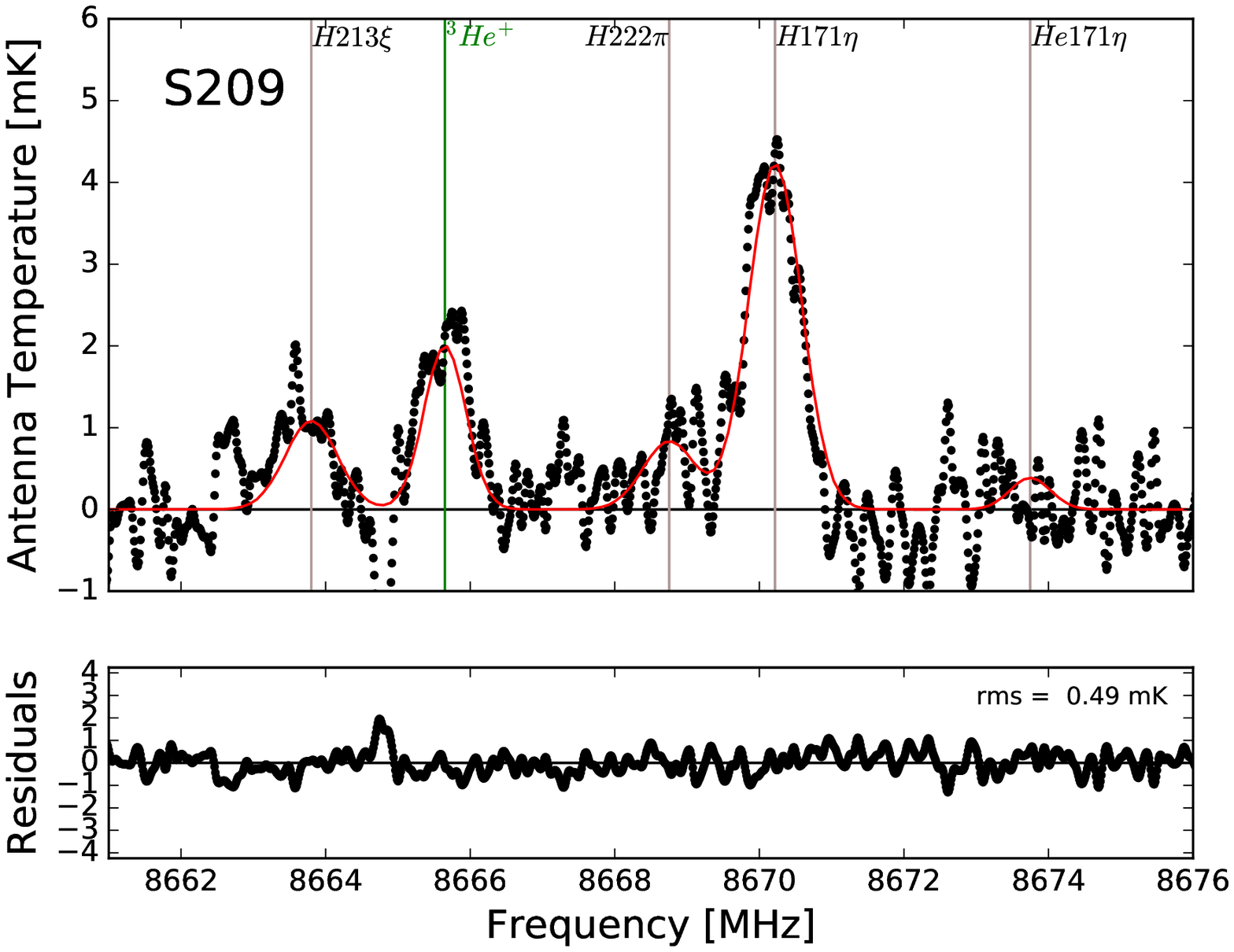} 
\includegraphics[angle=0,scale=0.45]{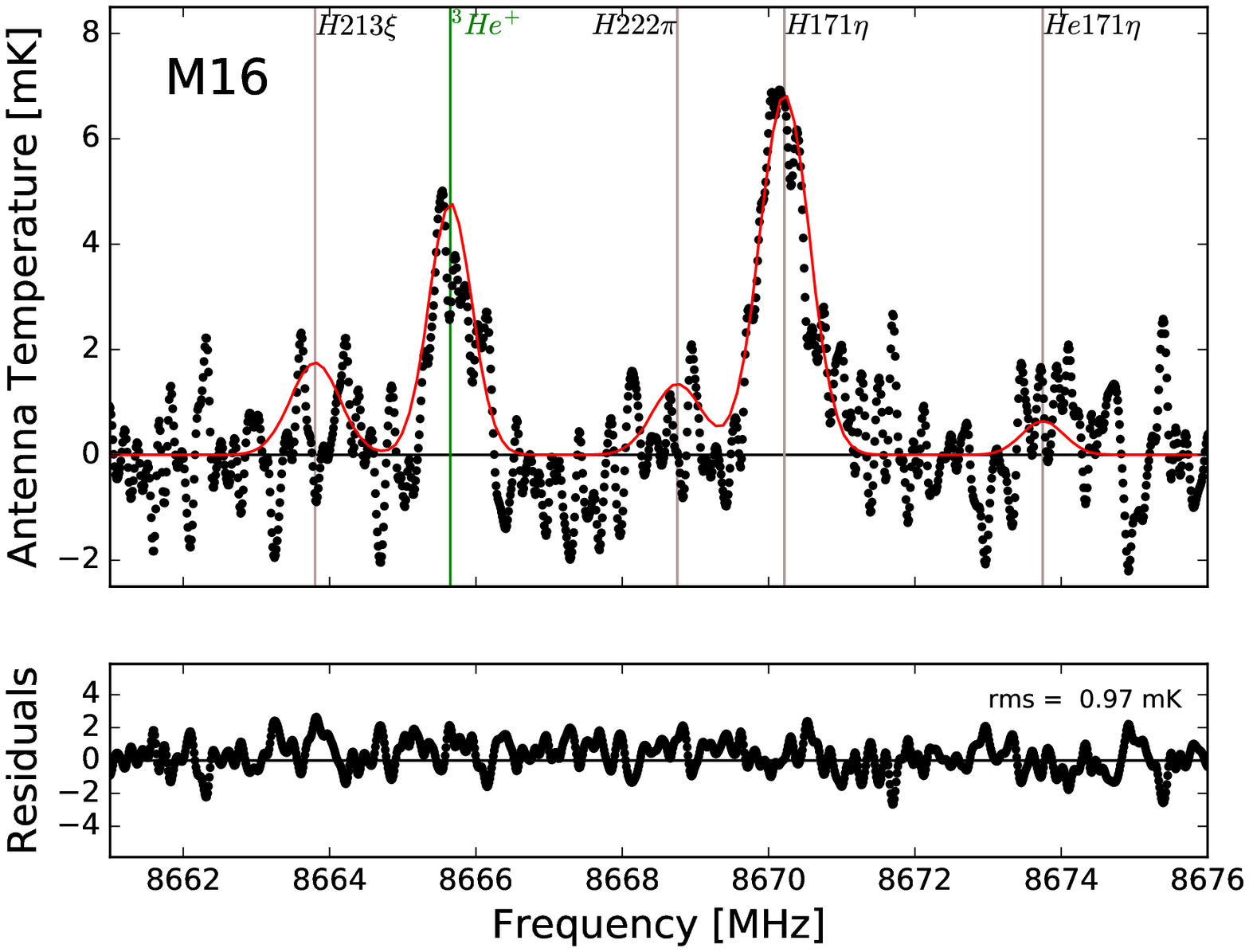} 
\includegraphics[angle=0,scale=0.45]{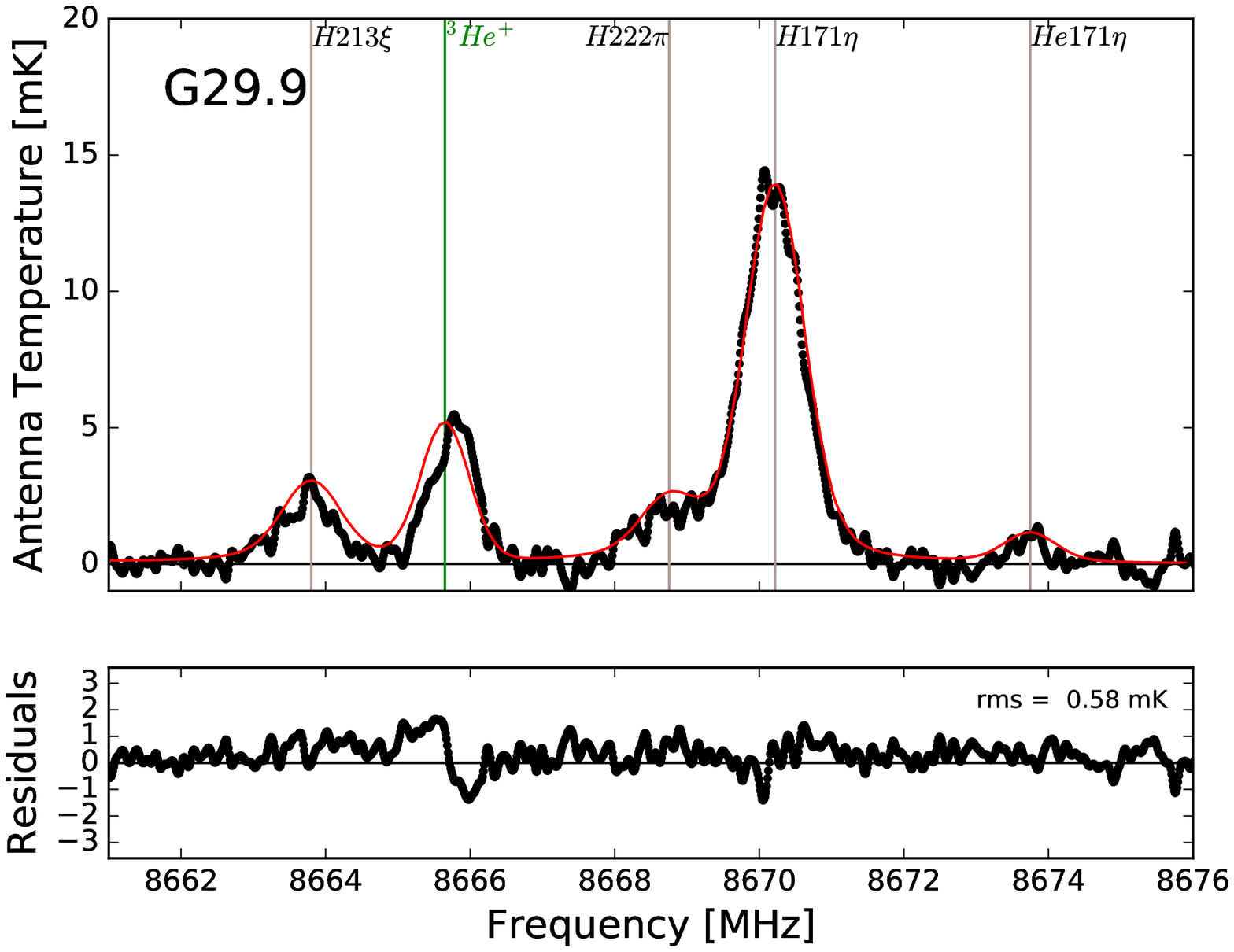} 
\includegraphics[angle=0,scale=0.45]{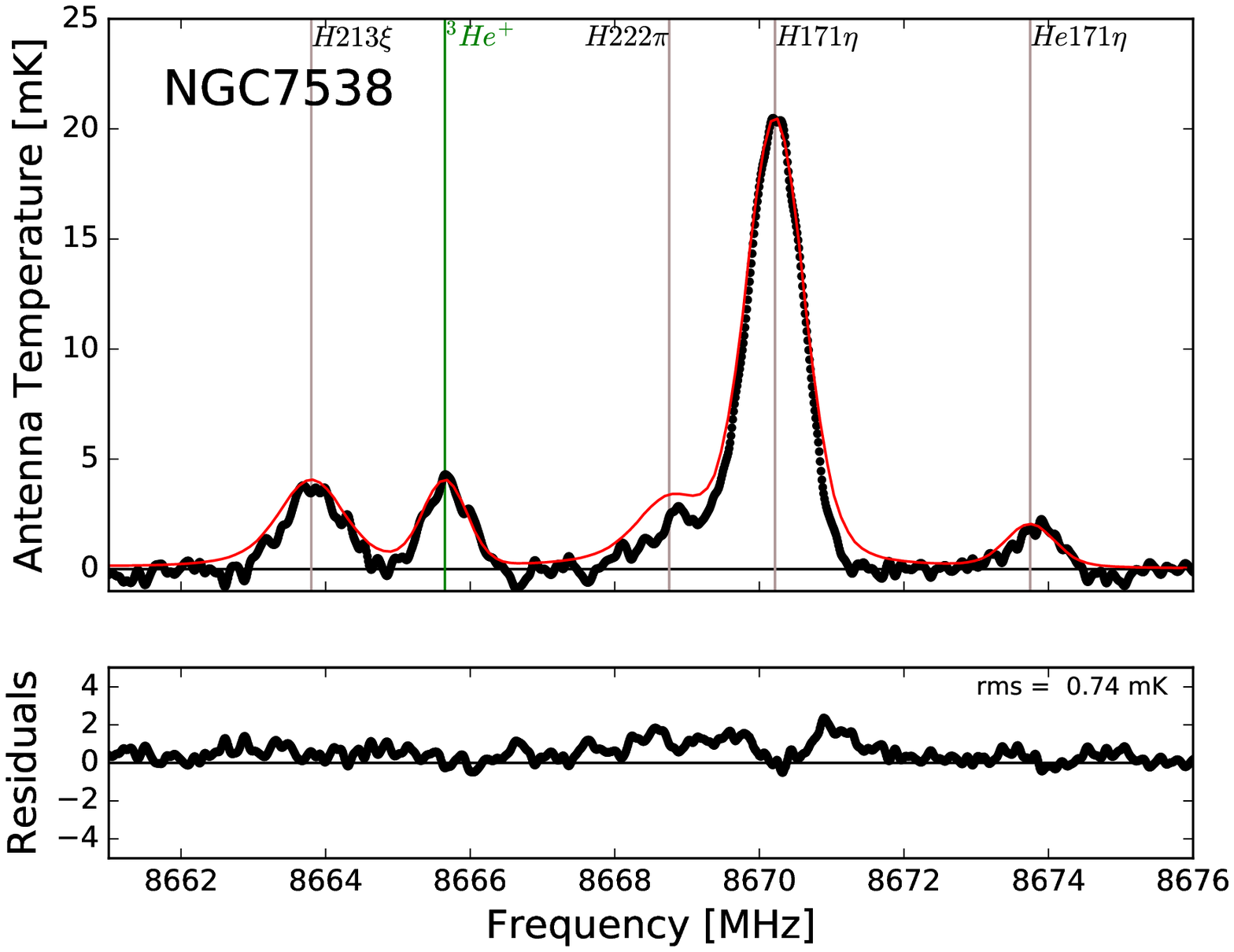} 
\caption{\hep3\ spectrum of S209 (top-left), M16 (top-right), G29.9
  (bottom-left), and \ngc{7538} (bottom-right). See
  Figure~\ref{fig:s206_he3} for details.}
\label{fig:four_he3}
\end{figure}

\subsection{S206}

S206 is a nearby, $D_{\rm sun} = 3.3$\kpc, diffuse \hii\ region, but
VLA continuum images do reveal some compact structures
\citep{balser95}.  The LTE NEBULA model matches the \hep3\ spectrum
very well with an ${\rm rms} = 0.26$\mk, the lowest value in our
sample (see Figure~\ref{fig:s206_he3}).  There is no systematic trend
in the RRL intensity with principal quantum number, n, indicating that
non-LTE effects and pressure broadening are negligible (see
Figures~\ref{fig:s206_rrl1}-\ref{fig:s206_rrl2}).  The nebula is
ionized by a single O4-O5 star and Fabry-Perot spectrophotometer data
suggest that the \hii\ region should contain no neutral helium
\citep{deharveng00}; therefore the ionization correction is
$\kappa_{\rm i} = 1.0$.  We derive a \her3\ abundance ratio of
\nexpo{1.89 \pm\ 0.16}{-5} by number.  This $1\,\sigma$ uncertainty
implies an 8.5\% accuracy in the abundance derivation.

\subsection{S209}

The large, diffuse \hii\ region S209 is the most distant in our sample
with a Galactocentric radius of $R_{\rm gal} = 16.2$\kpc.
Spectrophotometry reveals an O9 and two B1 stars exciting S209
\citep{chini84}.  The LTE NEBULA model contains several compact
components \citep[see][]{balser95} and is a good fit to the data with
a residual ${\rm rms} = 0.49$\mk\ in the \hep3\ sub-band
(Figure~\ref{fig:four_he3}). Some instrumental baseline structure
appears near the \hep3\ transition, however, increasing the
uncertainty of our measurement.  This structure appears to exist in
all scans and was not some intermittent feature (e.g., RFI).  RRL data
indicate that non-LTE effects are negligible (see
Figures~\ref{fig:s209_rrl1}-\ref{fig:s209_rrl2}).  A carbon line is
visible in the brighter RRL transitions and arises from a primarily
neutral photodissociation region (PDR) surrounding the ionized nebula
\citep[e.g.,][]{wenger13}.  Since the carbon emission line region is
not part of the \hii\ region, and we only consider hydrogen and helium
in our models, the NEBULA synthetic spectra do not include carbon
lines.  We determine a small ionization correction, $\kappa_{\rm i} =
1.10$, and derive a \her3\ abundance ratio of \nexpo{1.10 \pm\
  0.28}{-5} yielding a nominal uncertainty of 25\%.  The larger
uncertainty is primarily due to a weak \hep3\ line intensity (see
Table~\ref{tab:he3}).

\subsection{M16}

The Eagle nebula (M16) is a nearby, $D_{\rm sun} = 2.0$\kpc,
well-studied \hii\ region \citep{white99, indebetouw07}.  M16 is
ionized by several O5-type stars \citep{hillenbrand93}.  Using the VLA
continuum data published in \citet{white99}, we model the Eagle nebula
with several compact components and include a halo component from the
140 Foot continuum data.  The NEBULA LTE model produces high-n RRLs
that are a good fit to GBT spectra and therefore any non-LTE effects
or pressure broadening should be negligible.  Multiple velocity
components are present in the GBT RRL spectra, however, with a weaker
component at higher frequencies or lower velocities (see
Figures~\ref{fig:m16_rrl1}-\ref{fig:m16_rrl2}).  Nevertheless, the
\hep3\ sub-band residuals are well behaved with an ${\rm rms} =
0.97$\mk.  This is the largest \hep3\ sub-band rms in our sample
because of the short integration time on this source.  A carbon RRL is
also detected but as discussed above this transition arises from a PDR
and not within the \hii\ region.  Applying an ionization correction of
$\kappa_{\rm i} = 1.17$ yields a \her3\ abundance ratio of \nexpo{2.28
  \pm\ 0.48}{-5} and a nominal accuracy of 21\%.

\subsection{G29.9}

\begin{figure}
\includegraphics[angle=0,scale=0.8]{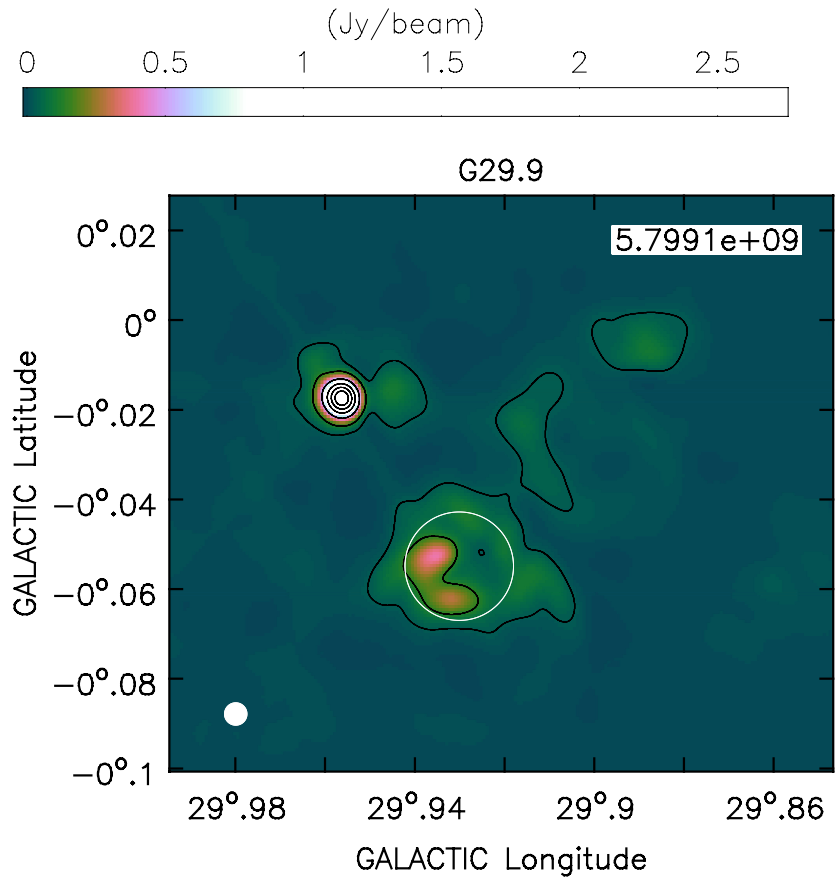} 
\caption{JVLA continuum image of G29.9 at 5.8\ghz.  The contour levels
  are at 0.01, 0.05, 0.2, 0.4, 0.6, and 0.8 times the peak value of
  2.76\jyb.  The rms noise in the image is 1\mjyb.  The synthesized
  beam size is shown in the bottom-left corner.  The white circle
  corresponds to the location and size of the GBT HPBW.}
\label{fig:g29.9_cont}
\end{figure}

The \hii\ region G29.9 is located near the large star formation
complex associated with W43 at the end of the Galactic bar.  There are
few \hii\ regions within the extent of the bar \citep{bania10}, and
moreover most \hii\ regions at these Galactic longitudes have
uncertain distances.  In our sample, G29.9 is the closest source to
the Galactic Center with $R_{\rm gal} = 4.4$\kpc.  We did not obtain
VLA continuum data for G29.9 in our previous work, so here we use a
recent image from the GLOSTAR survey with the Jansky VLA (JVLA) at
5.8\ghz\ (A. Brunthaler et al., in preparation).
Figure~\ref{fig:g29.9_cont} is the continuum image of G29.9 where the
circle corresponds to the GBT HPBW.

\begin{figure}
\includegraphics[angle=0,scale=0.45]{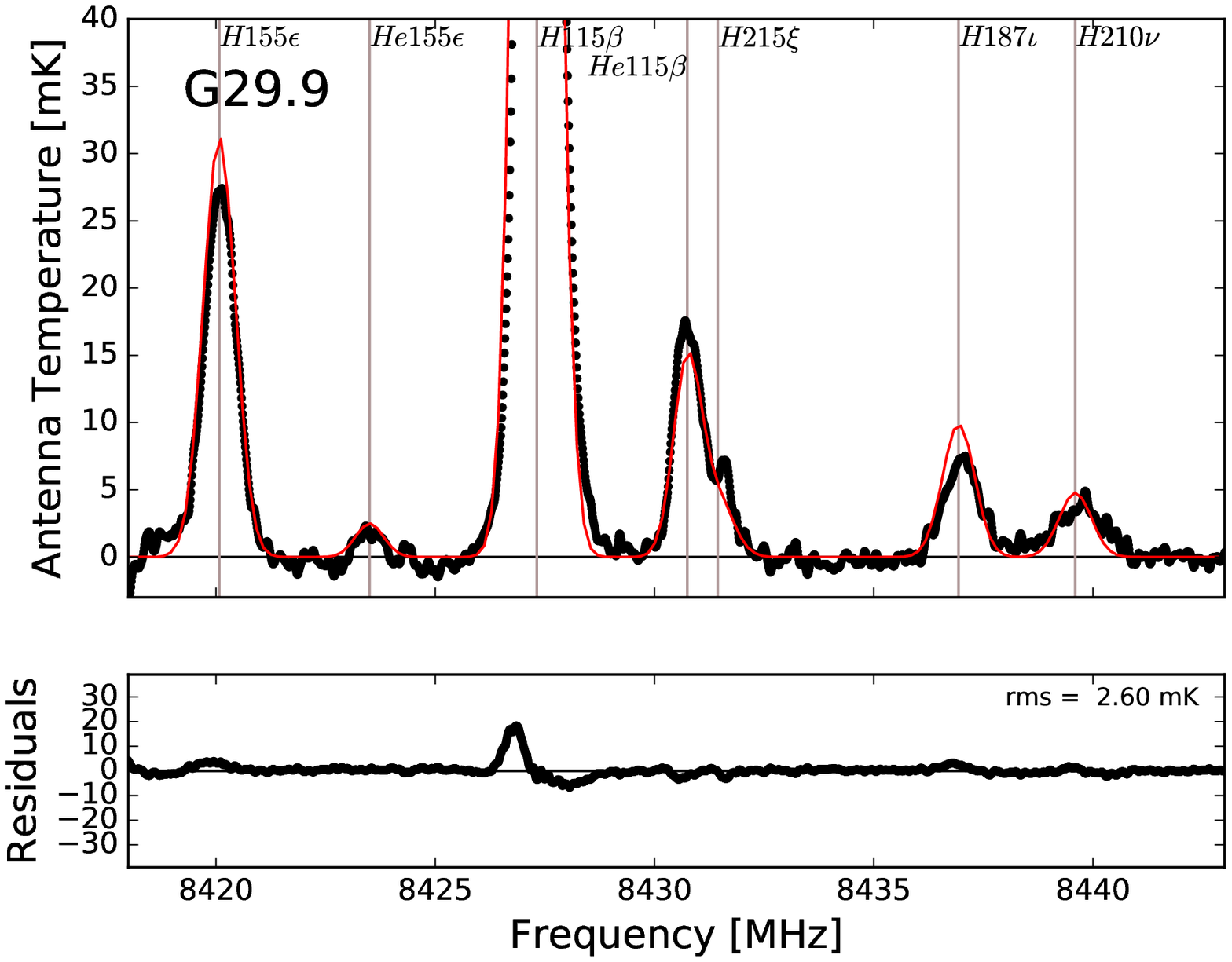} 
\includegraphics[angle=0,scale=0.45]{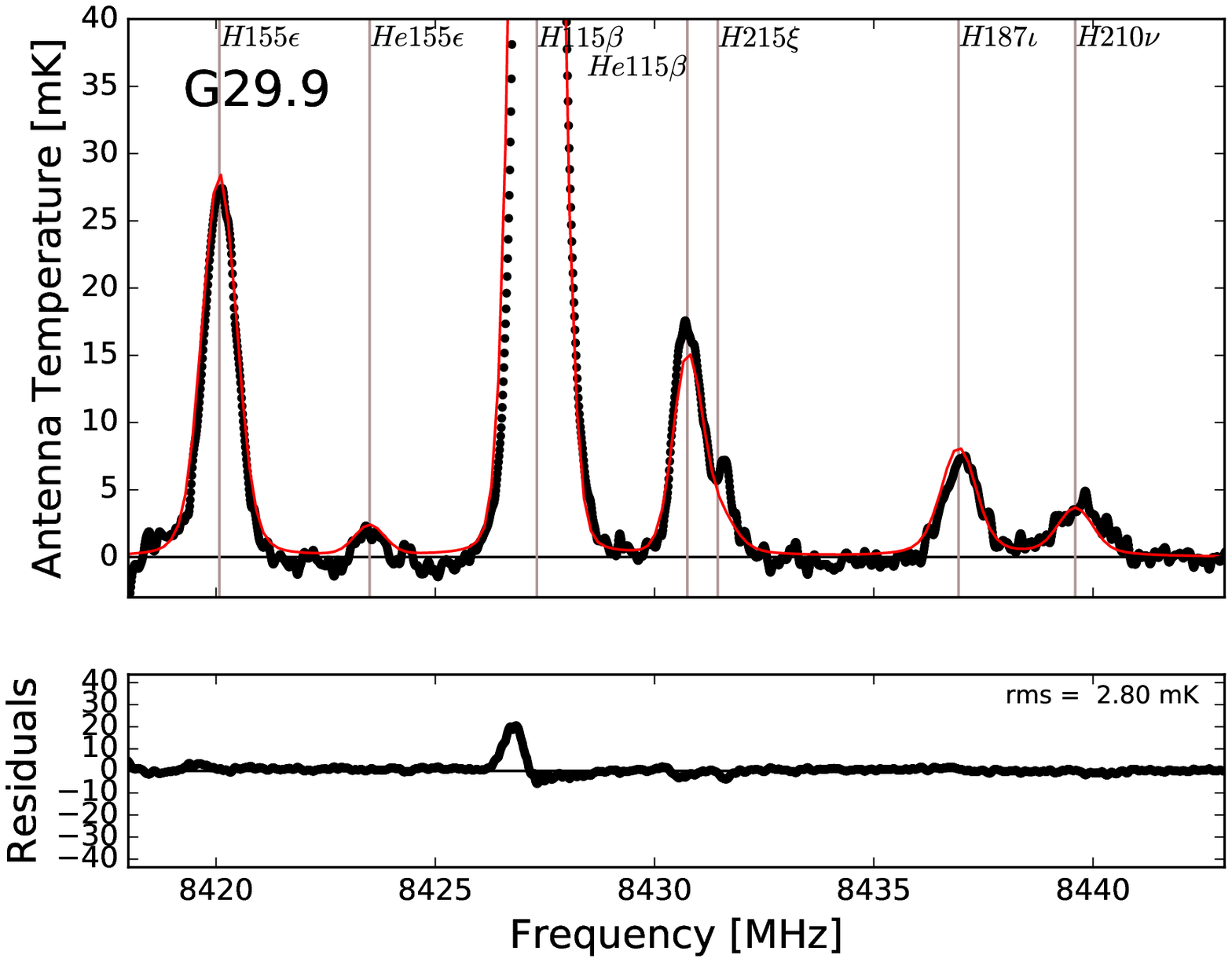} 
\caption{The H115$\beta$ sub-band for G29.9 assuming LTE (left) and
  including non-LTE effects (right).  The C115$\beta$ line is visible
  to the right of the H115$\beta$ in the GBT data but carbon is not
  included in the models.  See Figure~\ref{fig:s206_he3} for details.}
\label{fig:g29.9_nonlte}
\end{figure}

We model G29.9 as two compact components based on the JVLA data and a
halo component from the 140 Foot continuum (see
Table~\ref{tab:models}).  The LTE NEBULA model reveals non-LTE
effects: the high-n RRL intensities are overpredicted by the model
indicating pressure broadening.  Figure~\ref{fig:g29.9_nonlte} shows
the RRL spectrum for the H115$\beta$ sub-band.  The LTE NEBULA model
(left panel) predicts brighter, narrower profiles.  The non-LTE model
including pressure broadening reduces these discrepancies, but the
model RRL intensities for transitions with high n are still too large.
We therefore apply a filling factor of 0.01 in the compact components
to simulate higher local electron densities or additional density
structure.  The non-LTE models that include a clumpier medium are a
better fit to the data as shown by the right panel in
Figure~\ref{fig:g29.9_nonlte}.  The feature to the right of the
He115$\beta$ line corresponds to a blend between the C115$\beta$ and
H215$\xi$ lines.  Since we are not modeling carbon this feature is
stronger than the NEBULA prediction.  Simulations show that deriving
the \her3\ abundance ratio for sources with such structure have larger
uncertainties \citep{balser99a, bania07}.  To be conservative, we
therefore use 3 times the rms of the residuals for the uncertainty in
\hepr3.

The adopted NEBULA model for G29.9 is a reasonably good fit to the
data with an ${\rm rms} = 0.58$\mk\ in the residuals of the \hep3\
sub-band (Figure~\ref{fig:four_he3}).  There do appear to be multiple
velocity components detected in the RRLs (see
Figures~\ref{fig:g29.9_rrl1}-\ref{fig:g29.9_rrl2}).  The low \hepr4\
abundance ratio of 0.07 for G29.9 at $R_{\rm gal} = 4.4$\kpc\ produces
a significant ionization correction of $\kappa_{\rm i} = 1.39$.  We
derive a \her3\ abundance ratio of \nexpo{2.23 \pm\ 0.76}{-5} yielding
a nominal uncertainty of 34\%.

\subsection{\ngc{7538}}

\ngc{7538} is a nearby, $D_{\rm sun} = 2.8$, well-studied star
formation region \citep[e.g.,][]{fallscheer13, luisi16}.  The \hii\
region, also known as Sharpless 158 or S158, is a centrally diffuse
source with a bright rim to the west.  An O7 star located at the
center of the nebula provides the ionizing photons
\citep{deharveng79}.  The VLA continuum data show several compact
components with a very bright source to the south-east
\citep{balser95}.  The LTE NEBULA model does not provide a good fit to
the GBT data.  Similar to G29.9, non-LTE effects in the form of
pressure broadening are present.  Including pressure broadening and a
filling factor of 0.15 for the compact components produces a
reasonable fit to the data.  To be conservative, we therefore use 3
times the rms of the residuals for the uncertainty in \hepr3.

The GBT baseline structure appears to be slightly worse for this
source which explains some of the discrepancies between the model and
data (see Figures~\ref{fig:ngc7538_rrl1}-\ref{fig:ngc7538_rrl2}).
This is not unexpected since the total system temperature is higher
for this brighter source and we have demonstrated that the baseline
structure roughly scales with source intensity (see
\S{\ref{sec:baseline}}).  For example, the continuum intensity for
\ngc{7538} is almost two times the next brightest source in our
sample, G29.9.  Nevertheless, the residuals in the \hep3\ sub-band are
reasonable with ${\rm rms} = 0.74$\mk\ (Figure~\ref{fig:four_he3}).
We derive an ionization correction of $\kappa_{\rm i} = 1.06$ and a
\her3\ abundance ratio of \nexpo{1.97 \pm\ 1.08}{-5} giving a nominal
uncertainty of 55\%.

Based on the limited RRL data from the 140 Foot telescope,
\citet{bania02} suggested that G29.9 and \ngc{7538} were
morphologically simple and therefore included them in their sample of
sources with accurate \her3\ determinations.  The much richer RRL data
set observed with the GBT show that these sources are not as simple as
expected.  Nevertheless, our NEBULA models match the data reasonably
well and we include them in our analysis but with higher
uncertainties.

\section{The Time Evolution of the \he3\ Abundance}

During the Big Bang era of primordial nucleosynthesis the light
element \he3\ is predicted to be made in copious amounts with a final
primordial \her3\ abundance ratio of \her3\ $\sim 10^{-5}$ by number
\citep[see][and references within]{schramm77, boesgaard85, cyburt04}.
This primordial \he3\ abundance is further processed via stellar
nucleosynthesis, but it is difficult to predict the net effect of the
processes that produce and destroy \he3\ in stars.  There is evidence
that some stars do produce \he3, however, since detection of \hep3\ in
a few PNe reveal abundances of \her3\ $\sim 10^{-3}$, significantly
higher than the primordial value \citep[e.g.,][]{balser06b}.

Observations of \he3\ in Galactic \hii\ regions indicate a relatively
flat \her3\ radial abundance gradient across the Galactic disk, the
\he3\ Plateau \citep{bania02}.  Galactic chemical evolution models
predict a higher rate of star formation in the central regions of the
disk and we expect negative/positive radial gradients for elements
that have a net production/destruction in stars
\citep[e.g.,][]{chiappini97, schonrich09, lagarde12, minchev14}.  The
\he3\ Plateau, with zero radial gradient, was therefore interpreted as
representing the primordial abundance with \her3 = \nexpo{1.1 \pm\
  0.2}{-5} \citep{bania02}.  This value was independently confirmed
with higher accuracy by combining results from the WMAP with BBN
models \citep{romano03}.

The \he3\ Problem results from trying to reconcile the large \her3\
abundance ratios found in a few PNe, consistent with standard stellar
yields, with the essentially primordial values found in \hii\ regions
\citep{galli97}.  \citet{rood84} were the first to suggest that some
sort of slow extra-mixing might be occurring to reduce the \he3\
abundance and that such mixing could also explain the depletion of
\li7\ in main-sequence stars and the low abundance \cratio\ ratios
seen in low-mass RGB stars.  Over the past 30 years the \he3\ Problem
has galvanized a significant effort into improving stellar evolution
models with more realistic mixing physics \citep[e.g.,][]{zahn92,
  charbonnel95, eggleton06, charbonnel07a}.  \citet{lagarde12} provide
a potential solution to the \he3\ Problem by including the mixing
effects of the thermohaline instability and rotation.  Models that
include this physics significantly reduce the present day \he3\
abundance.  (They do not, however, deplete \he3\ below its primordial
value.)  Here we test these models by determining accurate \her3\
abundance ratios in five \hii\ regions over a range of Galactocentric
radii ($4.4\,{\rm kpc} < R_{\rm gal} < 16.2\,{\rm kpc}$).

\begin{figure}
\includegraphics[angle=0,scale=0.8]{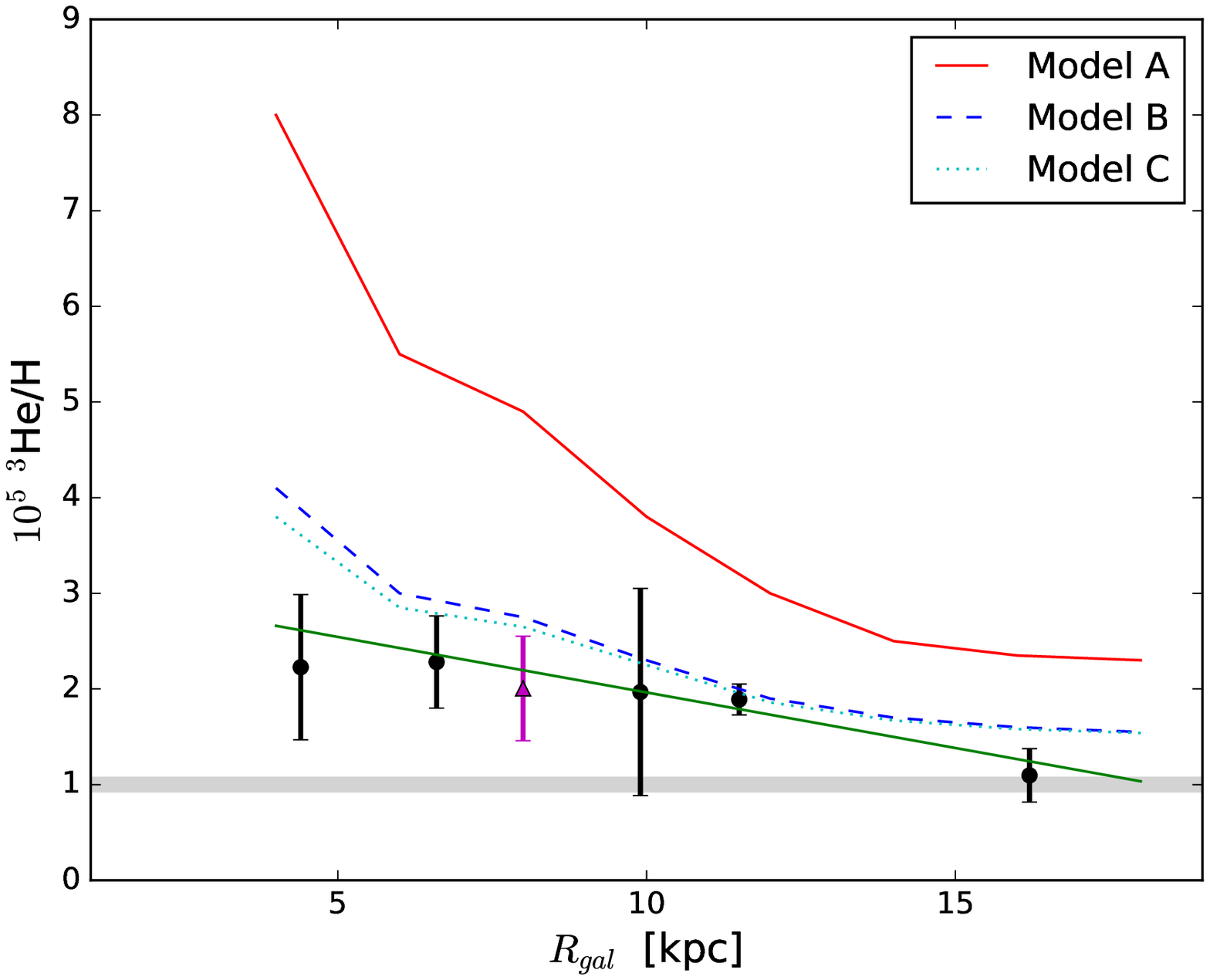} 
\caption{Evolution of \he3\ within the Milky Way disk.  Plotted is the
  \her3\ abundance ratio as a function of Galactocentric radius.  The
  GBT results derived here for \hii\ regions are shown as black
  circles.  The magenta triangle is the LISM abundance derived by
  \citet{gloeckler96} from their {\it Ulysses} measurement.  The
  curves correspond to three models from \citet{lagarde12}.  These
  models predict the \her3\ abundance ratio after 10\gyr\ of stellar
  processing has modified the primordial abundance produced during
  BBN.  {\it They are not fits to the \hii\ region $^{\it 3}$He/H
    abundances derived from observations shown here.}  Model A (solid
  red) uses standard yields for all stars; model B (blue dashed)
  considers thermohaline instability and rotation-induced mixing for
  96\% of low-mass stars ($M \le\ 2.5$\msun) and 100\% of high-mass
  stars ($M > 2.5$\msun); and model C (cyan dotted) considers
  thermohaline instability and rotation-induced mixing for all stars.
  The gray horizontal region is the primordial abundance range of
  \her3\ = \nexpo{1.00 \pm\ 0.07}{-5} from WMAP and BBN
  \citep{cyburt08}.  The green solid line is a linear fit to the data
  (\hii\ regions and LISM) using orthogonal distance regression which
  finds a slope of $-0.116 \pm\ 0.022\, \times
  \,$\expo{-5}$\,$kpc$^{-1}$.}
\label{fig:he3vsRgal}
\end{figure}

Figure~\ref{fig:he3vsRgal} summarizes our results.  Plotted is the
present day \her3\ abundance ratio as a function of $R_{\rm gal}$,
corresponding to $\sim 10$\gyr\ of stellar and Galactic evolution that
has occurred since the Milky Way formed.  Our GBT \hii\ region results
are shown as black circles.  The magenta triangle corresponds to the
\he3\ abundance in the local interstellar medium (LISM) from in situ
measurements on the spacecraft {\it Ulysses} \citep{gloeckler96}.  The
primordial \he3\ abundance, represented by the gray horizontal region,
is based on BBN and WMAP results and their uncertainties \citep[][also
see \citet{cyburt16, coc17}]{cyburt08}.  The models of
\citet{lagarde12} are shown as three different curves.  The top (solid
red) curve corresponds to standard \he3\ stellar yields and is
inconsistent with the present day \hii\ region \he3\ abundances.  The
two lower curves (dashed green and dotted cyan) include the
thermohaline instability and rotation-induced mixing.  They are
slightly higher than the data but consistent with the notion that
there is a small net production of \he3\ in stars.  The models do not
predict a linear relationship between \her3\ and $R_{\rm gal}$, but
for reference we show a linear fit to the data which has a slope of
$-0.116 \pm\ 0.022\, \times \,$\expo{-5}$\,$kpc$^{-1}$.

Clearly more accurate \her3\ abundances are needed for \hii\ regions
located at smaller $R_{\rm gal}$ where the models predict an upturn in
the \he3\ abundance.  The G29.9 \her3\ abundance derivation may suffer
from additional, unknown systematic error.  Previously the sample of
morphologically simple \hii\ regions at $R_{\rm Gal} \lsim\ 4$\kpc\
was extremely small due in part to the Galactic distribution of the
\hii\ region population.  New \hii\ region discovery surveys, however,
have the potential to provide new targets in this critical $R_{\rm
  gal}$ zone \citep{bania10, bania12, anderson11, anderson15,
  anderson18, brown17}.  Finally, a new generation of GCE models would
improve our understanding of the Galactic chemical evolution of \he3.
It may be that beside the stellar \he3\ yields other assumptions made
in the current GCE models are important.

\citet{charbonnel07b} proposed that strong magnetic fields in RGB
stars that evolved from Ap stars could inhibit mixing from the
thermohaline instability and thereby explain the few PNe with high
values of \her3.  \citet{lagarde12} simulates the effects of such
stars on the evolution of \he3\ (Model B), but the results are only
slightly different than if thermohaline mixing was occurring in all
stars (Model C).  We cannot distinguish between these models with our
\hii\ region data.  A high \her3\ abundance ratio derived for even a
single PN would indicate that some mechanism must be at play to
inhibit the extra mixing in this object. Extra-mixing should otherwise
occur in all low-mass stars.  There are three PNe with published
\hep3\ detections: \ngc{3242} \citep{rood92, balser97, balser99b},
J320 \citep{balser06b}, and \ic{418} \citep{guzman16}.  The \her3\
abundance ratio derived for these detections range from \nexpo{2}{-4}
to \nexpo{6}{-3}, an order of magnitude higher than the abundances
found in \hii\ regions.

Observations of the \hep3\ spectral transition are challenging.
Accurate measurement of these weak, broad lines requires significant
integration time and a stable, well-behaved spectrometer
\citep{balser94}.  This is particularly true when observing \hep3\ in
PNe since the lines are weaker and broader than for \hii\ regions.
Much effort went into detecting \hep3\ in \ngc{3242} using the
Max-Planck Institut f\"{u}r Radioastronomie (MPIfR) 100\m\ and the
NRAO 140 Foot telescopes.  Nevertheless, preliminary GBT observations
of \hep3\ in \ngc{3242} do not confirm the MPIfR 100\m\ result
\citep{bania10}.  One sign that there were problems with the
\ngc{3242} data was the large discrepancy between the model and
observed intensity of the H171$\eta$ RRL \citep[See Figure 3
in][]{balser99b}.  Additional GBT observations toward \ngc{3242} have
been taken and we plan to perform detailed modeling of this PNe.

\citet{guzman16} report a \hep3\ detection in \ic{418} with a SNR of
5.7 using the National Aeronautics and Space Administration (NASA)
Deep Space Station 63 (DSS-63).  They have simultaneously observed
many RRLs but do not provide any detailed models to assess their
accuracy.  For example, the \her4\ abundance ratio is 0.11 and 0.037
for the 91$\alpha$ and 92$\alpha$ RRLs, respectively.  These adjacent
RRLs should produce the same result yet they are a factor of 3
different.  We are therefore very suspicious of the claimed detection
of \hep3\ in \ic{418}.

The advantage of interferometers like the VLA is that much, but not
all, of the instrumental baseline structure is correlated out.  We
therefore deem that the detection in J320 with the VLA is more robust.
The SNR is only 4, but when averaging over a halo region the SNR
increases to 9.  Nevertheless, the limited bandwidth of the VLA did
not allow the simultaneous observation of many RRLs to constrain the
models and assess the spectral baseline stability.  Observations with
the much improved Jansky VLA would provide for a more robust
evaluation.

\section{Summary}

Studies of \he3\ provide important constraints to Big Bang
nucleosynthesis, stellar evolution, and Galactic evolution.  Standard
stellar evolution models that predict the production of copious
amounts of \he3\ in low-mass stars, consistent with the high \her3\
abundance ratios found in a few planetary nebulae, are at odds with
the approximately primordial values determined for Galactic \hii\
regions.  This inconsistency is called the \he3\ Problem.  Models that
include mixing from the thermohaline instability and rotation provide
a mechanism to reduce the enhanced \her3\ abundances during the RGB
stage \citep{charbonnel10, lagarde11}.  These yields together with GCE
models predict modest \he3\ production by stars over the lifetime of
the Milky Way \citep{lagarde12}.  The scatter of the \he3\ Plateau
abundances determined by \citet{bania02} is large and spans the range
of abundances predicted by \citet{lagarde12}.

Here we detect \hep3\ emission in five morphologically simple Galactic
\hii\ regions with the GBT over a wide range of Galactocentric radii:
$4.4\,{\rm kpc} < R_{\rm gal} < 16.2\,{\rm kpc}$.  Our goal is to
derive accurate abundance ratios for a small sample of sources to
uncover any trend in the \her3\ abundance with $R_{\rm gal}$, and to
compare our results with the predictions of \citet{lagarde12}.  We use
the radiative transfer program NEBULA, together with GBT measurements
of over 35 RRL transitions, to constrain \hii\ region models and
determine accurate \her3\ abundance ratios.  The RRLs are measured
with high SNRs and allow us to assess the quality of the spectral
baselines which is critical to measuring wide, weak lines accurately.
We find that S209, S209, and M16 are indeed simple sources that are
well characterized by LTE models.  G29.9 and \ngc{7538}, however,
contain density structure on spatial scales not probed by our VLA
observations and also require non-LTE models; the \her3\ abundance
ratios derived for these sources are therefore less accurate.  We
apply an ionization correction to convert \hepr3\ to \her3\ using
\he4\ RRLs since there may exist some neutral helium within the \hii\
region.

We determine a \her3\ radial gradient of $-0.116 \pm\ 0.022\, \times
\,$\expo{-5}$\,$kpc$^{-1}$, consistent with the overall trend
predicted by \citet{lagarde12}.  Our \her3\ abundance ratios, however,
are typically slightly less than the models that include thermohaline
mixing.  We do not have enough accuracy to determine whether or not
strong magnetic fields in some stars could inhibit the thermohaline
instability as predicted by \citep{charbonnel07b}.  More conclusive
measurements of \hep3\ in PNe would be useful to confirm that indeed
some stars do produce significant amounts of \he3.

\acknowledgments

We dedicate this paper to our late colleague Bob Rood who founded our \he3\
research team 35 years ago. We acknowledge with fondness and respect the support
and inspiration given by our colleagues of the international light elements
community who over the years have become our friends. We thank the GBT telescope
operators who made these observations.  Their expertise and diligence in running
our observing scripts were exceptional. We thank Bill Cotton for providing us
with the G29.9 JVLA image. This research was partially supported by NSF award
AST-1714688 to TMB.

\vspace{5mm}
\facilities{GBT}
\software{TMBIDL \citep{bania16}, NEBULA \citep{balser18}}

\appendix

\section{Galactic \hii\ Region Radio Recombination Line Spectra}

\begin{figure}
\includegraphics[angle=0,scale=0.45]{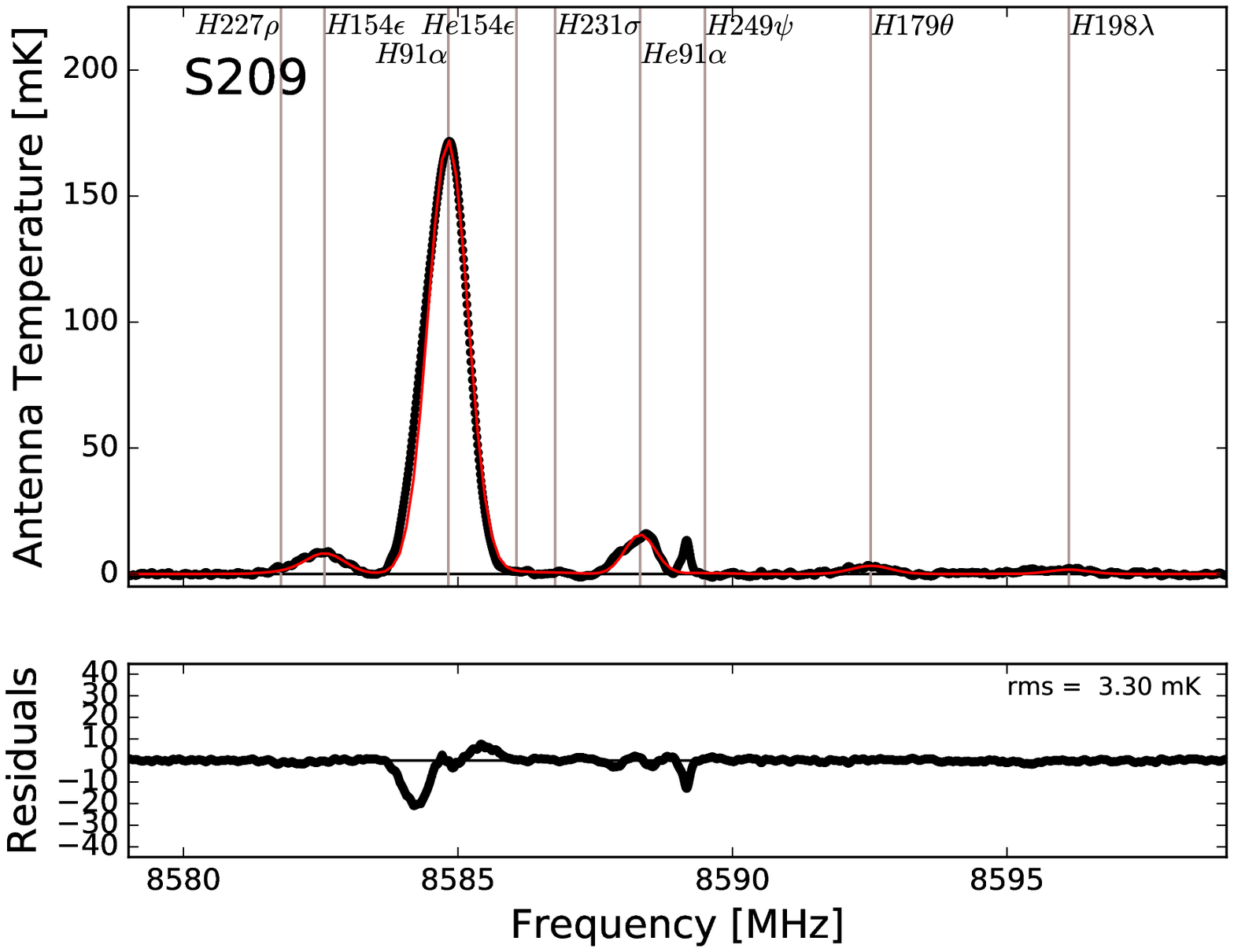} 
\includegraphics[angle=0,scale=0.45]{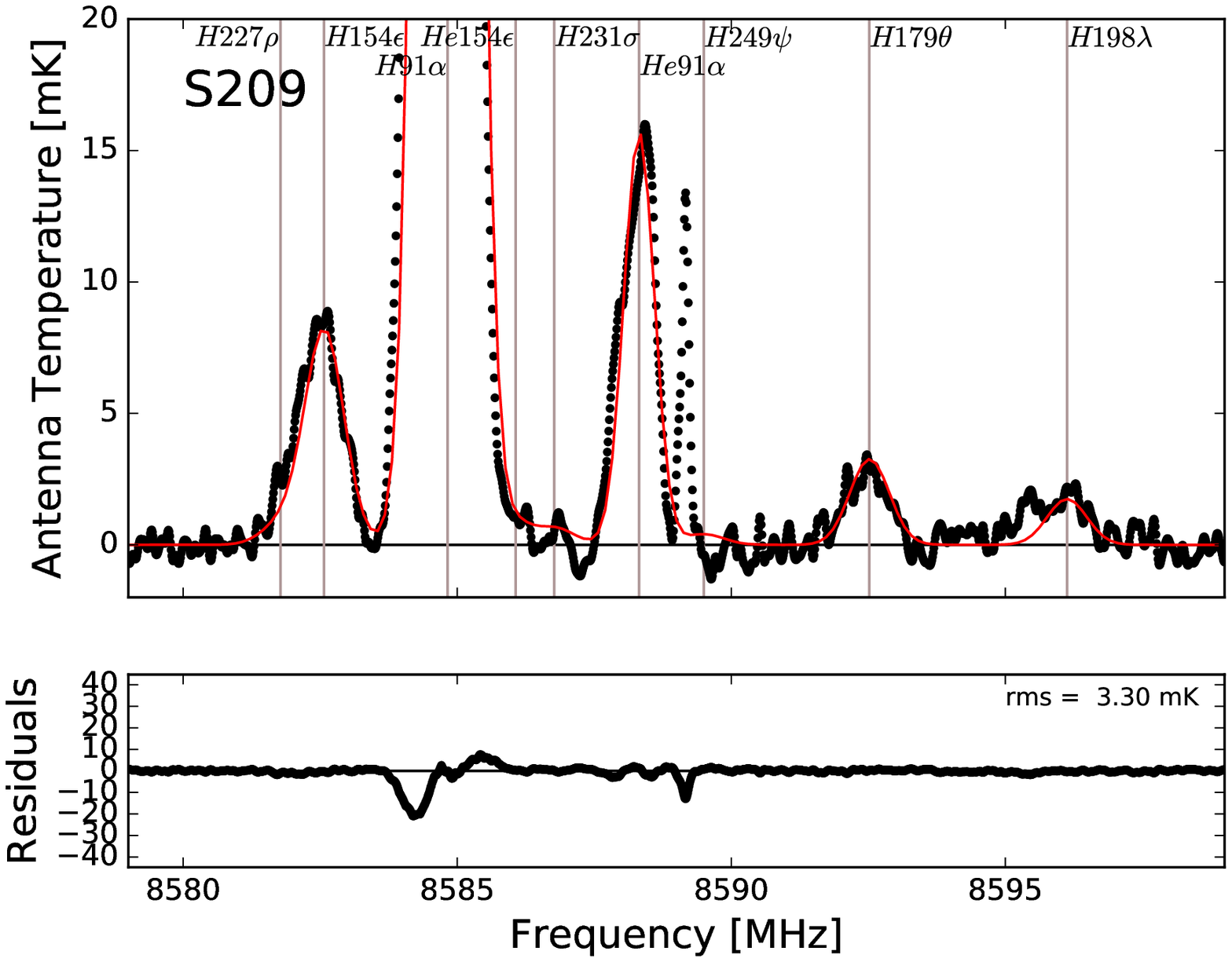} 
\includegraphics[angle=0,scale=0.45]{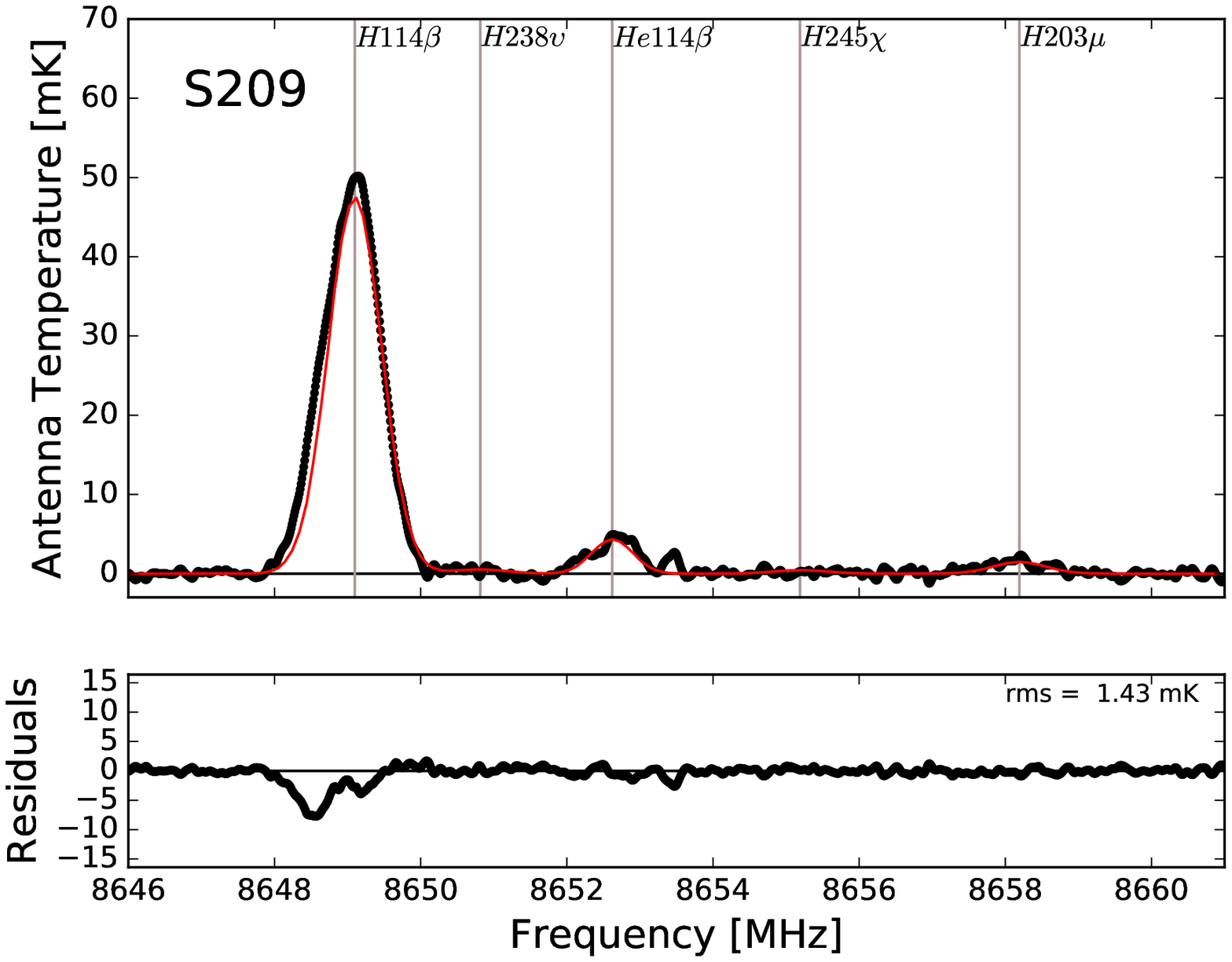} 
\includegraphics[angle=0,scale=0.45]{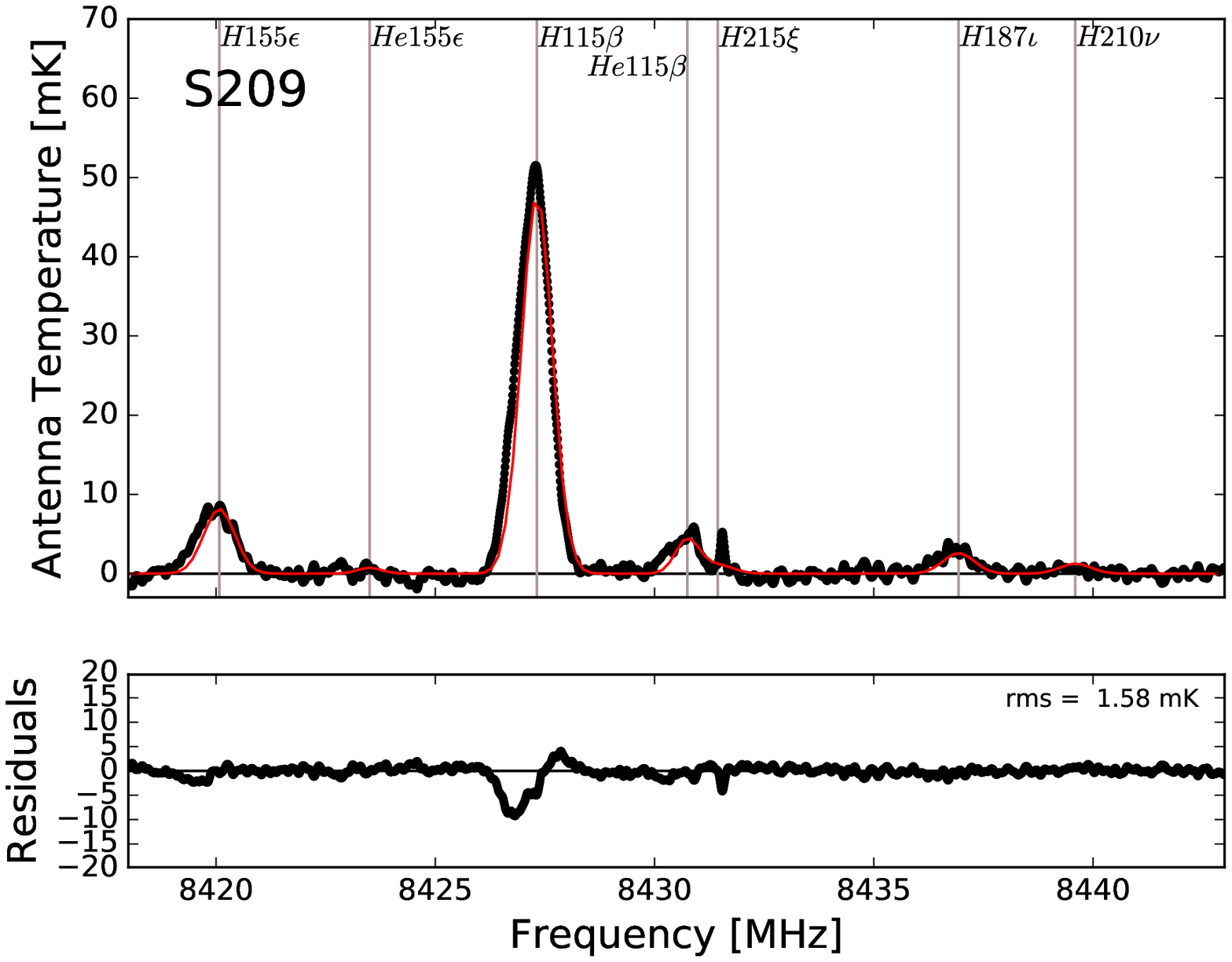} 
\caption{RRL spectra of S209 including the following sub-bands:
  H91$\alpha$ (top-left), expanded view of H91$\alpha$ (top-right),
  H114$\beta$ (bottom-left), and H115$\beta$ (bottom-right).  See
Figure~\ref{fig:s206_rrl1} for details.}
\label{fig:s209_rrl1}
\end{figure}

\begin{figure}
\includegraphics[angle=0,scale=0.45]{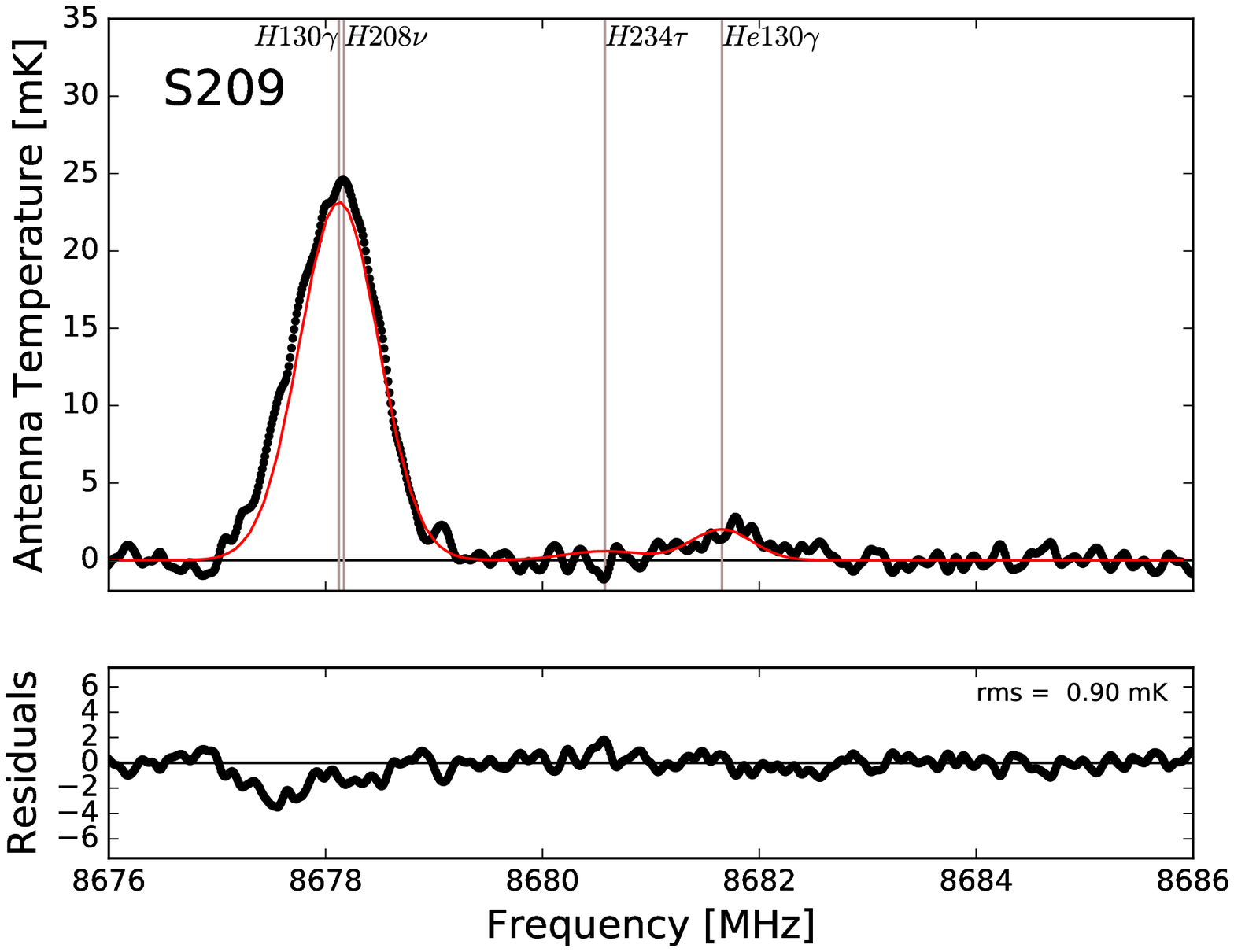} 
\includegraphics[angle=0,scale=0.45]{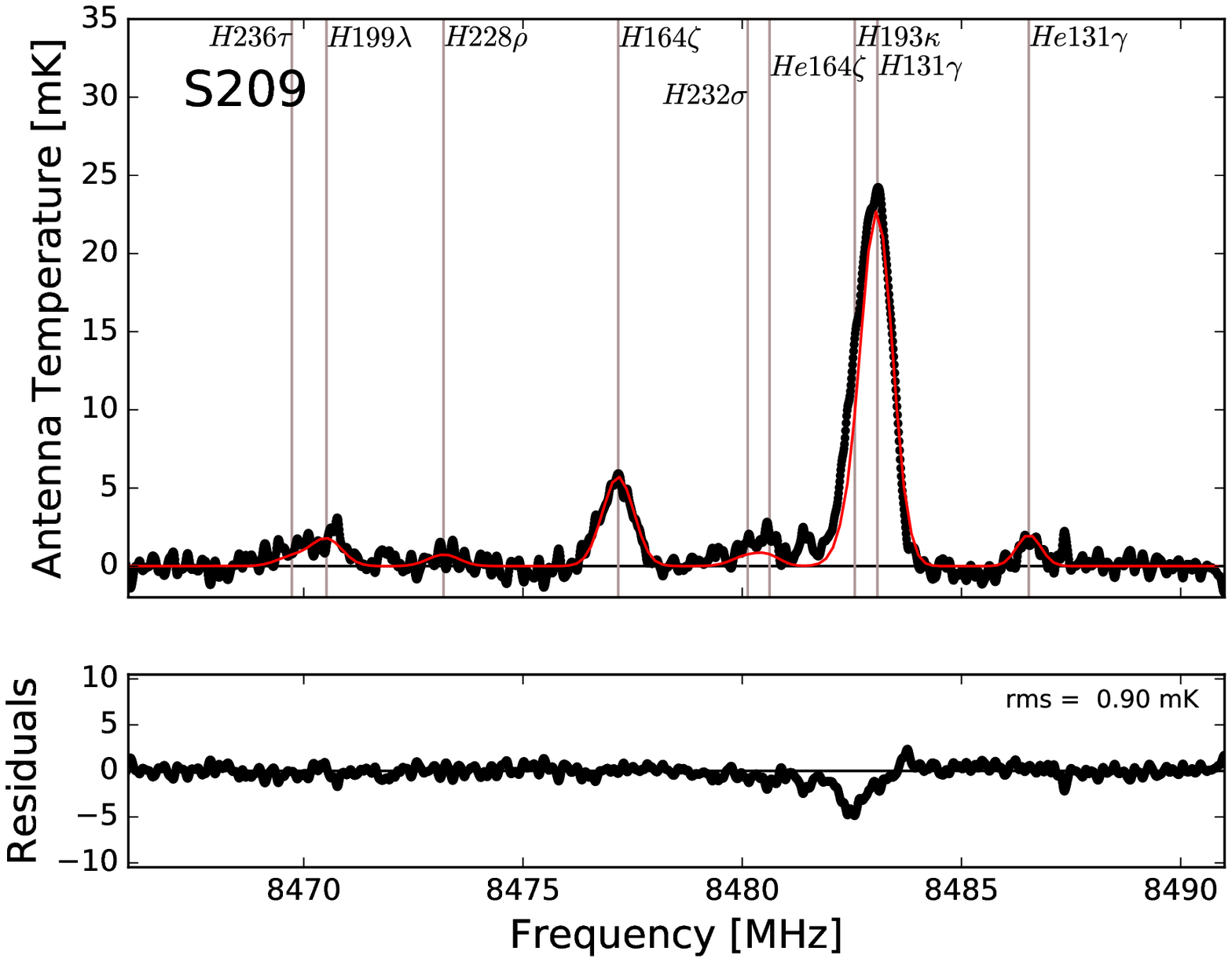} 
\includegraphics[angle=0,scale=0.45]{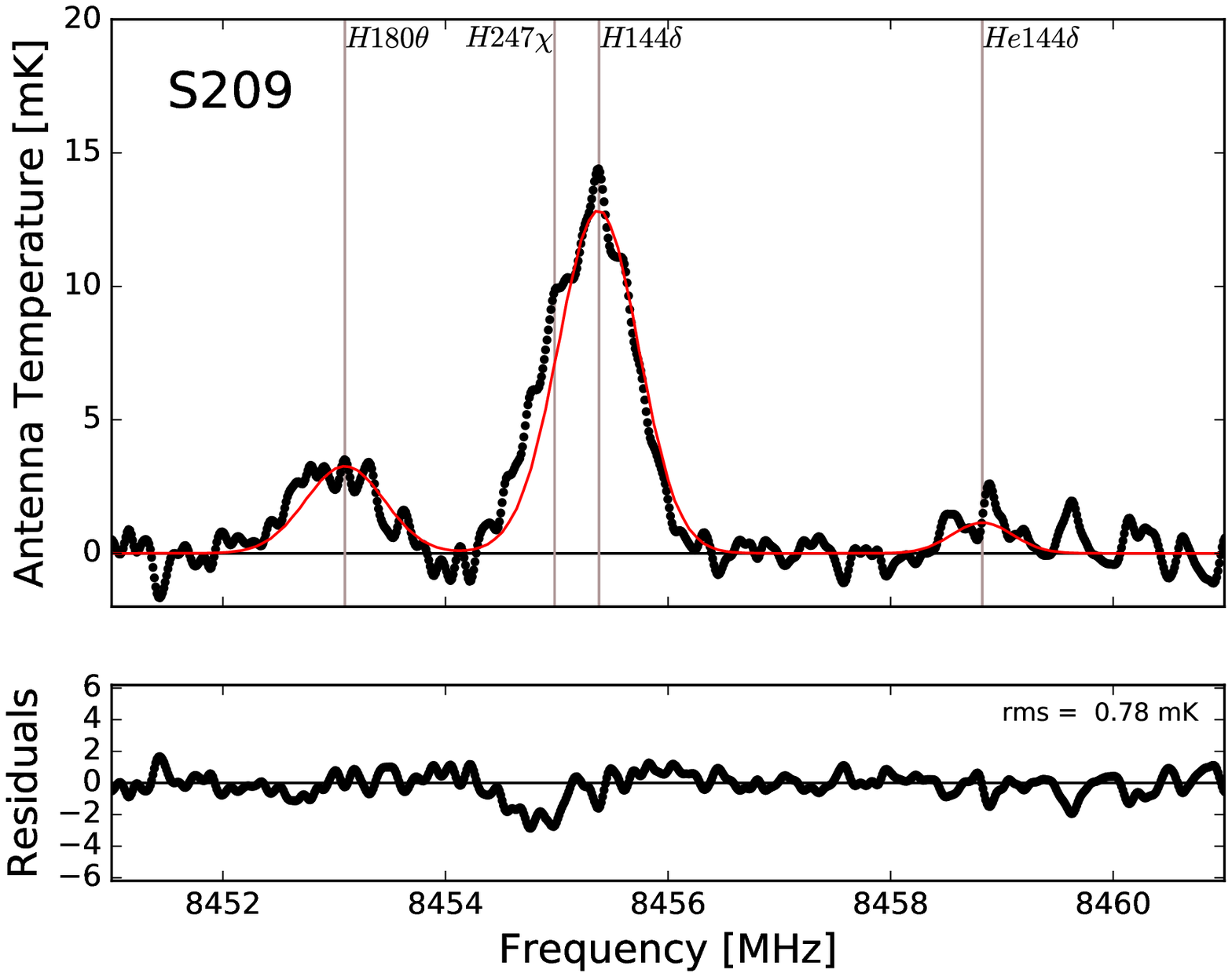} 
\includegraphics[angle=0,scale=0.45]{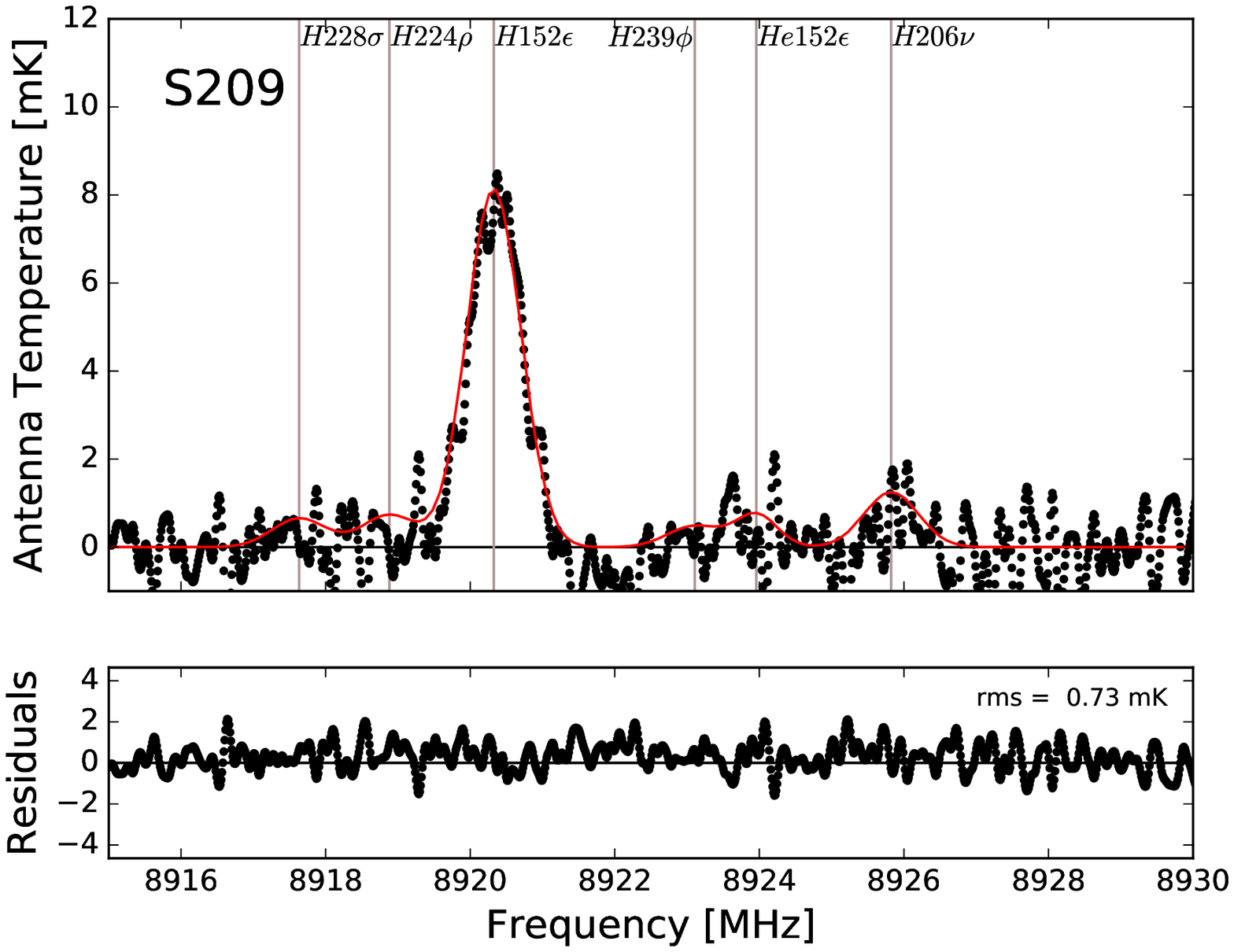} 
\caption{RRL spectra for S209 including the following sub-bands:
  H130$\gamma$ (top-left), H131$\gamma$ (top-right), H144$\delta$
  (bottom-left), and H152$\epsilon$ (bottom-right).  See
  Figure~\ref{fig:s206_rrl2} for details.}
\label{fig:s209_rrl2}
\end{figure}

\begin{figure}
\includegraphics[angle=0,scale=0.45]{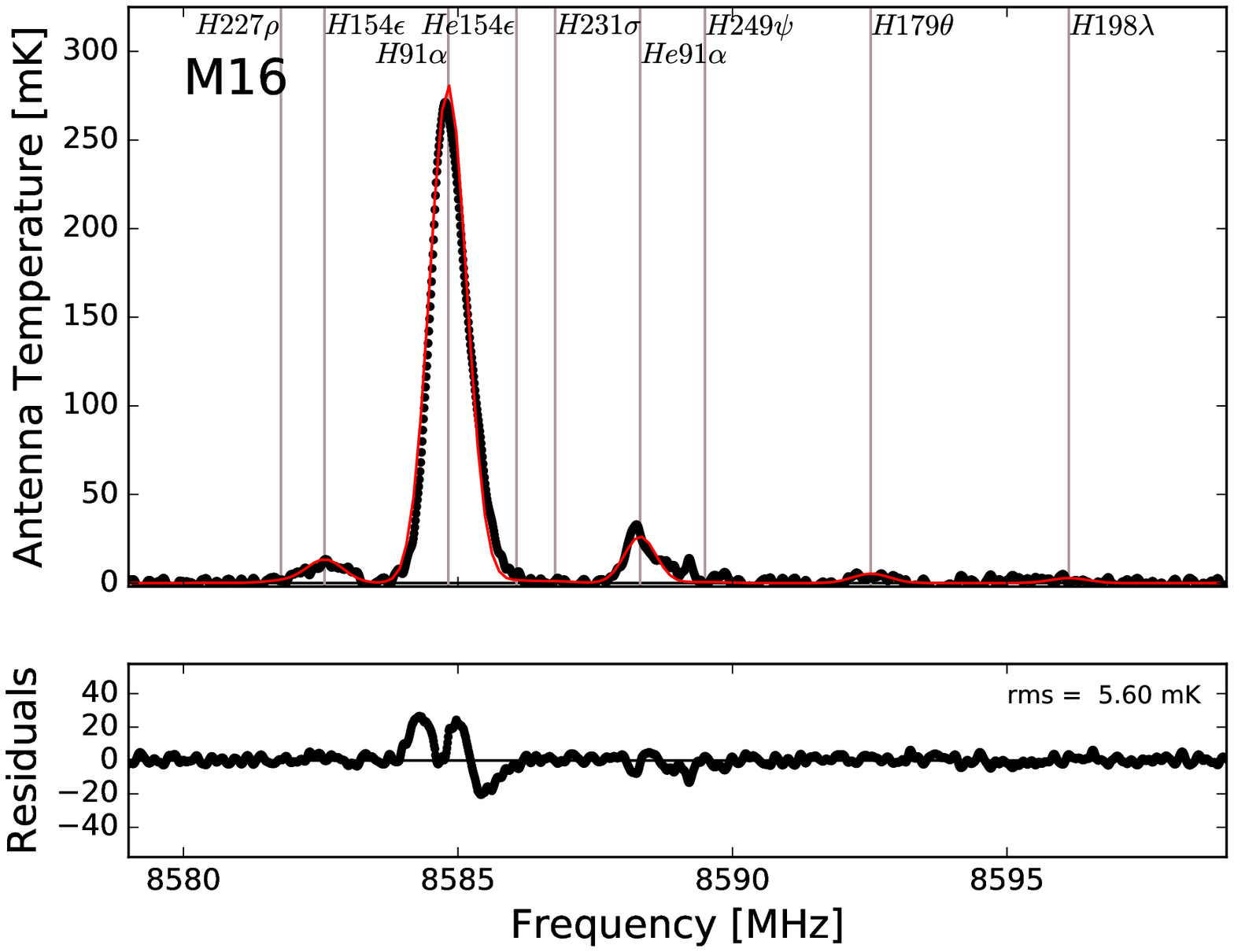} 
\includegraphics[angle=0,scale=0.45]{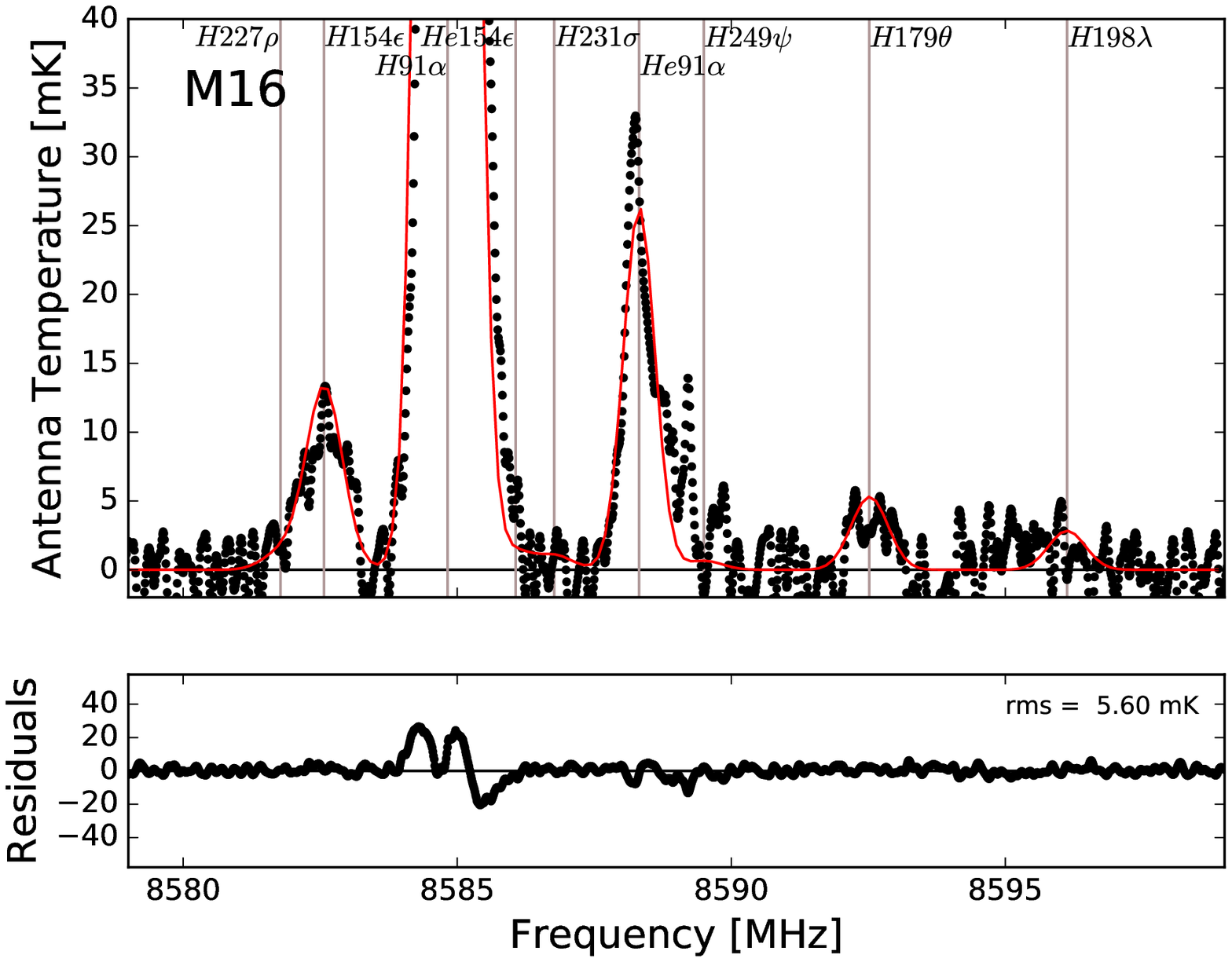} 
\includegraphics[angle=0,scale=0.45]{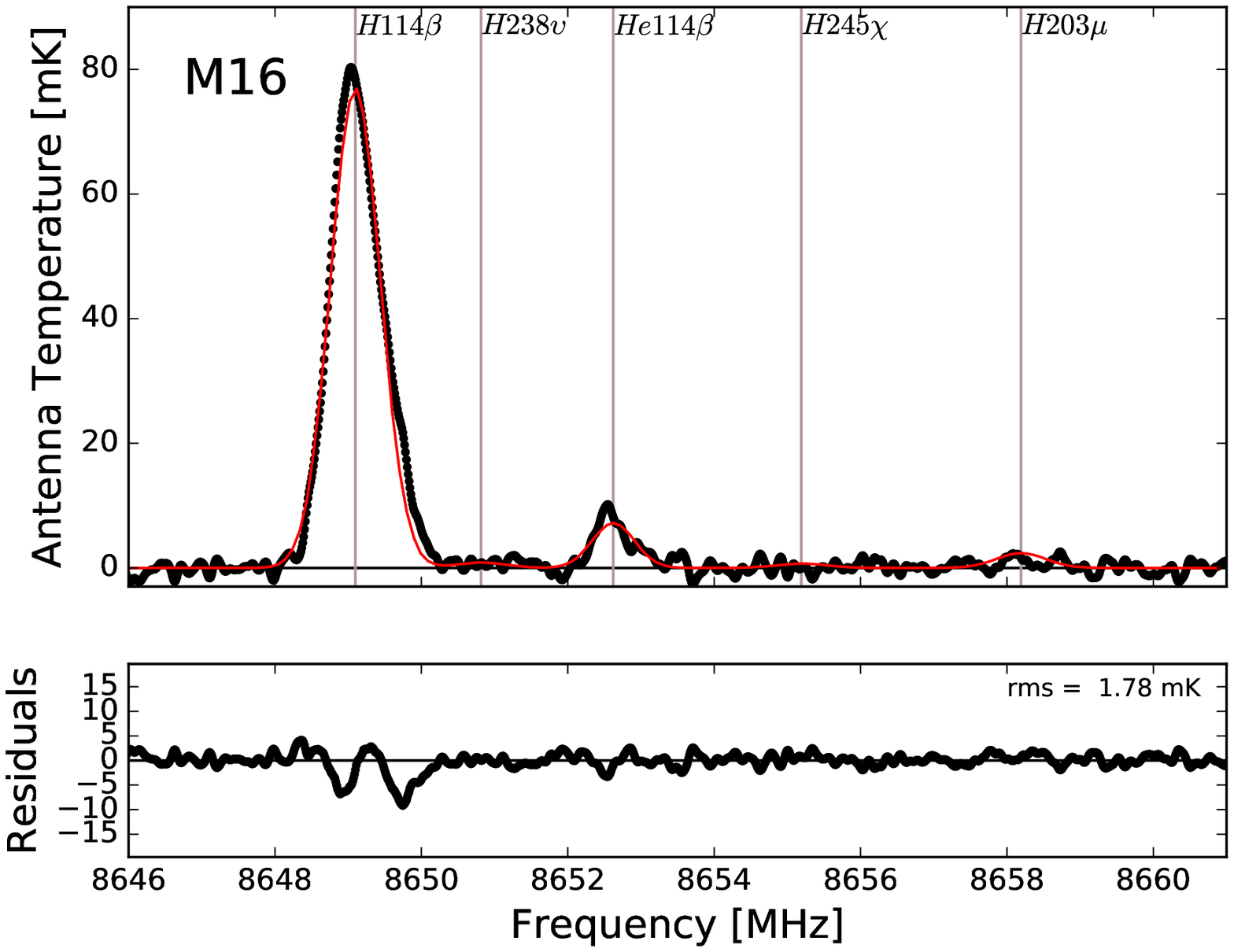} 
\includegraphics[angle=0,scale=0.45]{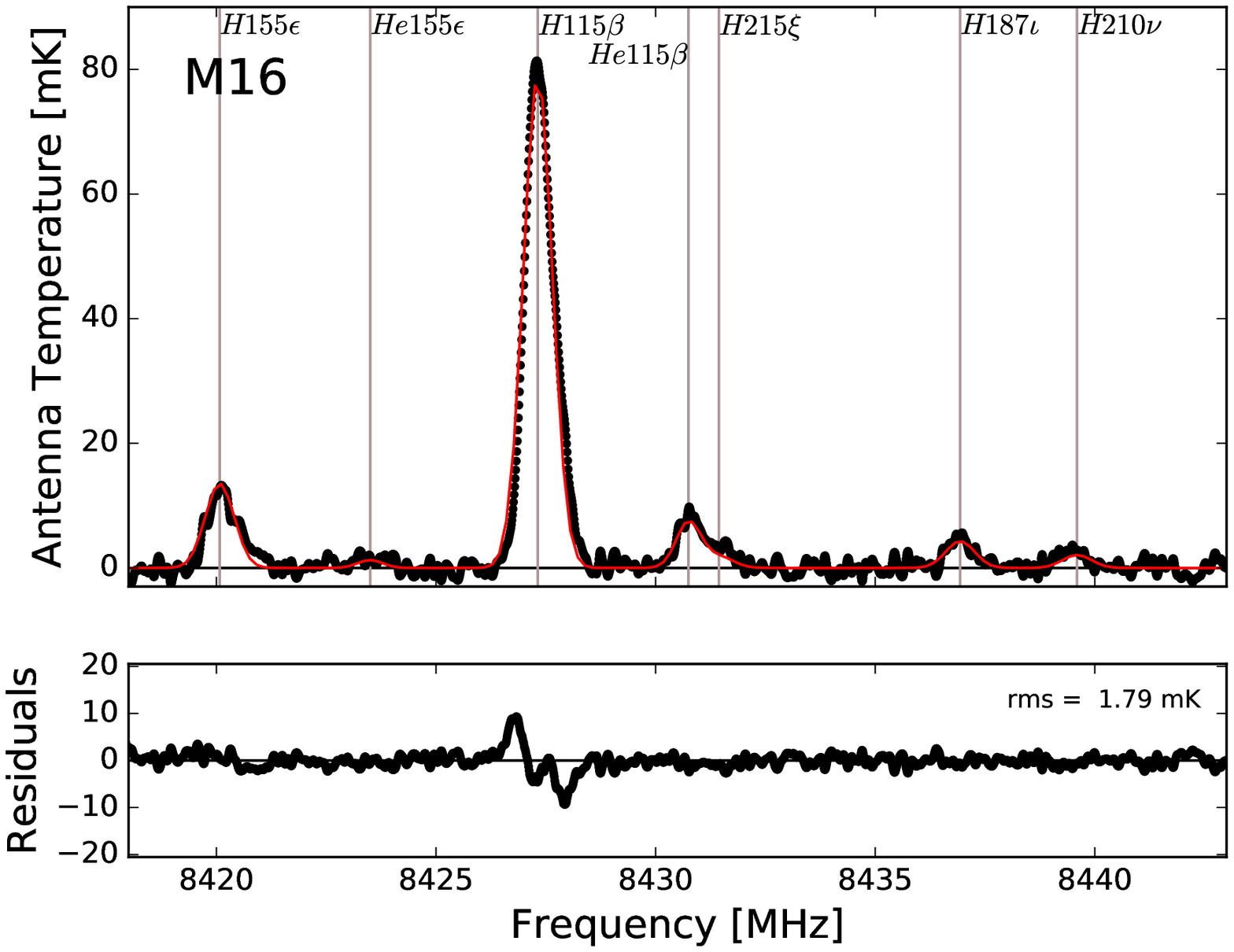} 
\caption{RRL spectra of M16 including the following sub-bands:
  H91$\alpha$ (top-left), expanded view of H91$\alpha$ (top-right),
  H114$\beta$ (bottom-left), and H115$\beta$ (bottom-right).  See
Figure~\ref{fig:s206_rrl1} for details.}
\label{fig:m16_rrl1}
\end{figure}

\begin{figure}
\includegraphics[angle=0,scale=0.45]{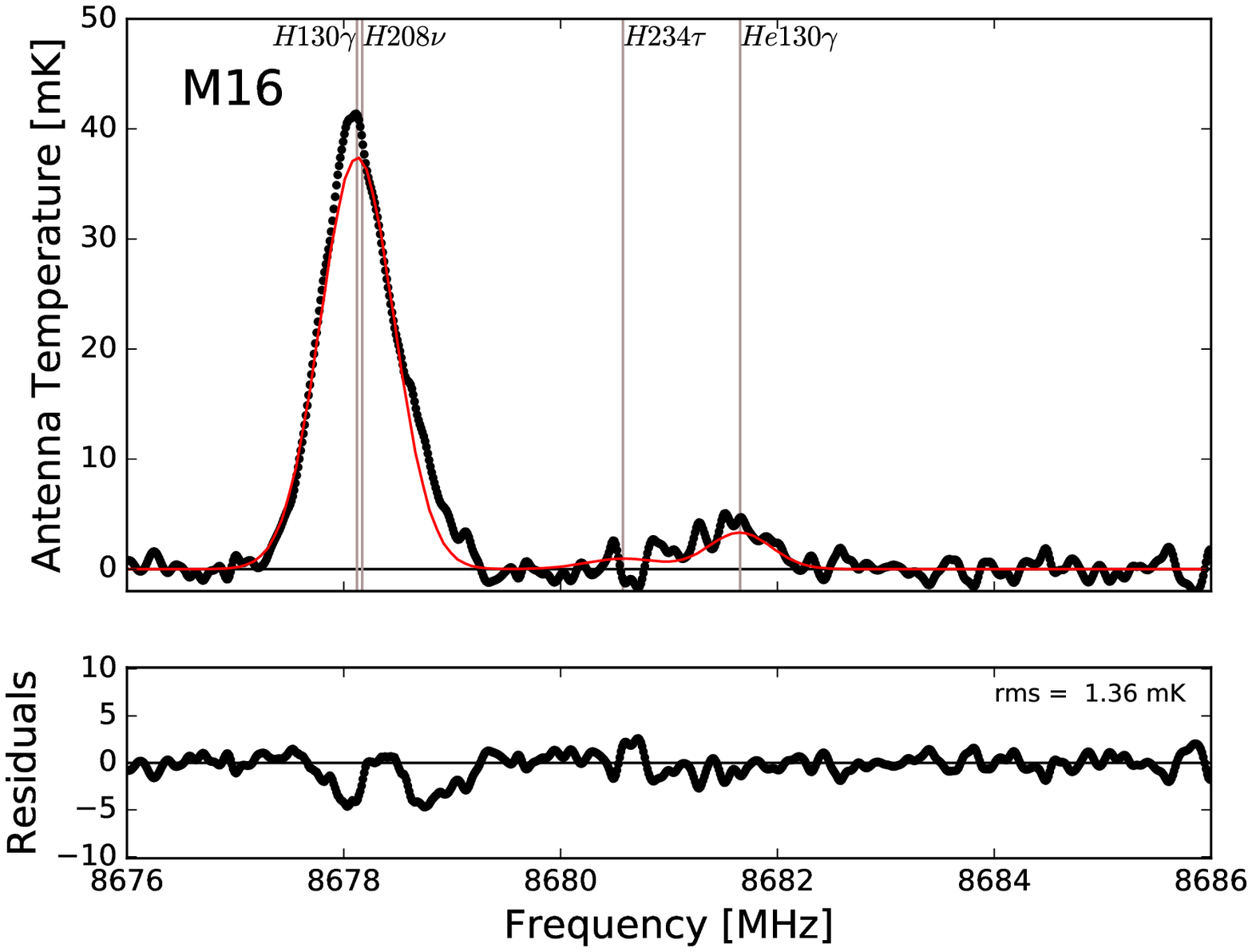} 
\includegraphics[angle=0,scale=0.45]{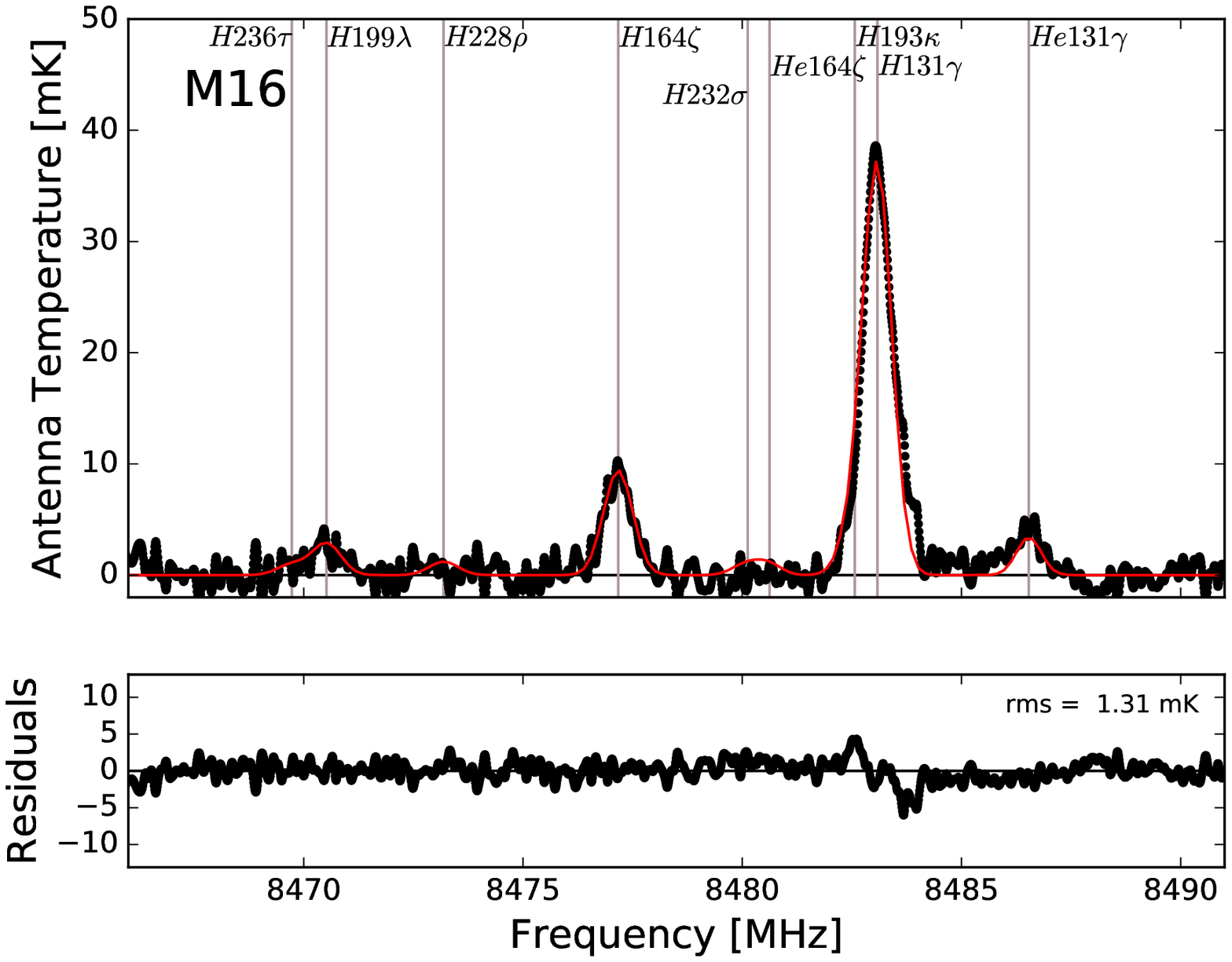} 
\includegraphics[angle=0,scale=0.45]{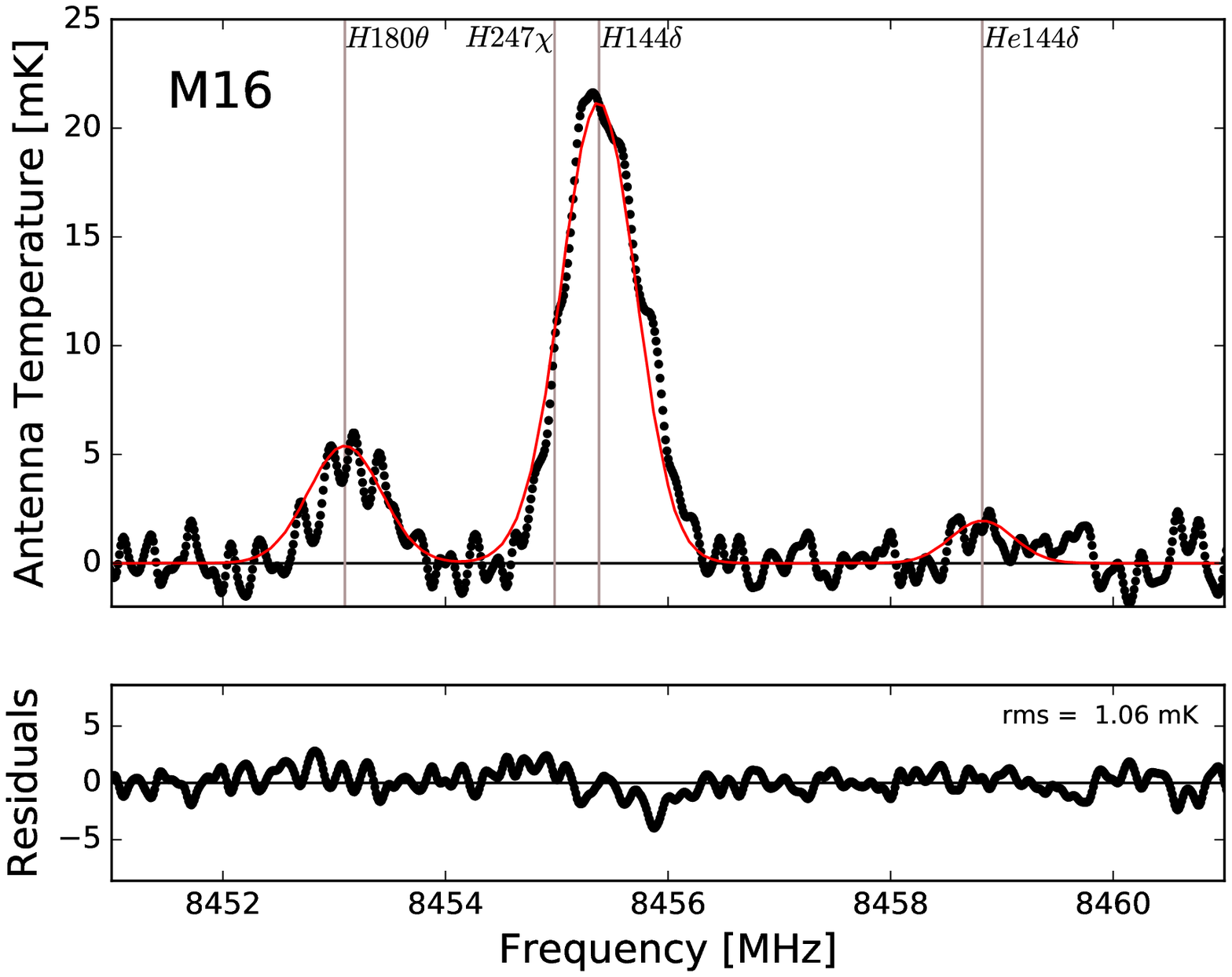} 
\includegraphics[angle=0,scale=0.45]{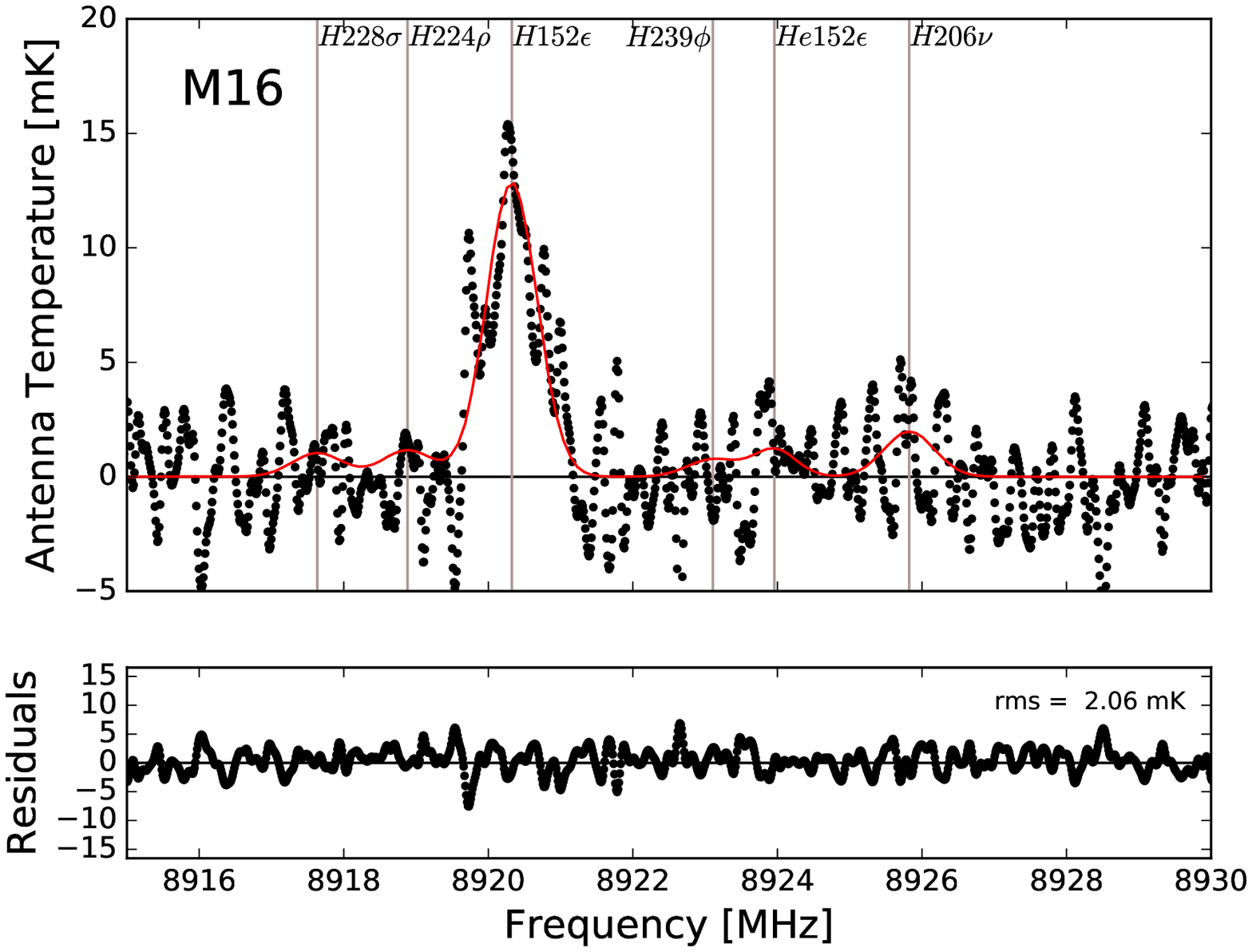} 
\caption{RRL spectra for M16 including the following sub-bands:
  H130$\gamma$ (top-left), H131$\gamma$ (top-right), H144$\delta$
  (bottom-left), and H152$\epsilon$ (bottom-right).  See
  Figure~\ref{fig:s206_rrl2} for details.}
\label{fig:m16_rrl2}
\end{figure}

\begin{figure}
\includegraphics[angle=0,scale=0.45]{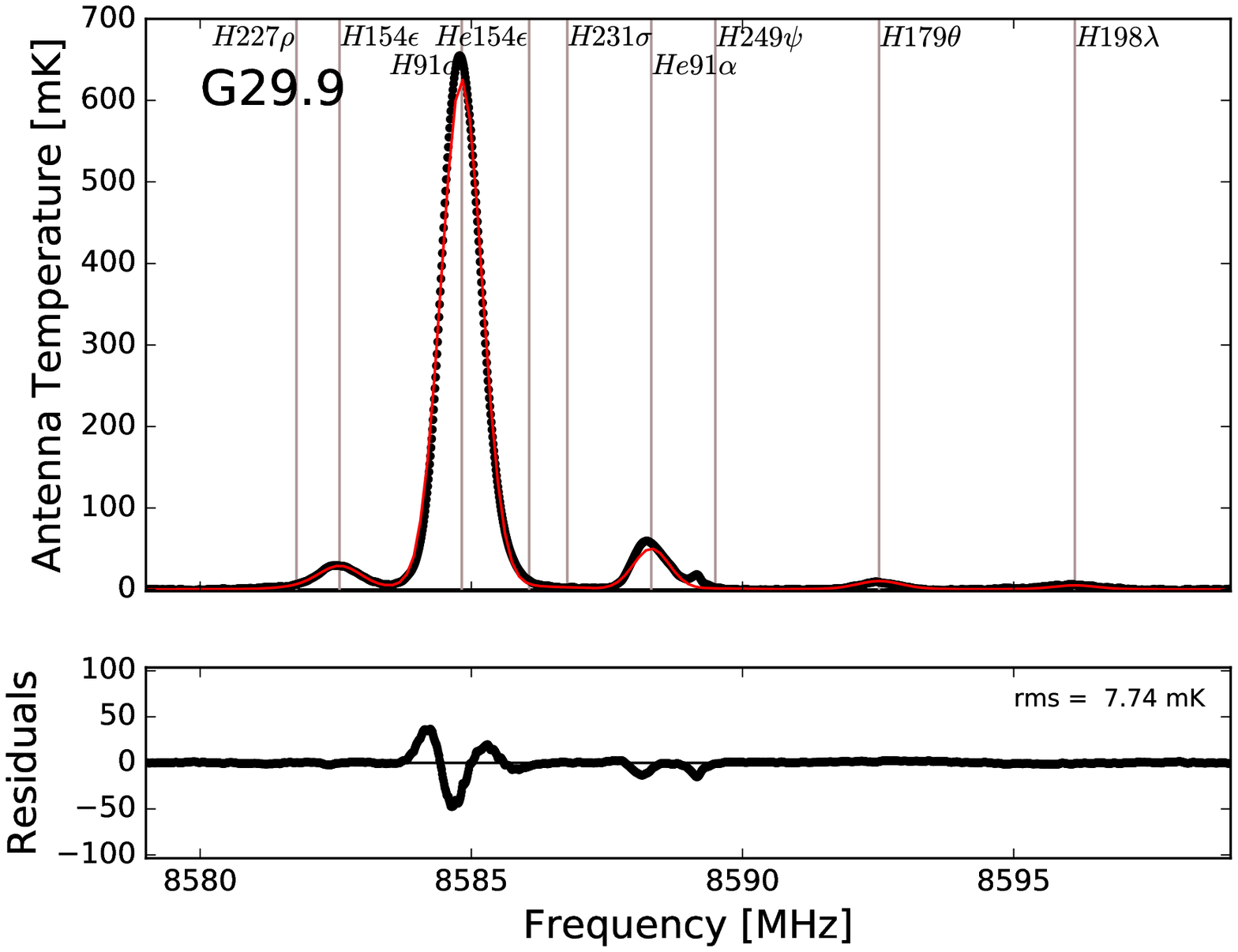} 
\includegraphics[angle=0,scale=0.45]{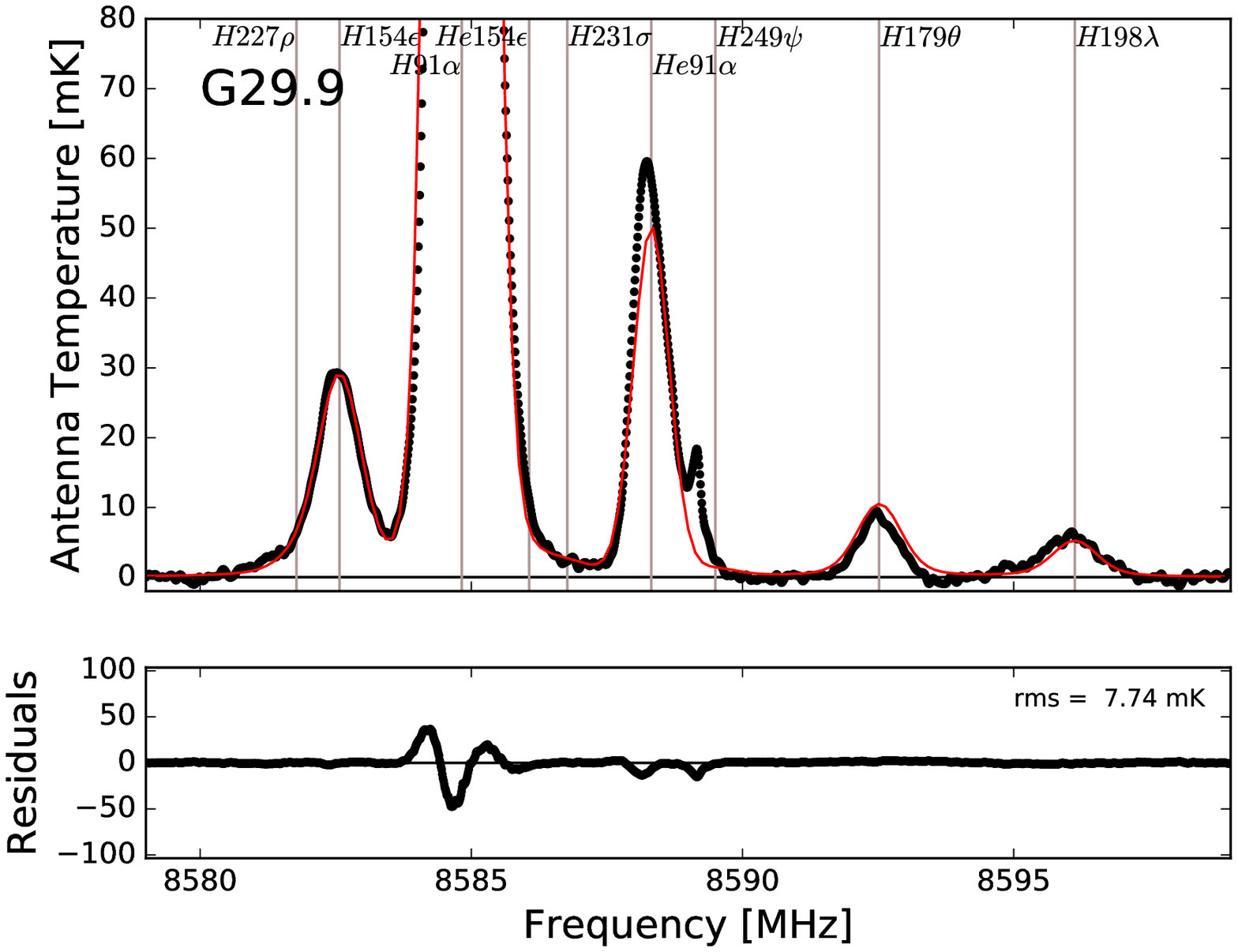} 
\includegraphics[angle=0,scale=0.45]{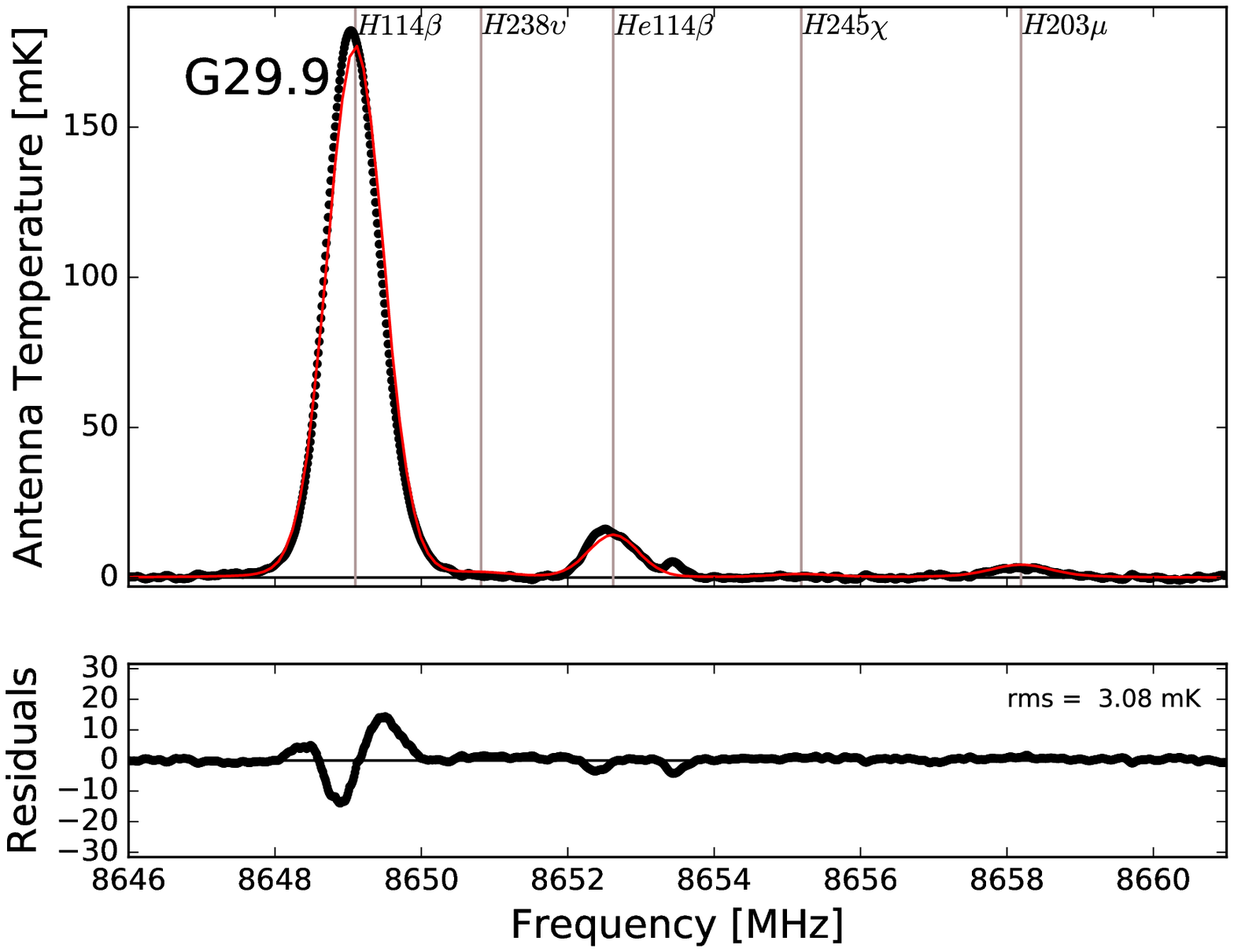} 
\includegraphics[angle=0,scale=0.45]{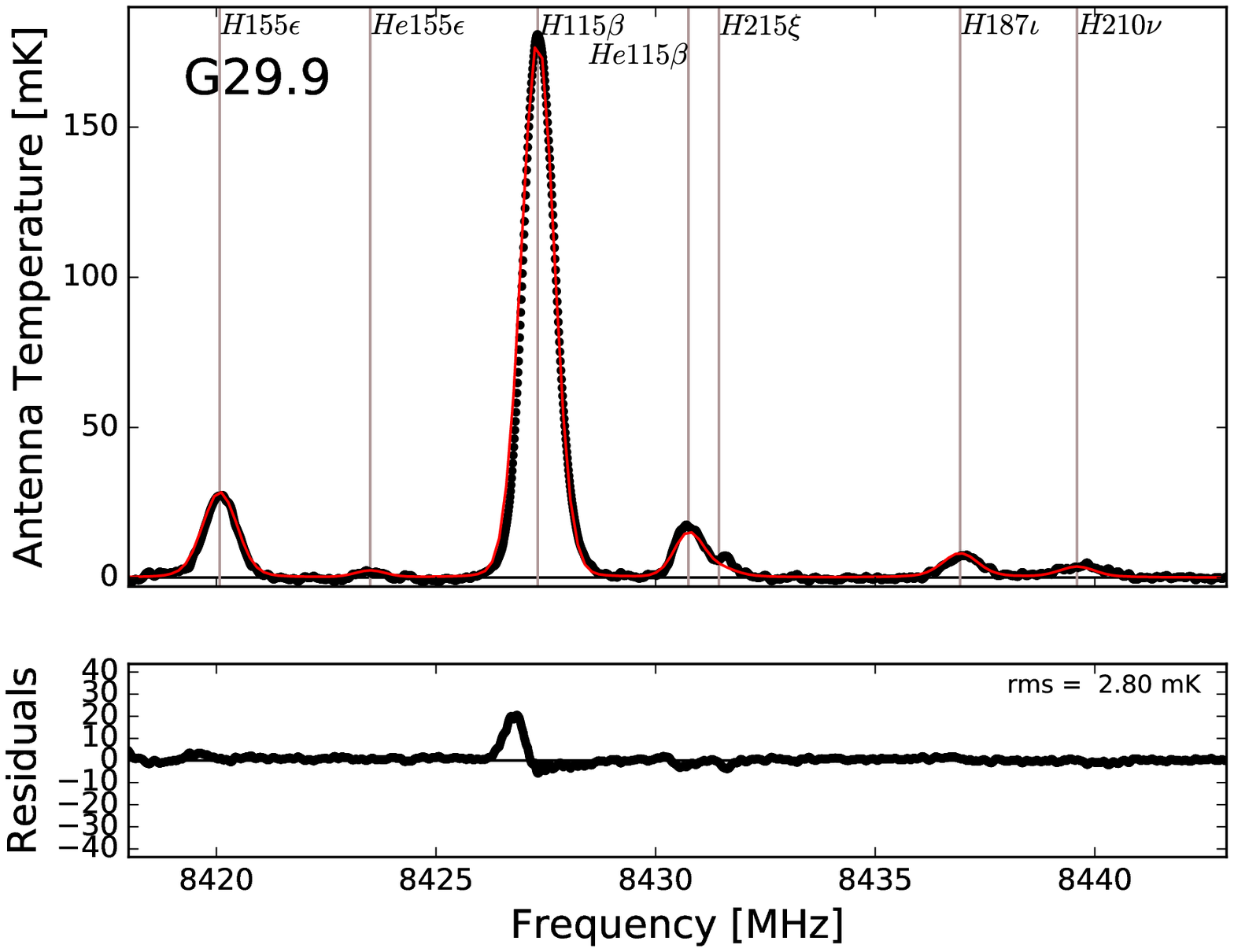} 
\caption{RRL spectra of G29.9 including the following sub-bands:
  H91$\alpha$ (top-left), expanded view of H91$\alpha$ (top-right),
  H114$\beta$ (bottom-left), and H115$\beta$ (bottom-right).  See
Figure~\ref{fig:s206_rrl1} for details.}
\label{fig:g29.9_rrl1}
\end{figure}

\begin{figure}
\includegraphics[angle=0,scale=0.45]{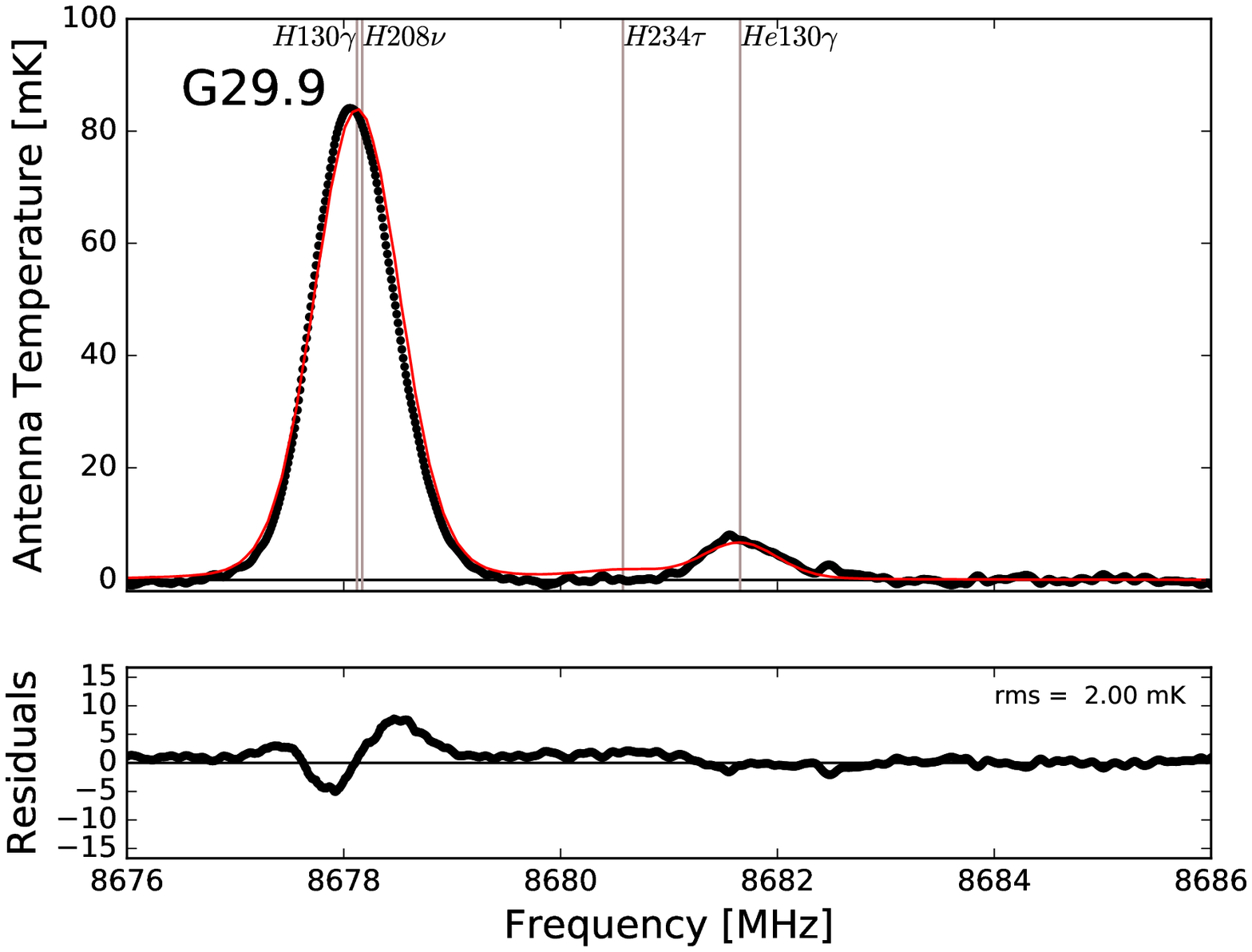} 
\includegraphics[angle=0,scale=0.45]{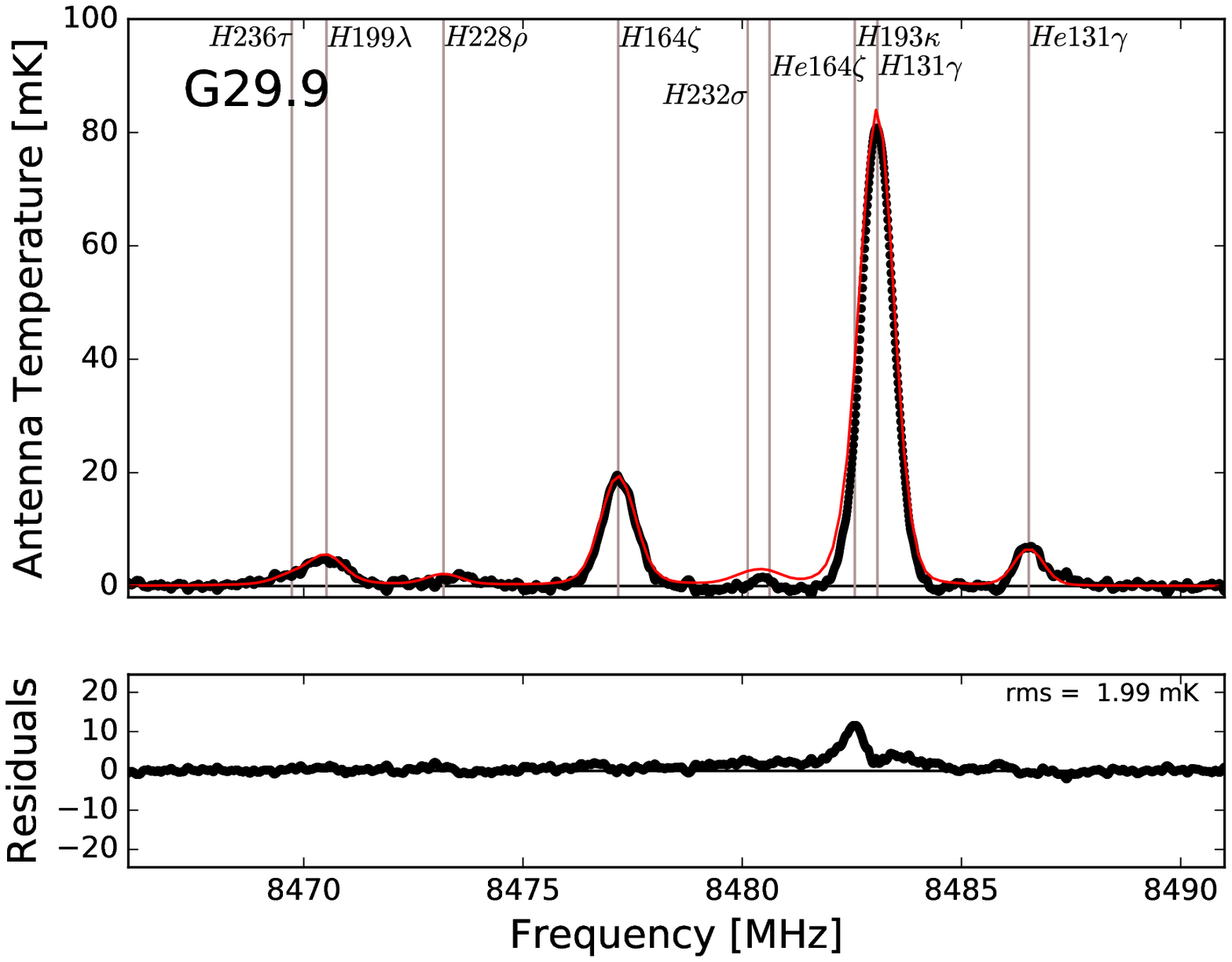} 
\includegraphics[angle=0,scale=0.45]{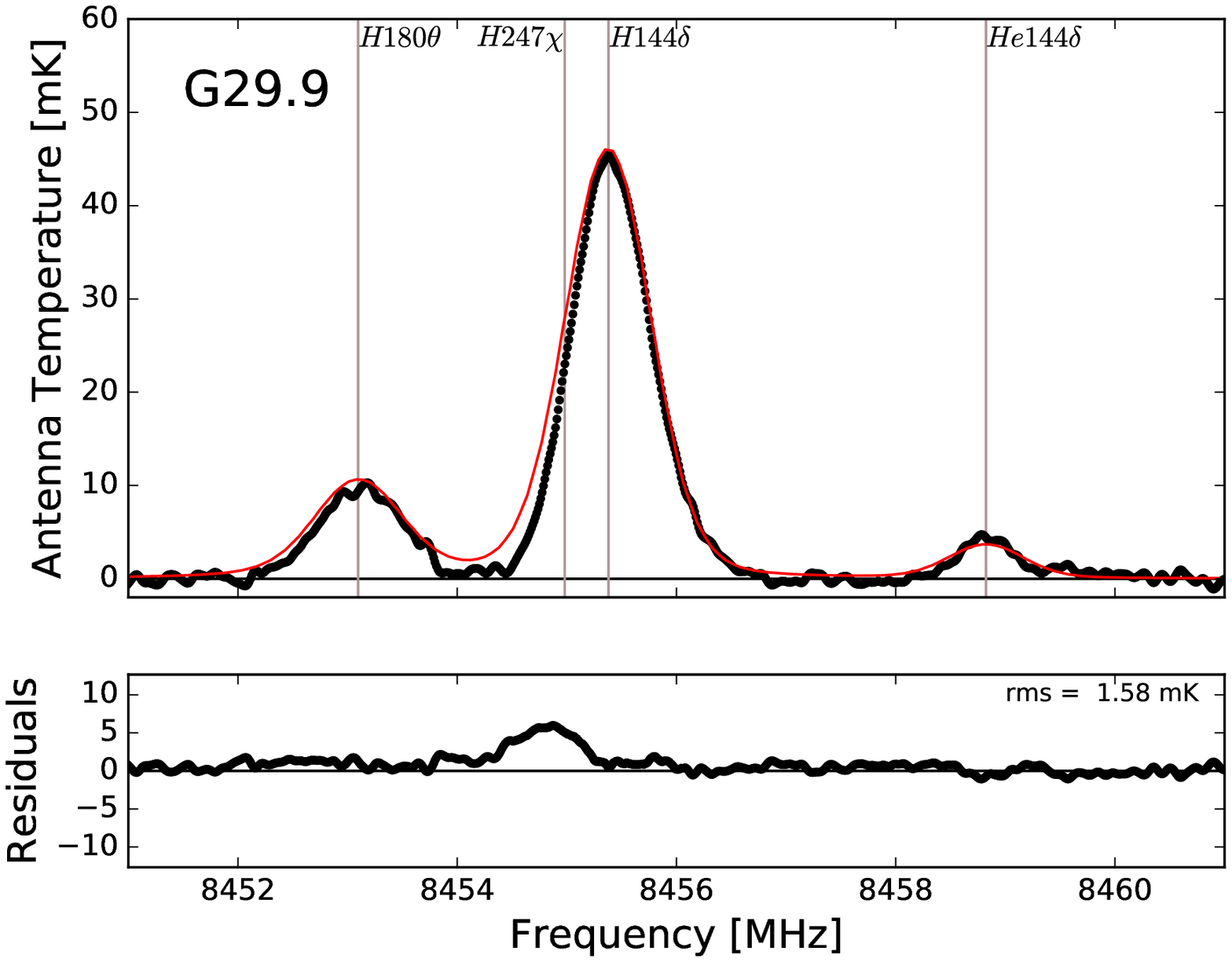} 
\includegraphics[angle=0,scale=0.45]{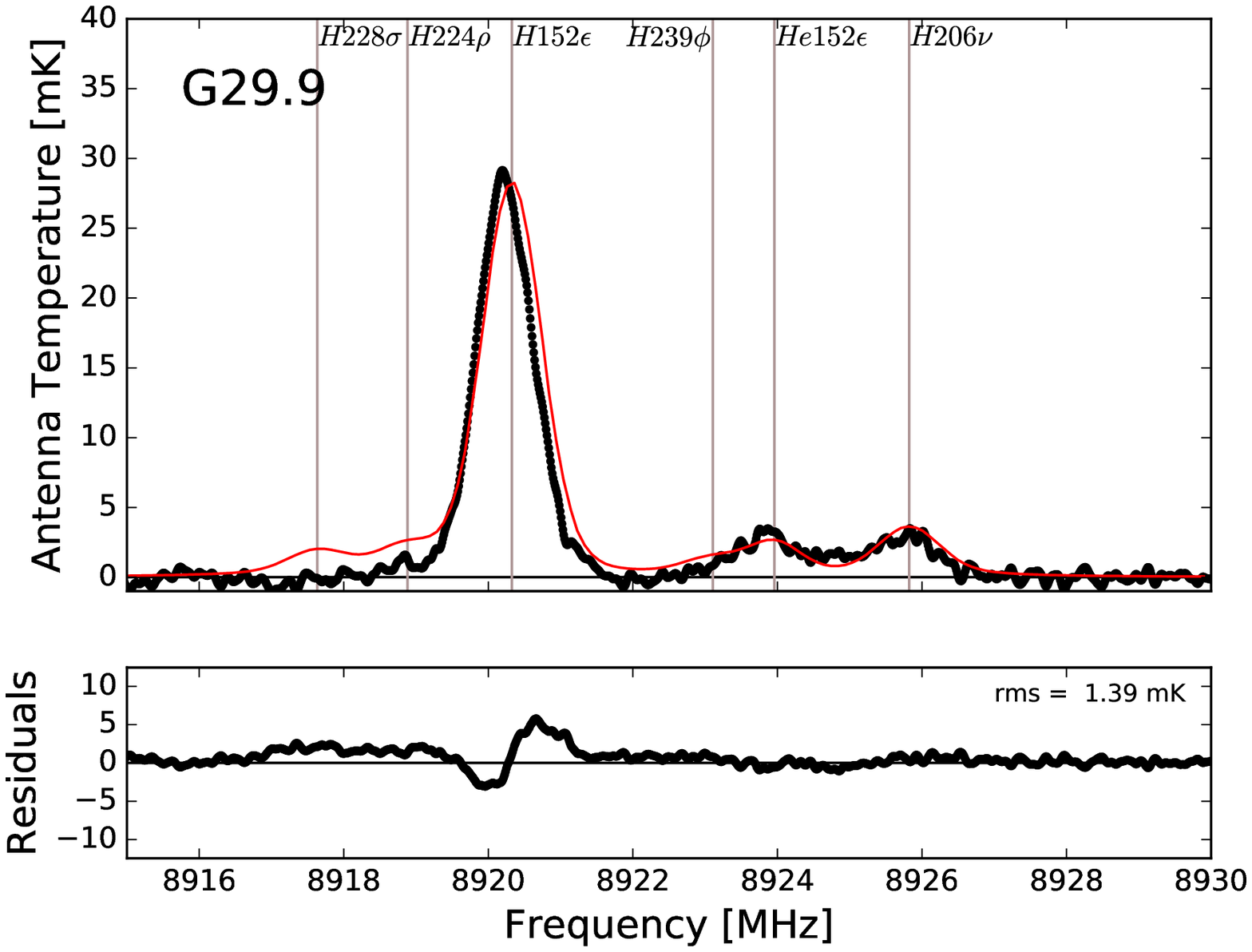} 
\caption{RRL spectra for G29.9 including the following sub-bands:
  H130$\gamma$ (top-left), H131$\gamma$ (top-right), H144$\delta$
  (bottom-left), and H152$\epsilon$ (bottom-right).  See
  Figure~\ref{fig:s206_rrl2} for details.}
\label{fig:g29.9_rrl2}
\end{figure}

\begin{figure}
\includegraphics[angle=0,scale=0.45]{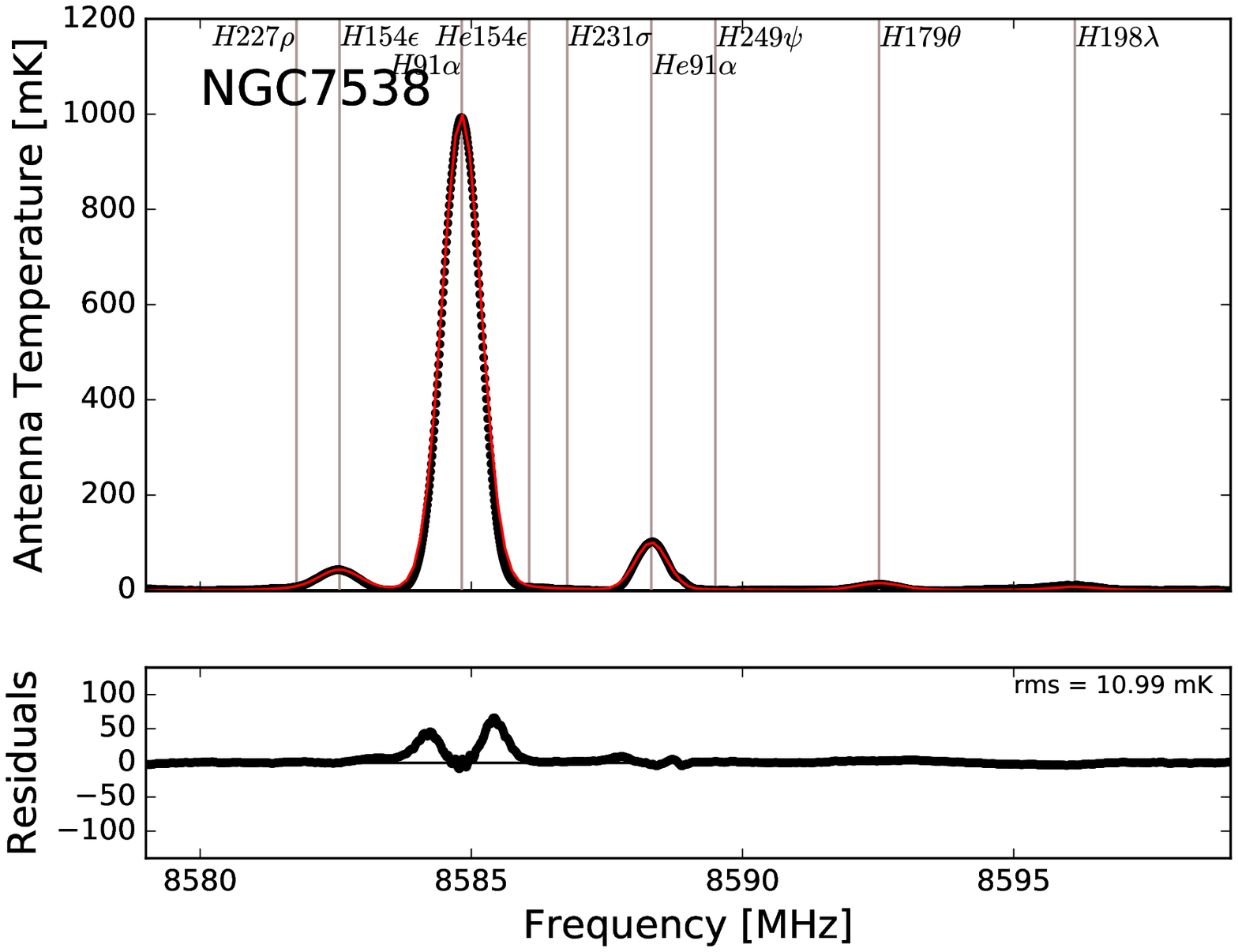} 
\includegraphics[angle=0,scale=0.45]{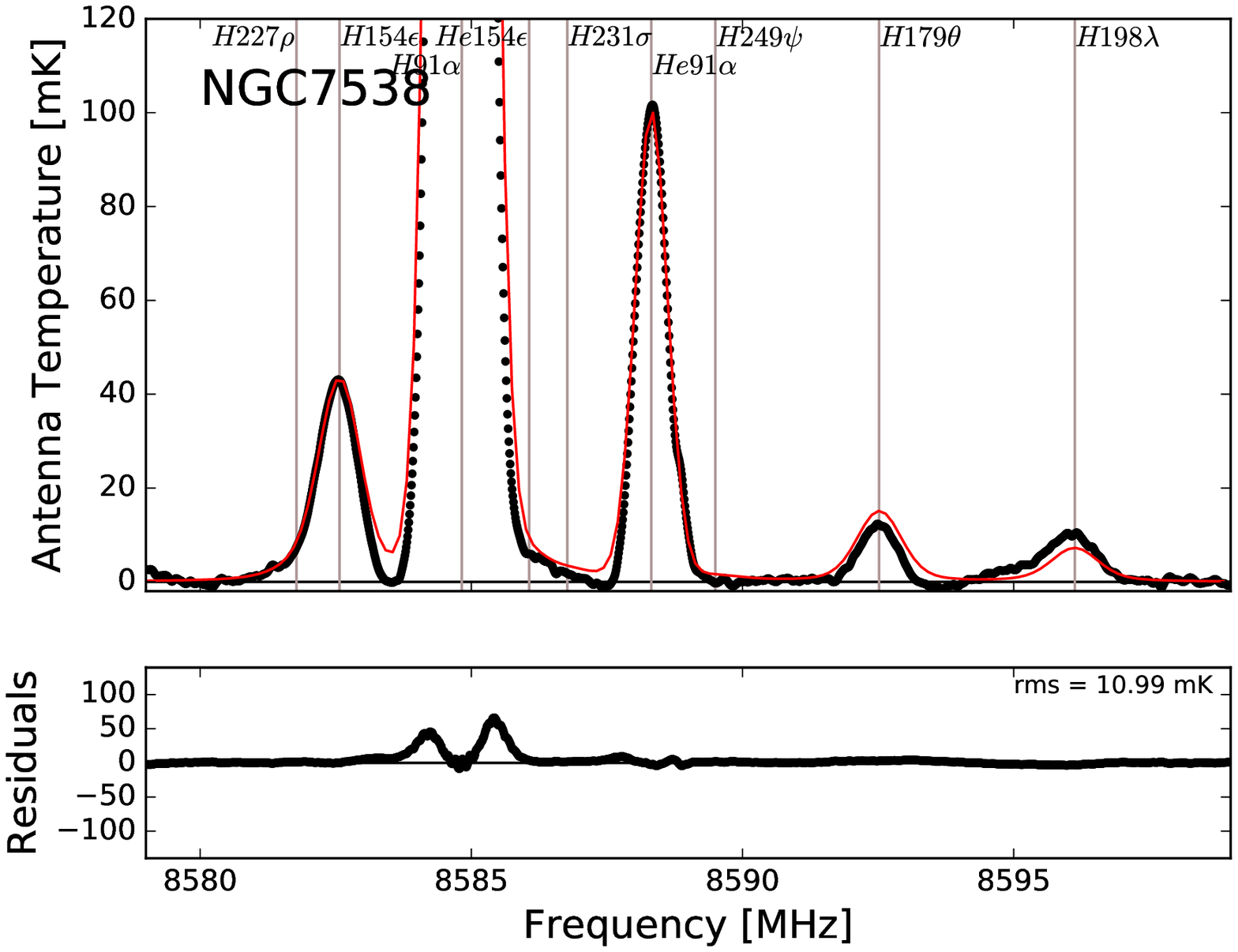} 
\includegraphics[angle=0,scale=0.45]{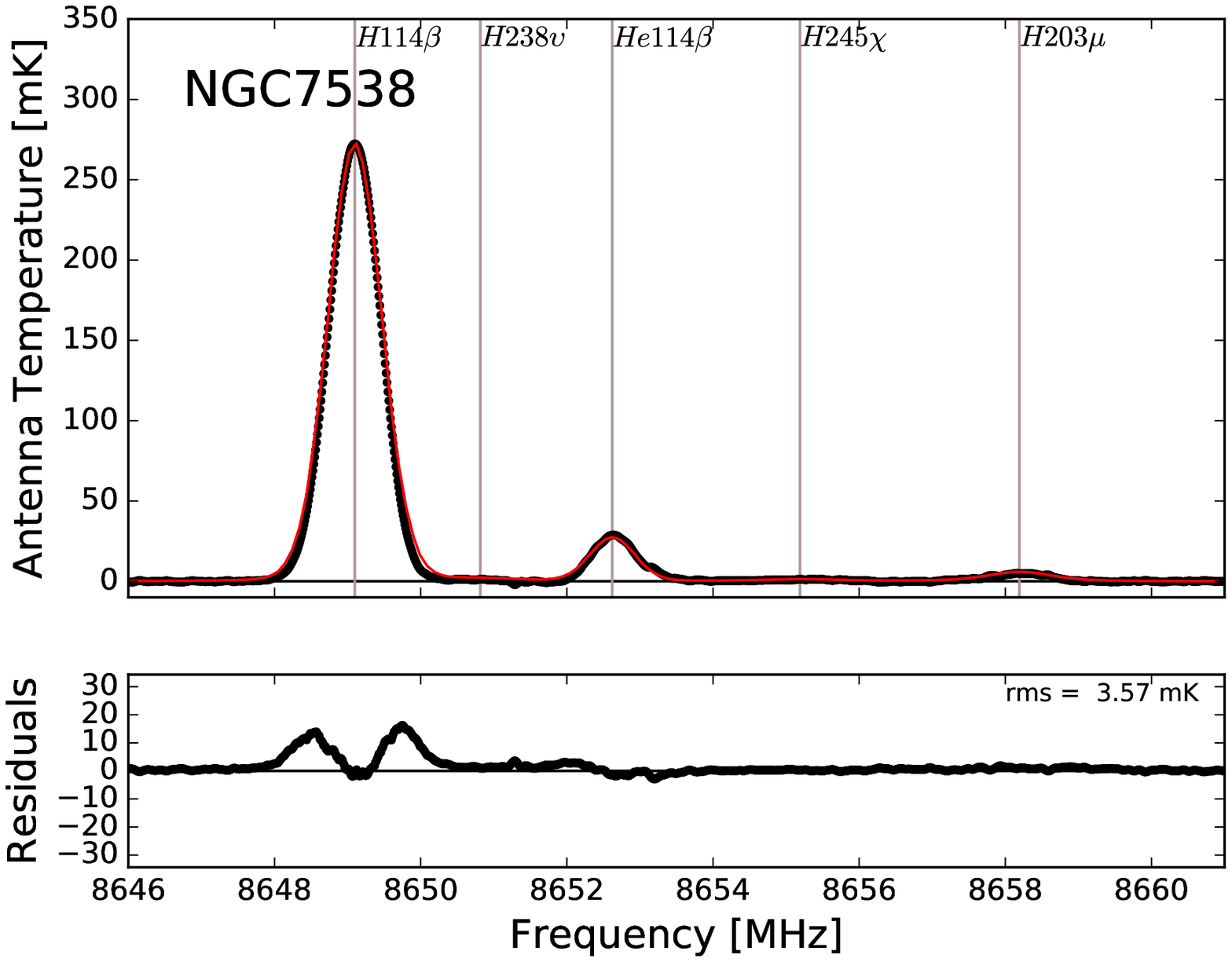} 
\includegraphics[angle=0,scale=0.45]{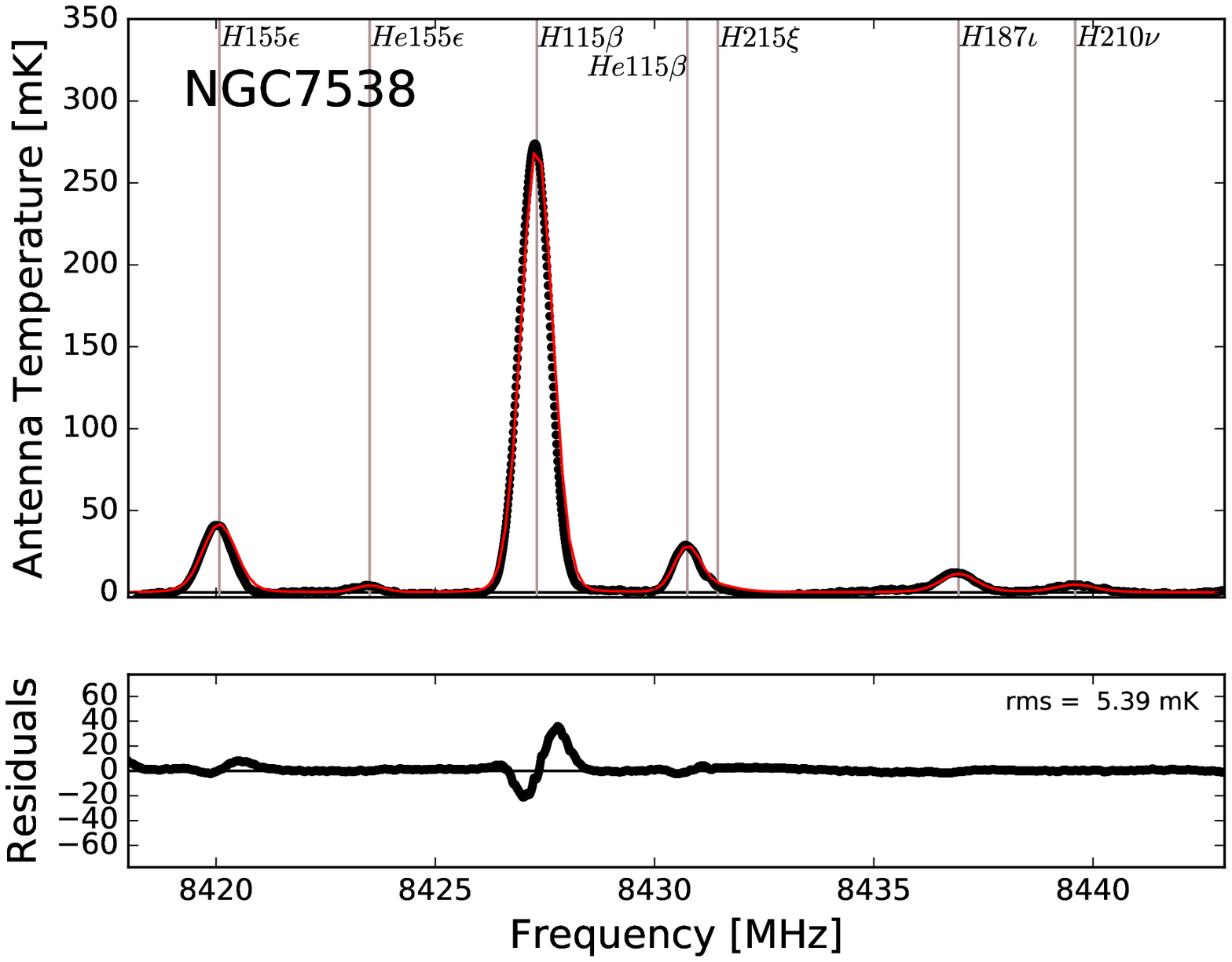} 
\caption{RRL spectra of NGC7538 including the following sub-bands:
  H91$\alpha$ (top-left), expanded view of H91$\alpha$ (top-right),
  H114$\beta$ (bottom-left), and H115$\beta$ (bottom-right).  See
Figure~\ref{fig:s206_rrl1} for details.}
\label{fig:ngc7538_rrl1}
\end{figure}

\begin{figure}
\includegraphics[angle=0,scale=0.45]{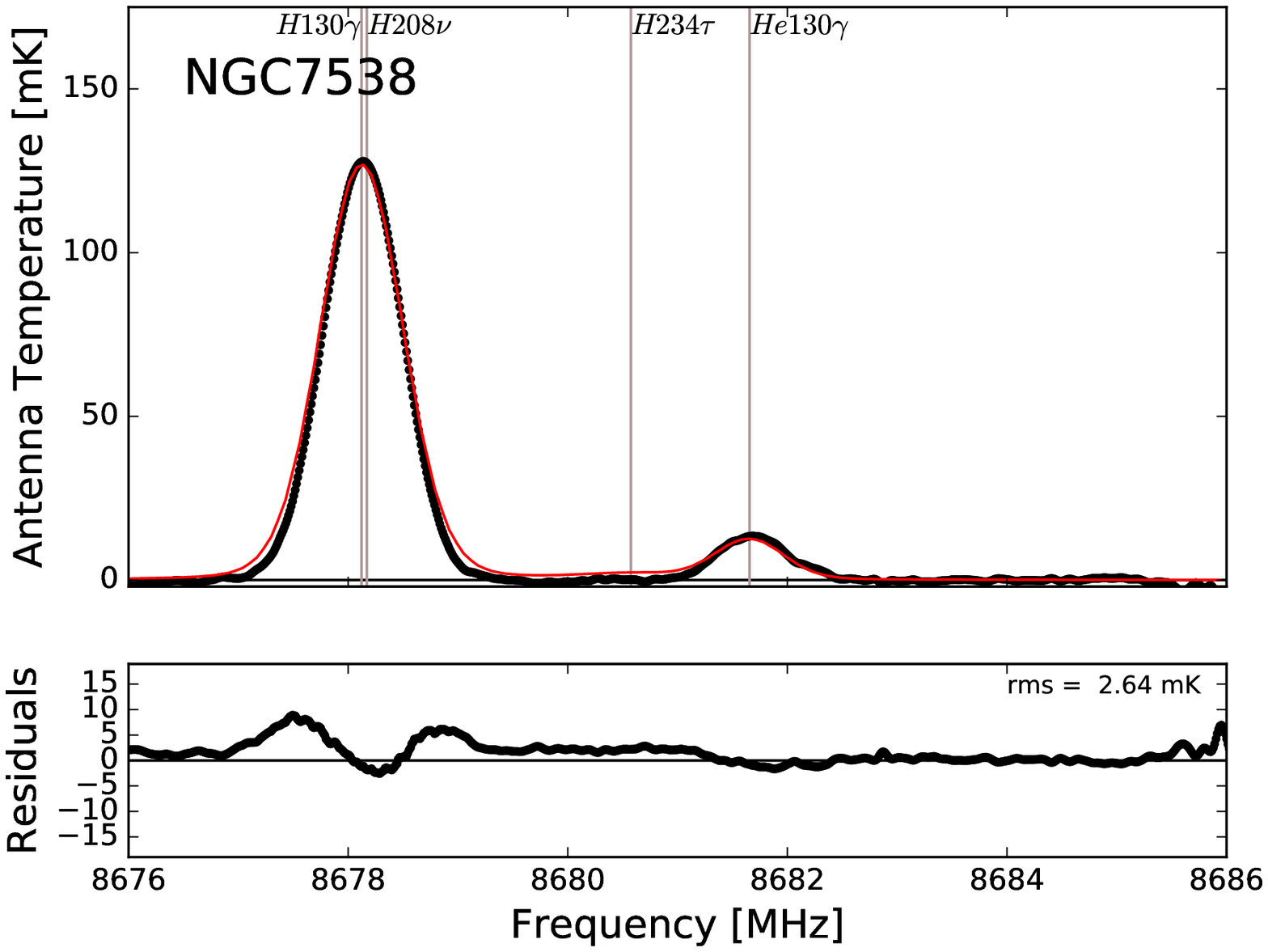} 
\includegraphics[angle=0,scale=0.45]{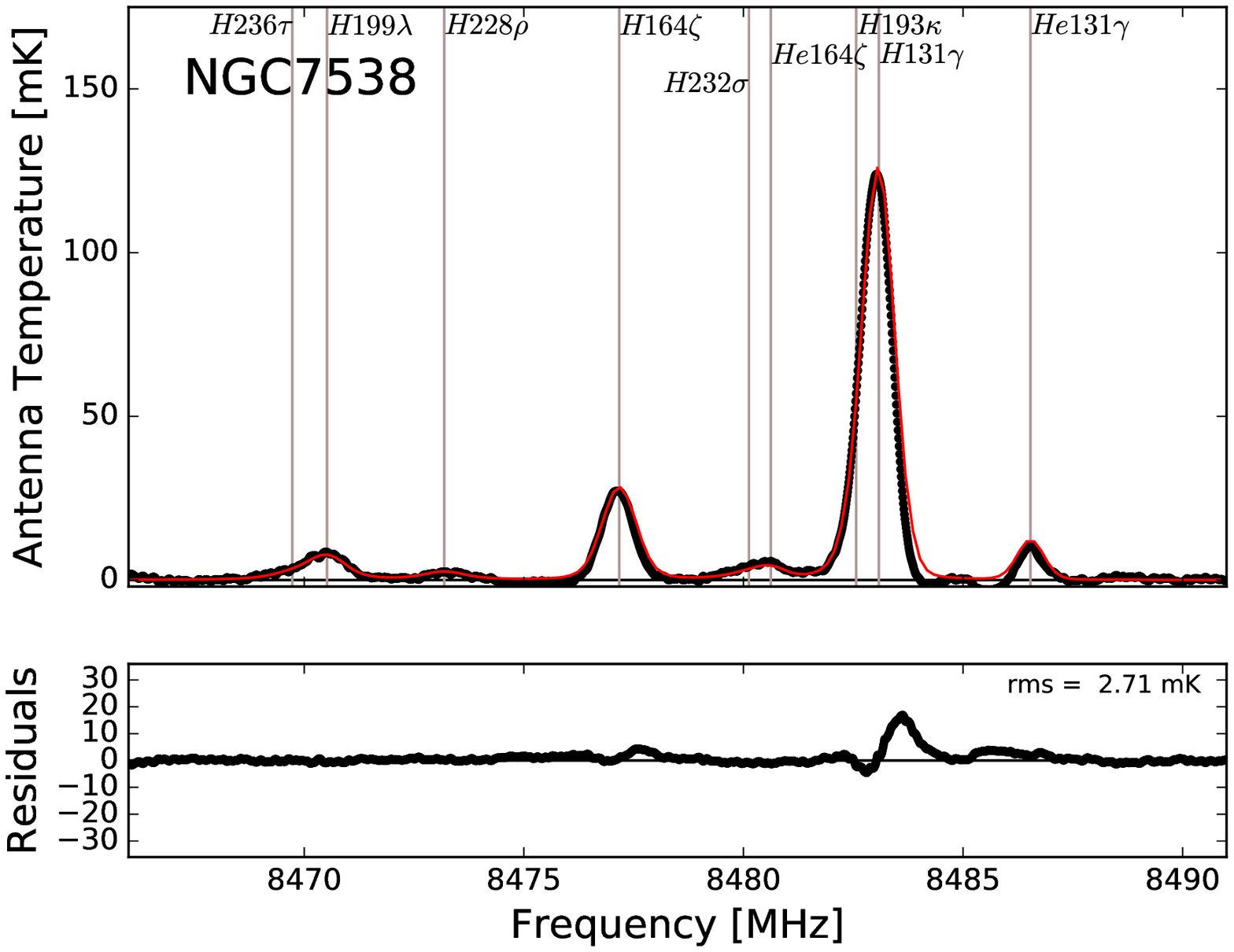} 
\includegraphics[angle=0,scale=0.45]{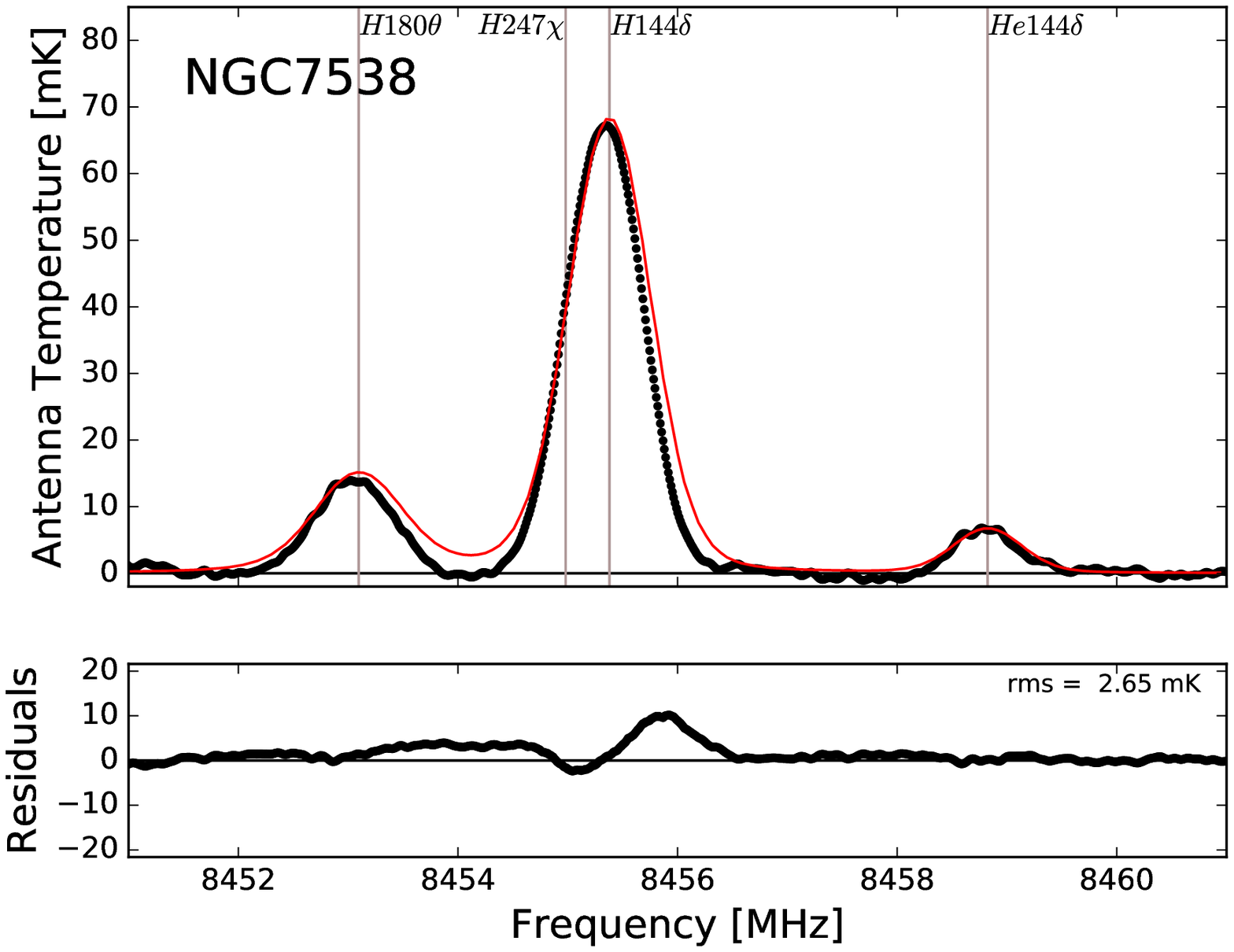} 
\includegraphics[angle=0,scale=0.45]{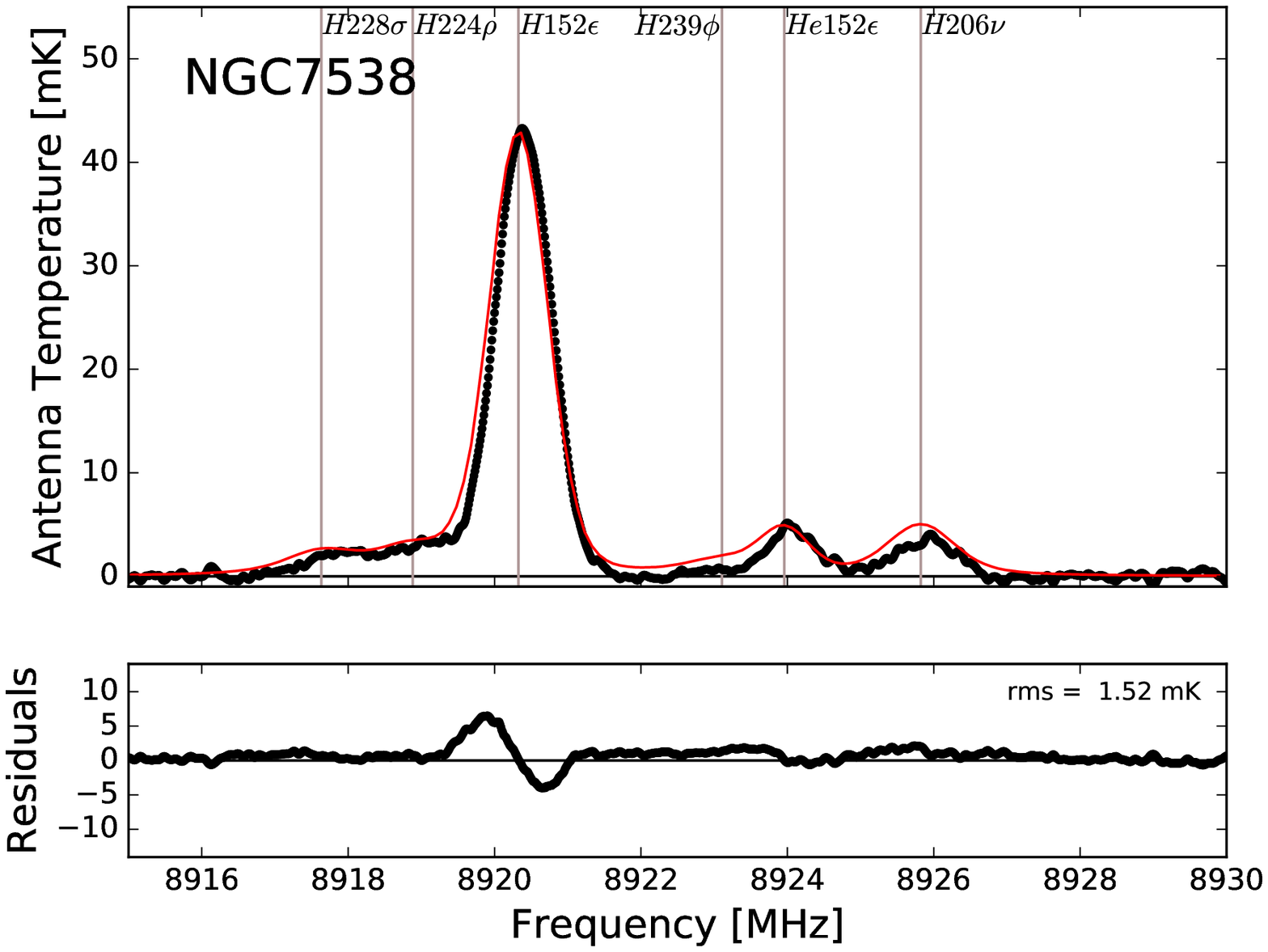} 
\caption{RRL spectra for NGC7538 including the following sub-bands:
  H130$\gamma$ (top-left), H131$\gamma$ (top-right), H144$\delta$
  (bottom-left), and H152$\epsilon$ (bottom-right).  See
  Figure~\ref{fig:s206_rrl2} for details.}
\label{fig:ngc7538_rrl2}
\end{figure}

For completeness we include RRL spectra for S209
(Figures~\ref{fig:s209_rrl1}-\ref{fig:s209_rrl2}), M16
(Figures~\ref{fig:m16_rrl1}-\ref{fig:m16_rrl2}), G29.9
(Figures~\ref{fig:g29.9_rrl1}-\ref{fig:g29.9_rrl2}), and \ngc{7538}
(Figures~\ref{fig:ngc7538_rrl1}-\ref{fig:ngc7538_rrl2}).  Plotted in
the top panel of each figure is the antenna temperature as a function
of rest frequency.  The black points display the observed spectrum and
the red curve is the NEBULA model.  The vertical lines mark the
location of various RRL transitions.  The residuals of the model and
data are shown in the bottom panel of each figure.

{}

\end{document}